\documentclass[aps,prd,superscriptaddress,floatfix,nofootinbib,notitlepage,12pt]{revtex4-2}

\usepackage{graphicx}
\usepackage{multirow}
\usepackage{slashed}
\usepackage{comment}
\usepackage{amsmath,amssymb,amsfonts}
\usepackage{verbatim}
\usepackage{bm}
\usepackage{mathtools}
\usepackage{colortbl}
\usepackage[dvipsnames,table]{xcolor}
\usepackage{braket}
\usepackage{adjustbox}
\usepackage{cancel}
\usepackage{makecell}
\usepackage{soul}
\usepackage{ulem}
\usepackage{enumitem}

\usepackage{tikz}
\usetikzlibrary{decorations.pathmorphing,decorations.markings,arrows.meta}
\tikzset{
 scalar/.style={dashed,line width=.55pt},
charged/.style={dashed,line width=.55pt,postaction={decorate},
decoration={markings,mark=at position .63 with {
\arrow{Latex[length=2.2mm,width=1.4mm]}}}},
chargedrev/.style={dashed,line width=.55pt,
postaction={decorate},
decoration={markings,mark=at position .63 with {
\arrowreversed{Latex[length=2.2mm,width=1.4mm]}}}},
vector/.style={decorate,decoration={snake,amplitude=1.1pt,
segment length=4pt},line width=.55pt},
vertex/.style={circle,fill=black,inner sep=1.35pt}}
\newcommand{\threeS}[5]{%
\begin{tikzpicture}[baseline=(current bounding box.center),scale=.9]
  \node[vertex] (v) at (0,0) {};
  \draw[scalar] (-1.45,0)--(v)
    node[midway,below=2pt] {$#1$};
  \draw[#2] (v)--(1.25,.75)
    node[midway,above=2pt] {$#3$};
  \draw[#4] (v)--(1.25,-.75)
    node[midway,below=2pt] {$#5$};
\end{tikzpicture}%
}

\newcommand{\ssv}[5]{%
\begin{tikzpicture}[baseline=(current bounding box.center),scale=.88]
  \node[vertex] (v) at (0,0) {};
  \draw[#4] (-1.45,0)--(v)
    node[midway,below=2pt] {$#1$};
  \draw[#5] (v)--(1.16,.72)
    node[midway,above=2pt] {$#2$};
  \draw[vector] (v)--(1.17,-.72)
    node[midway,below=2pt] {$#3$};
\end{tikzpicture}%
}
\newcommand{\ssvv}[6]{%
\begin{tikzpicture}[baseline=(current bounding box.center),scale=.86]
  \node[vertex] (v) at (0,0) {};
  \draw[#5] (-1.25,.72)--(v)
    node[midway,above=2pt] {$#1$};
  \draw[#6] (-1.25,-.72)--(v)
    node[midway,below=2pt] {$#2$};
  \draw[vector] (v)--(1.22,.72)
    node[midway,above=2pt] {$#3$};
  \draw[vector] (v)--(1.22,-.72)
    node[midway,below=2pt] {$#4$};
\end{tikzpicture}%
}

\newcommand{\vvv}[3]{%
\begin{tikzpicture}[baseline=(current bounding box.center),scale=.88]
\node[vertex] (v) at (0,0) {};
\draw[vector] (-1.45,0)--(v)
node[midway,below=2pt] {$#1$};
\draw[vector] (v)--(1.18,.72)
node[midway,above=2pt] {$#2$};
\draw[vector] (v)--(1.18,-.72)
node[midway,below=2pt] {$#3$};
\end{tikzpicture}%
}

\usepackage[colorlinks]{hyperref}
\hypersetup{
	colorlinks=true,
	linkcolor=blue,
	filecolor=blue,
	urlcolor=blue,
	citecolor=magenta,
	pdfpagemode=FullScreen
}

\newcommand{\orcid}[1]{\hspace{1mm}\href{https://orcid.org/#1}{%
  \includegraphics[height=0.3cm,keepaspectratio]{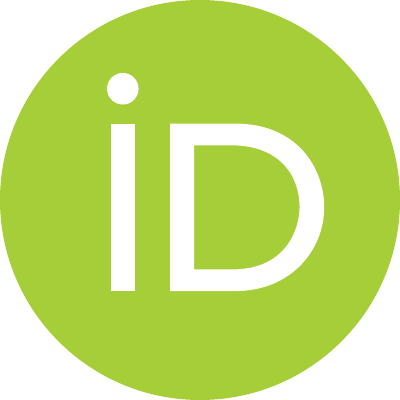}}}

\definecolor{deepgreen}{rgb}{0.2,0.8,0.2}
\definecolor{deepblue}{rgb}{0.2,0.4,0.8}
\definecolor{deepred}{rgb}{0.8,0.2,0.2}

\newcommand{\gev}{\text{GeV}\,}
\newcommand{\ttop}      {{t\bar{t}}}

\newcommand{\br}{\text{Br}}
\newcommand{\tev}{\text{TeV}}

\newcommand{\bea}{\begin{eqnarray}}
\newcommand{\eea}{\end{eqnarray}}

\usepackage[
left=1.8cm,
right=1.8cm,
top=2.2cm,
bottom=2.2cm
]{geometry}

\begin{document}

\title{Probing the Inert Scalar Sector of the Inert Doublet Model via Vector-Boson Fusion at a Muon Collider}

\author{Abdesslam Arhrib\orcid{0000-0001-5619-7189}}
\email{aarhrib@uae.ac.ma}
\affiliation{FST, Abdelmalek Essaadi University, B.P.\ 416, Tangier, Morocco}

\author{Es-said Ghourmin\orcid{0009-0007-1597-8537}}
\email{s.ghourmin123@gmail.com}
\affiliation{Laboratory of Theoretical and High Energy Physics (LPTHE), Faculty of Sciences, Ibnou Zohr University, B.P.\ 8106, Agadir, Morocco}

\author{Ayoub Hmissou\orcid{0000-0002-2548-3403}}
\email{ayoub1hmissou@gmail.com}
\affiliation{Laboratory of Theoretical and High Energy Physics (LPTHE), Faculty of Sciences, Ibnou Zohr University, B.P.\ 8106, Agadir, Morocco}

\author{Larbi Rahili\orcid{0000-0002-1164-1095}}
\email{rahililarbi@gmail.com}
\affiliation{Laboratory of Theoretical and High Energy Physics (LPTHE), Faculty of Sciences, Ibnou Zohr University, B.P.\ 8106, Agadir, Morocco}

\author{Bassim Taki\orcid{0009-0009-2642-1288}}
\email{taki.bassim@edu.uiz.ac.ma}
\affiliation{Laboratory of Theoretical and High Energy Physics (LPTHE), Faculty of Sciences, Ibnou Zohr University, B.P.\ 8106, Agadir, Morocco}

\date{\today}

\begin{abstract}
We investigate how a future high-energy muon collider (MuC) probe the inert scalar sector of the Inert Doublet Model (IDM) through a comprehensive set of $2\to4$ electroweak gauge-boson fusion processes. The scalar spectrum consists of two $CP$-even neutral scalars, $h$ and $H$, one $CP$-odd neutral scalar $A$, and a pair of charged scalars $H^\pm$. We assume the neutral scalar $H$ is the lightest inert state and the dark matter (DM) candidate, while we identify the observed $125$~GeV Higgs boson with the scalar from the Standard-Model-like doublet. We consider charged Higgs pair production, $\mu^+\mu^-\to H^+H^-\nu\bar{\nu}$, induced by $WW$ fusion, as well as the associated channels $\mu^+\mu^-\to HH^\pm\mu^\mp\nu$ and $\mu^+\mu^-\to AH^\pm\mu^\mp\nu$, induced by  $WZ$ and $W\gamma$ fusion.
Owing to their Vector Boson Fusion (VBF) origin, the corresponding cross sections increase significantly with collider energy, making them particularly well suited for exploration at a high-energy MuC. We first study  the $2\to 2$ subprocesses $WW\to H^+ H^-$ and $WV\to  H^+ S_i$, with $V=Z, \gamma$ and $S_i=H,A$, and establish the presence of destructive interference among the contributing diagrams. After imposing theoretical consistency requirements and current experimental constraints, including those from DM and collider searches, we evaluate the corresponding signal rates at $\sqrt{s}=3$, $10$, and $20$~TeV. We further perform a detector-level Monte Carlo analysis combined with a cut-based signal-background selection. Our results demonstrate that the sensitivity to inert scalar production grows significantly with collision energy, establishing the $10$ and $20$~TeV stages of a MuC as powerful probes of the IDM scalar sector.
\end{abstract}

\maketitle

\section{Introduction}
The discovery of the Higgs boson by the ATLAS and CMS collaborations at the Large Hadron Collider (LHC) has completed the particle spectrum of the Standard Model (SM) \cite{ATLAS:2012yve,CMS:2012qbp}. 

Nevertheless, the SM cannot account for several observations, most notably the nature of dark matter (DM) \cite{ParticleDataGroup:2024cfk}, among others \cite{Super-Kamiokande:1998kpq,Crivellin:2023zui}. Moreover, it leaves open the possible existence of additional scalar degrees of freedom. From a phenomenological standpoint, extended Higgs sectors remain among the best-motivated frameworks for addressing these issues \cite{Georgi:1985nv,Kanemura:1993hm,Aoki:2009ha,Branco:2011iw,Arhrib:2011uy,Ouazghour:2018mld,Benbrik:2022bol,Robens:2025kaa,Hmissou:2025uep}, while being subject to stringent flavour constraints \cite{Misiak:2020vlo} and offering possible explanations for some experimental anomalies \cite{ALEPH:2006tnd,CMS:2021yci,CMS:2022goy,Biekotter:2023oen,ATLAS:2024bjr,Benbrik:2025hol,Benbrik:2025wkz}. 

In parallel with the ongoing LHC experiments, future collider initiatives aim to significantly extend both the energy and precision frontiers. In this landscape, future muon colliders (MuCs) \cite{AlAli:2021let,InternationalMuonCollider:2025sys} have emerged as a particularly attractive option: they combine the advantages of a lepton collider environment with the potential to reach multi-TeV center-of-mass energies \cite{Delahaye:2019omf,Han:2020uid,Long:2020wfp}. At such energies, electroweak radiation off the initial-state muons becomes increasingly important, enhancing production mechanisms driven by Vector Boson Fusion (VBF) topologies \cite{Dawson:1984gx,Kane:1984bb,Kunszt:1987tk,Chanowitz:1984ne,Gunion:1986gm,Chen:2016wkt,Buttazzo:2018qqp,Ruiz:2021tdt,Costantini:2020stv,Han:2020uid,Garosi:2023bvq,Denner:2024yut}. This feature has stimulated extensive studies of the MuC sensitivity to electroweak-scale new physics (NP), in particular highlighting the prospects for probing electroweak DM and, more broadly, discovering heavy Beyond-the-SM (BSM) states, including dedicated analyses of additional neutral Higgs bosons and charged scalar signatures \cite{Braathen:2024ckk,Han:2020uak,Han:2021udl,Vignaroli:2023rxr,Akeroyd:1999xf,Hashemi:2012nz,Akeroyd:2000zs,Ouazghour:2023plc,Ouazghour:2024twx,Ouazghour:2024twx,BrahimAit-Ouazghour:2025xap,Ghosh:2025dcv,Coleppa:2026esg}.   

Extending the SM scalar sector with an additional inert $SU(2)$ doublet is particularly attractive, and the resulting model, referred to as the Inert Doublet Model (IDM), has already been studied in great detail (see, e.g., Refs. \cite{Dolle:2009fn,LopezHonorez:2010eeh,Arhrib:2013ela,Belyaev:2016lok,Abouabid:2023cdz}). The Higgs spectrum contains, in addition to the observed $125$~ GeV Higgs boson $h$, two electrically charged Higgs scalars $H^+\equiv(H^-)^\ast$ 
together with two neutral scalars: one $CP$-even scalar $H$ and one $CP$-odd scalar $A$. Owing to the exact $\mathbb{Z}_2$ symmetry, the lightest inert state is stable and can therefore serve as a DM candidate. This make the IDM particularly appealing, as it provides a minimal and predictive framework in which collider signatures are directly tied to the existence of a stable dark sector state \cite{Deshpande:1977rw,Ma:2006km,Barbieri:2006dq,Gustafsson:2007pc,Hambye:2007vf,Agrawal:2008xz,andreas2009neutrinos,Abouabid:2025whn}.  

Another important feature of the IDM is that its collider signatures are predominantly driven by the electroweak gauge interactions, making VBF processes particularly relevant at high-energy lepton colliders such as a MuC. In this regard, as the phenomenologically viable parameter space is already shaped by relic-density, direct-detection, electroweak-precision, and collider constraints, several conventional search channels may remain difficult at hadron machines, in contrast to a multi-TeV MuC, where the enhancement of electroweak boson radiation opens a particularly promising avenue to access the inert sector through fusion topologies. 

Indeed, recent work \cite{Braathen:2024ckk} has shown that the neutral inert-scalar pair-production process $\mu^+\mu^- \to AA \nu \bar{\nu}$ can provide significant sensitivity to the IDM at high energies, thereby motivating a wider exploration of such VBF channels in this model. More broadly, while the High-Luminosity Large Hadron Collider (HL-LHC) and future $e^+e^-$ colliders such as the electron-positron Future Circular Collider (FCC-ee) and the Compact Linear Collider (CLIC) provide complementary constraints \cite{Belyaev:2018ext,Bal:2025nbu,Kalinowski:2018kdn}, the direct production reach of these machines depends on the masses of the inert scalars. For the mass configurations considered in this work, the $HA$ and $H^+H^-$ production thresholds are well above the maximum FCC-ee center-of-mass energy of approximately $365$~GeV \cite{FCC:2025lpp}. At the HL-LHC, the sensitivity likewise depends on production rates, mass splittings, decay chains, and other factors, rather than on a universal inert scalar mass limit. A multi-TeV MuC would therefore provide a complementary opportunity to follow up possible hints of new physics or, in their absence, to explore the inert scalar sector at higher energies.

In this work, we pursue this strategy and investigate a broad class of VBF-driven $2 \to 4$ processes in the IDM involving both charged and neutral scalars. For charged inert-scalar production, we study the pair-production process $\mu^+\mu^- \to H^+ H^- \nu \bar{\nu}$ driven by $WW$ fusion, extending earlier studies of charged Higgs pair production in the IDM \cite{Aoki:2013lhm}. We further consider the associated production channels $\mu^+\mu^- \to H H^{\pm} \,\mu^{\mp}\nu$ and $\mu^+\mu^- \to A H^{\pm} \,\mu^{\mp}\nu$, which receive contributions from mixed $WZ$ and $W\gamma$ fusion topologies.
Taken together, these processes allow us to probe the neutral, charged, and mixed inert sectors of the IDM within a common high-energy framework.

A key feature of our analysis is that all these production modes are dominated by $t$-channel electroweak exchange, with no leading resonant $s$-channel contribution. As a consequence, their cross sections exhibit the characteristic high-energy behaviour of VBF mechanisms, receiving a logarithmic enhancement, $\sim\log(s/M_{V}^2)$, with  increasing  center-of-mass energy. This makes them particularly relevant in the multi-TeV regime and naturally motivates a comparative study at different collider energies. For this reason, we evaluate all channels at $\sqrt{s}=3$, $10$ and $20$~TeV, in order to track the evolution of the signal rates from the lower multi-TeV range to the regime where electroweak fusion becomes especially effective.

The collider phenomenology is further shaped by the dominant decay pattern of the heavier inert states. We focus here on the characteristic bosonic
decay pattern in which the charged scalar decays predominantly through $H^{\pm} \to H W^\pm$, while the neutral scalar $A$ mainly decays as $A \to H Z$. Since the stable DM candidate $H$ escapes the detector, the resulting signatures are characterised by sizeable missing energy accompanied by leptons and/or jets from the subsequent leptonic or hadronic decays of the electroweak gauge bosons. These features lead to a wide variety of experimentally relevant final states and offer several complementary handles for signal-background discrimination.

In the phenomenologically allowed parameter space of the IDM, we compute the signal rates for all the production modes listed above after imposing the relevant theoretical and experimental constraints. We then perform a full detector-level Monte Carlo (MC) analysis with a cut-based event selection to estimate the discovery sensitivity.

The rest of this paper is structured as follows: in Section~\ref{sec:model}, we provide a concise overview of the IDM, outlining the relevant constraints used to determine the viable parameter space. Section~\ref{sec:comp-step} is devoted to the study of the $2\to 2$ subprocesses, where we show the destructive interference responsible for the smallness of the cross section and explain how we calculate the $2\to 4$ cross sections at a MuC. Section~\ref{sec:num-res} presents the numerical results for the various VBF production channels considered in this work. In Section~\ref{sec:monteCarlosetup}, we present the detector-level MC setup and the cut-based signal-background analysis for the considered IDM signal processes at a future MuC. Finally, we summarise our findings and conclude in Section~\ref{sec:conclusion}. Technical details for the $2\to 2$  subprocesses are presented in the appendices.


\section{IDM in a nutshell} 
\label{sec:model}
The IDM is an extension of the SM in which an additional Higgs doublet $\Phi_2$ is added to the SM doublet $\Phi_1$, 
providing a stable dark matter candidate. Both fields have the form
\begin{eqnarray}
	\Phi_1 = \left (\begin{array}{c}
		G^\pm \\
		\frac{1}{\sqrt{2}}(v + h + i G) \\
	\end{array} \right)
	\qquad , \qquad
	\Phi_2 = \left( \begin{array}{c}
		H^\pm\\ 
		\frac{1}{\sqrt{2}}(H + i A) \\ 
	\end{array} \right) 
\end{eqnarray} 
where $G$ and $G^\pm$ denote the Nambu-Goldstone bosons absorbed by the
longitudinal components of the $Z$ and $W^\pm$ bosons, respectively, $v$ is the
vacuum expectation value (VEV) of the SM-like doublet $\Phi_1$, $h$ is the SM-like Higgs boson, while $H^\pm$, $H$, and $A$ are the inert scalars contained in the $\Phi_2$ doublet. The tree-level, gauge-invariant, and CP-conserving potential reads:
\begin{eqnarray}
	V(\Phi_1,\Phi_2) &=& \mu_1^2 |\Phi_1|^2 + \mu_2^2 |\Phi_2|^2  + \lambda_1 |\Phi_1|^4
	+ \lambda_2 |\Phi_2|^4 +  \lambda_3 |\Phi_1|^2 |\Phi_2|^2 + \lambda_4
	|\Phi_1^\dagger \Phi_2|^2 \nonumber \\
	&+&\frac{\lambda_5}{2} \left\{ (\Phi_1^\dagger \Phi_2)^2 + {\rm h.c}. \right\} 
	\label{eq:pot-idm} 
\end{eqnarray}
where, by hermiticity of the potential, all $\lambda_i$ ($i = 1, \cdots, 5$) parameters are real.\\
Assuming electroweak symmetry breaking (EWSB), the masses of the scalar particles can be expressed in terms of the potential parameters as
\begin{eqnarray}
&& m_h^2 = - 2 \mu_1^2 = 2 \lambda_1 v^2 \nonumber \\
&& m_{H}^2 = \mu_2^2 + \frac{1}{2} (\lambda_3 + \lambda_4 + \lambda_5) v^2 = \mu_2^2 + \frac{1}{2} \lambda_{345} v^2 \nonumber \\
&&  m_{A}^2 = \mu_2^2 +\frac{1}{2} (\lambda_3 + \lambda_4 - \lambda_5) v^2 = \mu_2^2 + \frac{1}{2} \bar{\lambda}_{345} v^2 \nonumber \\
&&  m_{H^{\pm}}^2 = \mu_2^2 + \frac{1}{2} \lambda_3 v^2.
\label{spect.IHDM}
\end{eqnarray}

It is worth mentioning that the scalar doublet $\Phi_2$ does not couple to the SM fermions. Moreover, throughout the present study, we consider the scenario in which $H$ is the lightest inert particle, with $m_H < m_A,\, m_{H^\pm}$.\\
Additionally, instead of the potential parameters (${\mu_1^2,\mu_2^2,\lambda_1,\lambda_2,\lambda_3,\lambda_4,\lambda_5}$), it is convenient to describe the IDM Higgs sector by the physical parameter basis consisting of
\begin{equation}
\label{eq:setofparameters}
\mathcal{P}=\{v=246\,\gev,\, m_h=125\,\gev,\,m_{H^\pm},\,m_H,\,m_A,\,\lambda_2,\,\lambda_{345} \}
\end{equation}
while $\lambda_2$ and $\lambda_{345}$ remain as independent couplings.

\noindent
From both theoretical and experimental perspectives, the IDM, like many BSM extensions, is subject to many constraints, which we briefly recall below:
\begin{enumerate}
\renewcommand{\labelenumi}{(\alph{enumi})}
\item Vacuum stability at tree level~\cite{Deshpande:1977rw}, which guarantees the boundedness from below of the IDM potential in all the directions of the field space.
\item Tree-level unitarity~\cite{Lee:1977eg}.
\item The electroweak-symmetry-breaking inert vacuum is required to be the global minimum of the scalar potential~\cite{Espinosa:2011ax,Branchina:2018qlf}.
\item Electroweak $S$ and $T$ parameters~\cite{Peskin:1990zt,Peskin:1991sw}, which probe quantum effects on electroweak parameters, such as gauge couplings and gauge-boson masses, and are therefore important for assessing the validity of NP scenarios. We require $\chi^2_{ST} < 5.99$ for consistency with the current best-fit values \cite{ParticleDataGroup:2024cfk}:  
\begin{eqnarray}
S = -0.05 \pm 0.07, \quad
T = 0.00 \pm 0.06 \quad {\rm and} \quad \rho_{ST} = 0.93, 
\end{eqnarray}
while assuming $U=0$.
\item Exotic SM Higgs decays: as no significant deviation from the SM Higgs width has been observed within the current experimental resolution, 
the branching ratio of the Higgs boson into non-standard final states is strongly constrained by~\cite{ATLAS:2023tkt}
\begin{eqnarray}
\br(h\to \text{inv})=\frac{\Gamma(h \to \text{inv})}{\Gamma^{\text{pred}(h \to \text{SM})}+\Gamma(h \to \text{inv})} < 0.13,
\end{eqnarray}
where the predicted total decay width of the SM Higgs boson, for $m_{h} \simeq 125$~GeV, is given by \cite{LHCHiggsCrossSectionWorkingGroup:2016ypw,ParticleDataGroup:2024cfk},
\begin{equation}
\label{eq:pred-width-SM}
\Gamma^{\text{pred}}(h \to \text{SM})=4.10 \pm 0.06\,\text{MeV}.
\end{equation}
\item LEP direct searches yield lower bounds on neutral inert scalar masses, $m_{A, H} > $ 80-90 GeV, and on charged boson masses, $m_{H^{\pm}} > 70$ GeV \cite{ParticleDataGroup:2010dbb,Pierce:2007ut,ALEPH:2013htx,Arbey:2017gmh}.
\item DM constraints are evaluated with {\tt micrOMEGAs\_5.0.4} \cite{Alguero:2023zol} using the IDM implementation. We impose the observed relic density as an upper bound, $\Omega_H h^2 \leq \Omega_{\rm DM}^{\rm obs}h^2 = 0.12$, thereby allowing $H$ to constitute only a fraction of the observed DM abundance. Accordingly, the spin-independent direct detection (DD) cross section is rescaled by the relic-density fraction and required to satisfy the experimental upper limits, in particular those from the LUX-ZEPLIN (LZ) experiment~\cite{LZ:2024zvo}, i.e.,
\begin{equation}
\xi\,\sigma_{\rm SI}(m_H)
\leq
\sigma_{\rm SI}^{\rm LZ}(m_H),
\qquad
\xi \equiv
\frac{\Omega_H h^2}{\Omega_{\rm DM}^{\rm obs}h^2}.
\end{equation}
\item Additionally,  in the IDM, the Higgs decays into SM particles have rates similar to those in the SM, except for $h\to \gamma \gamma $, $h\to \gamma Z$ which are sensitive to the charged Higgs contributions \cite{Arhrib:2012ia}. Since only the $h \to \gamma\gamma$ amplitude is modified, $\mu_{\gamma\gamma} \simeq \kappa_{\gamma}^2$. We impose the  current global average $\mu_{\gamma\gamma}=1.10\pm0.07$~\cite{Heo:2024cif} at $95\%$ C.L. For comparison, the projected precisions on $\kappa_{\gamma}$ are $1.9\%$ at the HL-LHC~\cite{deBlas:2019rxi} and $0.7\%$ at the HL-LHC+$10$~TeV MuC \cite{Castelli:2025mqk}.
\end{enumerate}

\section{Computational Procedure Steps}
\label{sec:comp-step}
The phenomenology of charged Higgs bosons can be explored through kinematically accessible production channels at a MuC. These include $2\to2$ processes such as $\mu^+ \mu^- \to H^+ H^-$ and $\mu^+ \mu^- \to W^\pm H^\mp$, $2\to3$ processes like $\mu^+ \mu^- \to \tau^+ \nu_{\tau} H^-$ and $\mu^+ \mu^- \to t\bar{b} H^-$, as well as $2\to4$ processes including $\mu^+ \mu^- \to H^+ H^- \nu \bar{\nu}$ and $\mu^+ \mu^- \to S_i H^{\pm} \mu^+ \bar{\nu}$, with $S_i=H$ or $A$. The first four channels have been studied in detail in Ref.~\cite{Ouazghour:2023plc} within the 2HDM framework.

In this section, we study $\mu^+\mu^- \to W^\ast W^\ast\to H^+ H^- \nu \bar{\nu}$, $\mu^+\mu^- \to W^\ast V^\ast\to H H^{\pm} \mu^{\pm} \bar{\nu}$, and $\mu^+\mu^- \to W^\ast V^\ast\to A H^{\pm} \mu^{\pm} \bar{\nu}$, where $H$ is a stable dark matter particle and $V=Z, \gamma$.  
To further illustrate the dynamical origin of the enhancement of the processes $\mu^+\mu^- \to H^+ H^- \nu\bar{\nu}$ and 
 $\mu^+\mu^- \to H^+ S_i \nu\bar{\nu}$ ($S_i=H, A$)
 at high energies, we first study  the partonic subprocesses $W^+W^- \to H^+ H^-$ and $W^+V \to H^+ H, H^+ A$, with $V=Z,\gamma$. 

\subsection{ $W^+W^- \to H^+H^-$}
\begin{figure}[!h]
\centering
\includegraphics[scale=0.4]{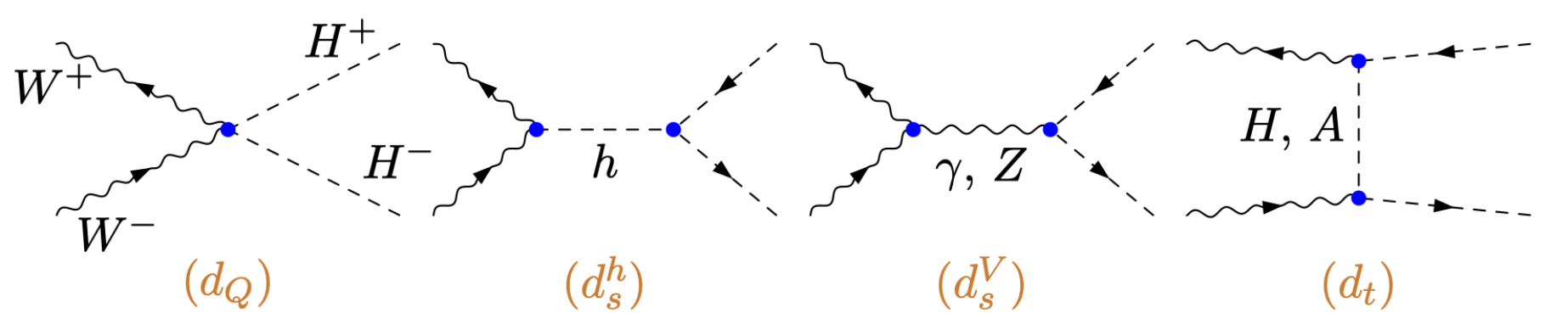}
\caption{Feynman diagrams within the IDM contributing to $W^+W^- \to H^+ H^-$ in the Feynman gauge: $(d_Q)$ refers to the contact diagram, both $(d_{s}^{h})$ and $(d_{s}^{V})$ represent the $s$-channel contributions related to the Higgs boson $h$ and to $V=\gamma,Z$, respectively, while $(d_{t})$ stands for the $H$ or $A$ $t$-channel contributions.}
\label{mesfig:fig1}
\end{figure}

For the subprocess $W^+W^- \to H^+H^-$, the leading-order diagrams are summarised in Fig.~\ref{mesfig:fig1},
while the amplitudes as well as the squared amplitudes for each individual diagram have been evaluated, in the Feynman gauge, taking into account the sum over the polarisations of the incoming gauge bosons. Details of the calculation are presented 
in Appendix~\ref{app:VBF_HpHm} together with the Feynman rules used for the involved couplings. To quantify the contributions of the individual polarisation states, we denote by $\sigma_{\lambda_1\lambda_2}$ the partonic cross section for fixed helicities $\lambda_i=0,\pm1$ of the incoming $W$ bosons, and define
\begin{equation}
\sigma_{LL}=\sigma_{00},\quad
\sigma_{LT}=\frac{\sigma_{0+}+\sigma_{0-}}{2},\quad
\sigma_{TL}=\frac{\sigma_{+0}+\sigma_{-0}}{2},\quad
\sigma_{TT}=\frac{\sigma_{++}+\sigma_{+-}+\sigma_{-+}+\sigma_{--}}{4},
\label{diff_polarisations}
\end{equation}
so that the unpolarised partonic cross section reads
\begin{equation}
\sigma_{UU}=\frac{1}{9}\sum_{\lambda_1,\lambda_2}\sigma_{\lambda_1\lambda_2}
=\frac{\sigma_{LL}+2\sigma_{LT}+2\sigma_{TL}+4\sigma_{TT}}{9}.
\label{uupolarisation_res}
\end{equation}
We mention here that all helicity combinations are retained in the numerical evaluation, i.e. the automatic helicity selection of FormCalc is disabled.
\begin{figure}[!h]
\centering
\includegraphics[width=0.495\textwidth]{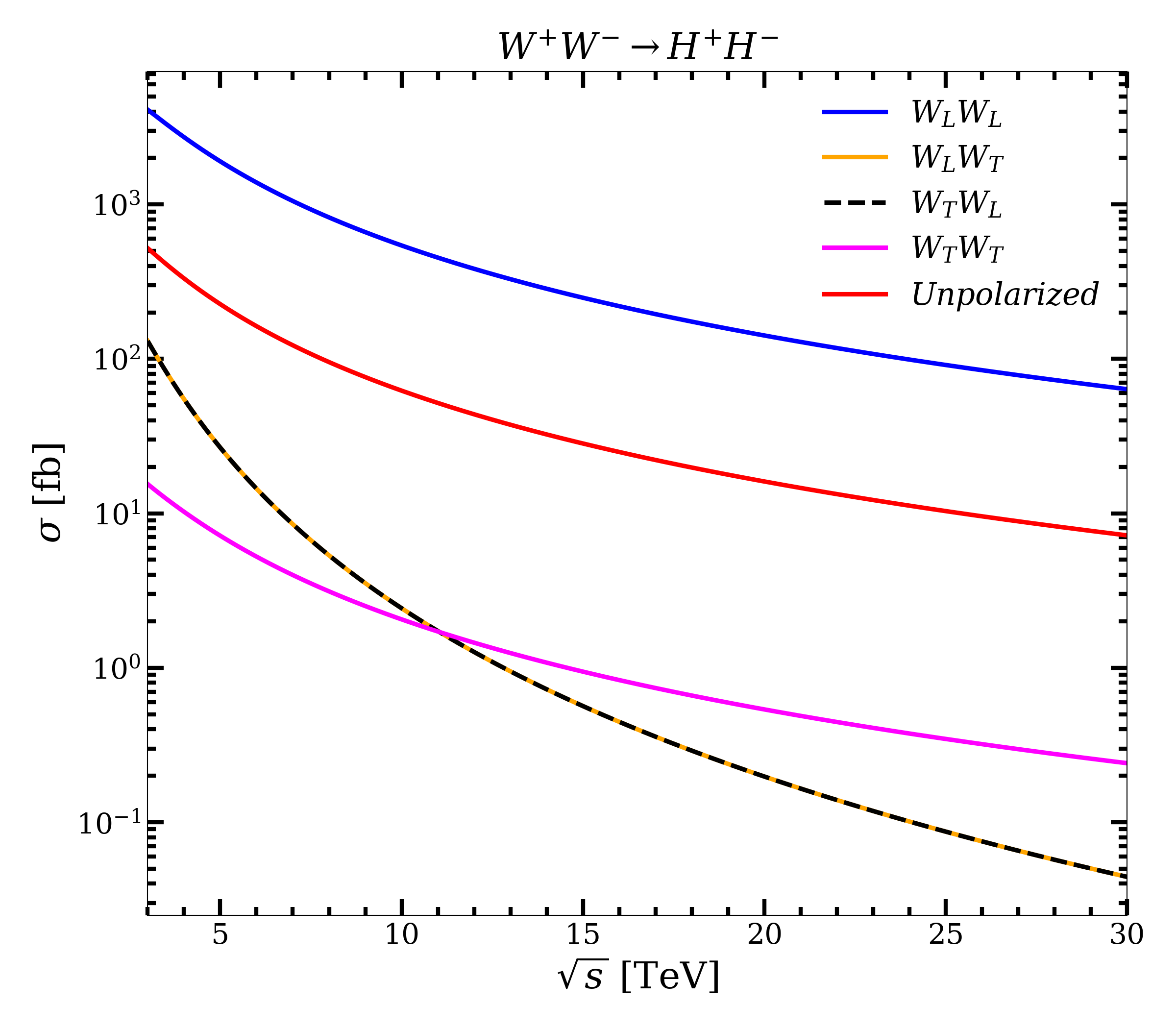}
\includegraphics[width=0.495\textwidth]{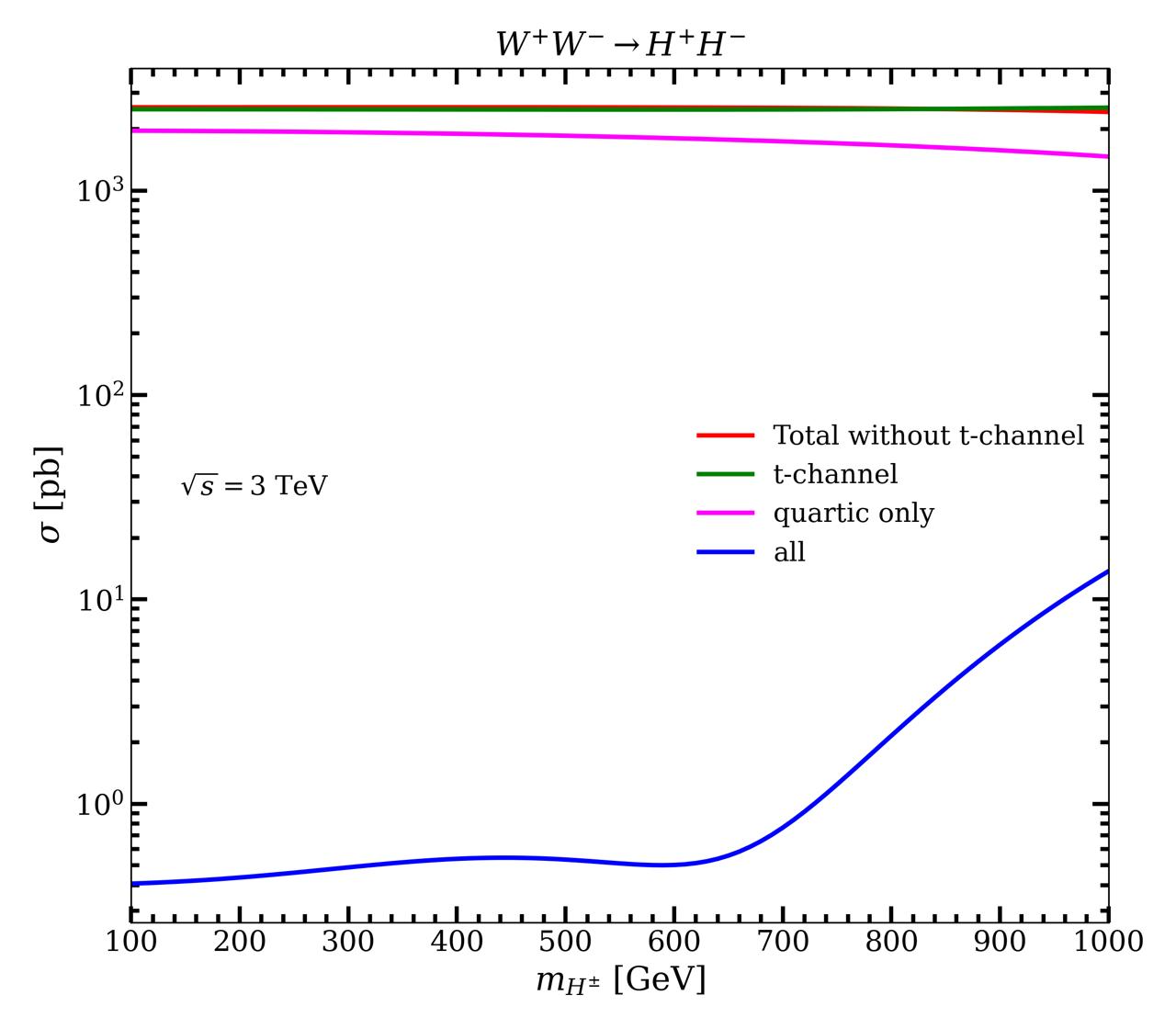}
\caption{Partonic cross section for the subprocess $W^+W^- \to H^+H^-$. Left: as a function of the $WW$ center-of-mass energy $\sqrt{s}$ for $m_{H^\pm}=628$ GeV, shown for the polarisation configurations defined in Eq.(\ref{diff_polarisations}), together with the unpolarised result of Eq. (\ref{uupolarisation_res}). Right: as a function of the charged Higgs boson mass $m_{H^\pm}$ at $\sqrt{s}=3$~TeV, comparing the individual tree-level contributions and their  destructive interference in the full tree-level amplitude.}
\label{mesfig:fig2}
\end{figure}
 
We illustrate in Fig.~\ref{mesfig:fig2} the cross section for the subprocess $W^+W^- \to H^+ H^-$. In the left panel, where the cross section is displayed separately for various polarisation states of the incoming $W$ bosons, it is evident that the longitudinal configuration $W_LW_L$ dominates over the transverse contributions throughout the considered energy range, exceeding the mixed and purely transverse modes by more than one order of magnitude at $\sqrt{s}=3$ TeV, and by more than two orders at the highest energies. Such behaviour is a characteristic feature of EWSB dynamics and confirms that charged inert scalar production at a multi-TeV MuC is primarily sensitive to the longitudinal component of the weak gauge bosons. Among the subleading contributions, the mixed configurations $W_LW_T$ and $W_TW_L$ decrease faster than the purely transverse one, and become smaller than it above $\sqrt{s} \simeq 11$ TeV. We also find destructive interference among the contributing Feynman diagrams within the longitudinal $LL$ channel, which affects its high-energy behaviour and slightly reduces the corresponding cross section.

On the other hand, the right panel in Fig.~\ref{mesfig:fig2} exhibits the striking difference between the individual contributions and the full result. This separation demonstrates 
the existence of destructive interference among the contributing diagrams. Indeed, while each individual contribution is of the order of $10^3$~pb, the resulting cross section is reduced by roughly three orders of magnitude due to destructive interference among the contributing amplitudes\footnote{We stress that such cancellations should not be identified diagram by diagram with gauge invariance: individual Feynman diagrams are generally gauge dependent, whereas the complete matrix element is gauge invariant.}.
The observed cancellations reflect the coherent interplay of the different contributions required to ensure the proper high-energy behaviour of the amplitude. Despite these cancellations, the VBF topology becomes increasingly important at multi-TeV center-of-mass energies, where the $W$-boson luminosity is enhanced, making it a primary production mechanism at a high-energy MuC.

\subsection{ $W^+V \to H^+S_i$, $V=Z, \gamma$, $S_i=H,A$}

\begin{figure}[!h]
\centering
\includegraphics[scale=0.4]{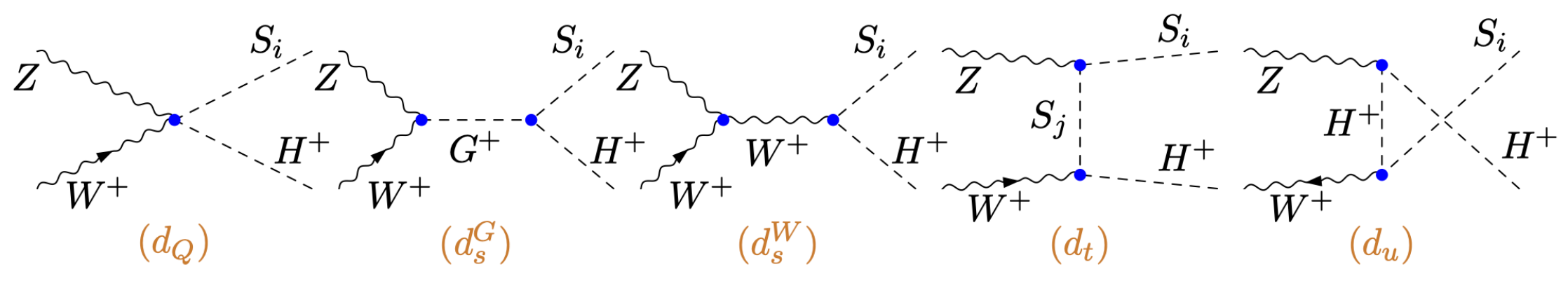}\\
\includegraphics[scale=0.36]{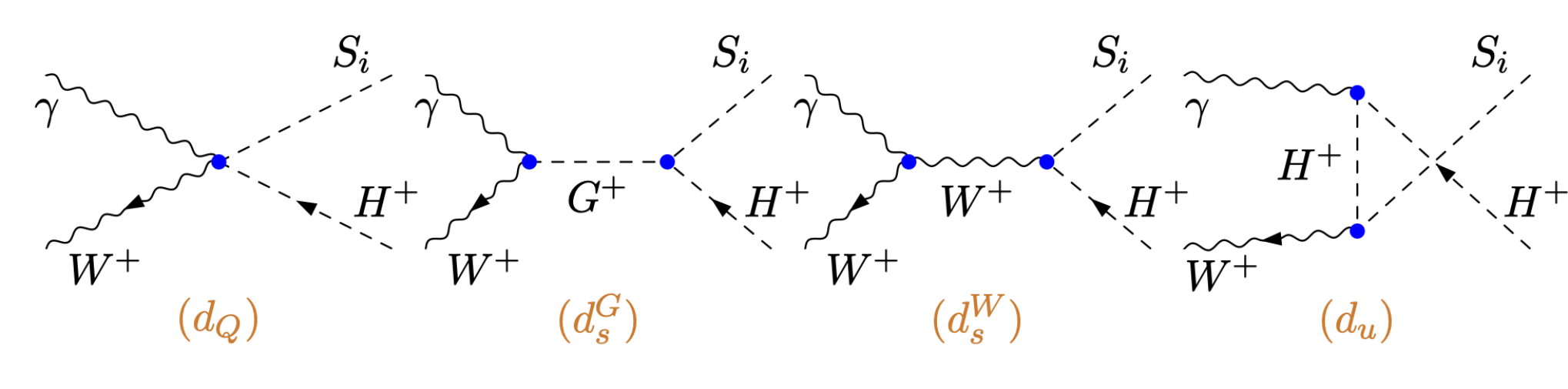}
\caption{Feynman diagrams within the IDM contributing to $W^+V \to H^+ S_i$, for $V=Z$ (upper panel) and $V=\gamma$ (lower panel), with $S_i=H$ or $A$ (and $S_j=A$ or $H$ in diagram $(d_{t})$), shown in the Feynman gauge. Diagram $(d_Q)$ is the contact term arising from the quartic coupling; $(d_{s}^{G})$ and $(d_{s}^{W})$ are the $s$-channel exchanges of the Goldstone $G^\pm$ and the gauge boson $W^\pm$, respectively; while $(d_{t})$ and $(d_{u})$ stand respectively for the $S_i$ $t$-channel and $H^+$ $u$-channel contributions.}
\label{mesfig:fig3}
\end{figure}

Similarly to the previous case, we have computed in the Feynman gauge\footnote{We verified gauge invariance by comparing Feynman and unitarity gauge calculations, and show the unitarity-gauge diagrams here for clarity.} the tree level amplitudes for $W^+V \to H^+S_i$, with $V=Z, \gamma$, and $S_i=H,A$. The corresponding Feynman diagrams are illustrated in Fig.~\ref{mesfig:fig3}. Likewise, Appendix~\ref{app:VBF_WZ} provides a detailed derivation of the analytic expressions for each individual contribution to the subprocess $W^+Z \to H^+S_i$, and shows the squared amplitude and the interference terms after summing over the polarisations of the $W$ and $Z$ bosons. Appendix~\ref{app:VBF_Wga} gives only the analytic expressions for each individual contribution to the subprocesses $W^+\gamma \to H^+S_i$, together with the total tree-level cross sections. 

\begin{figure}[!h]
\centering
\includegraphics[width=0.495\textwidth]{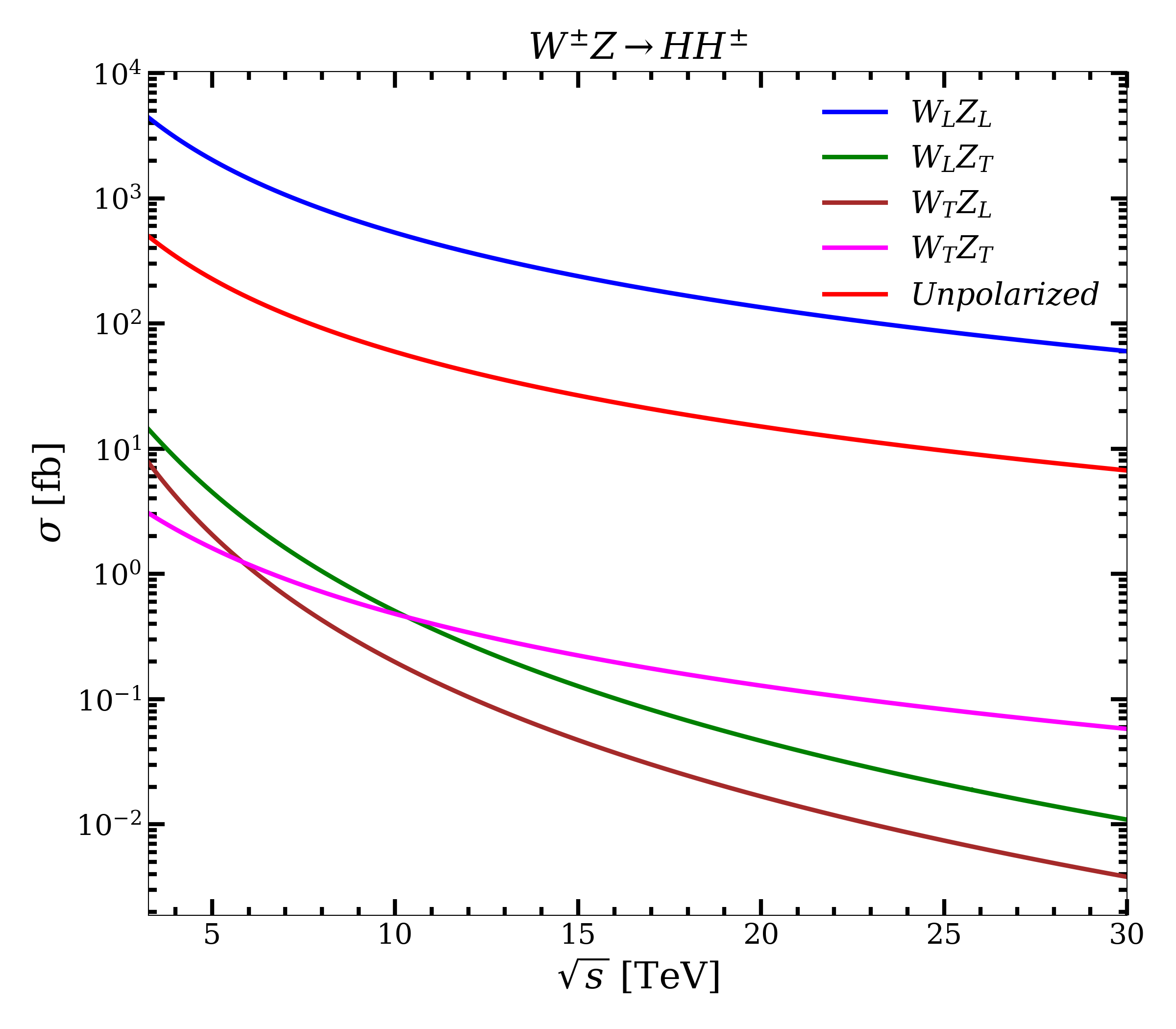}
\includegraphics[width=0.495\textwidth]{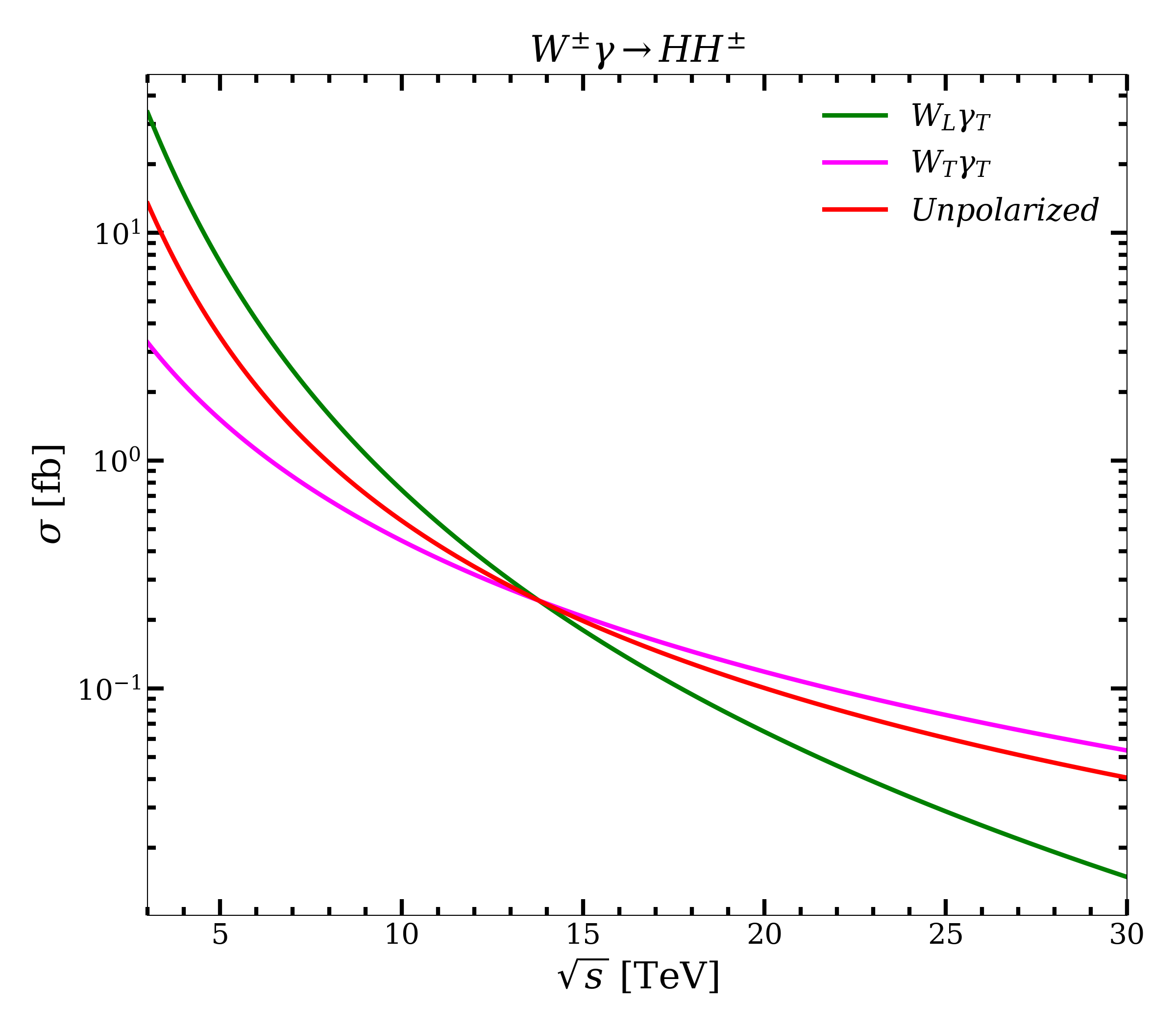}\\
\includegraphics[width=0.495\textwidth]{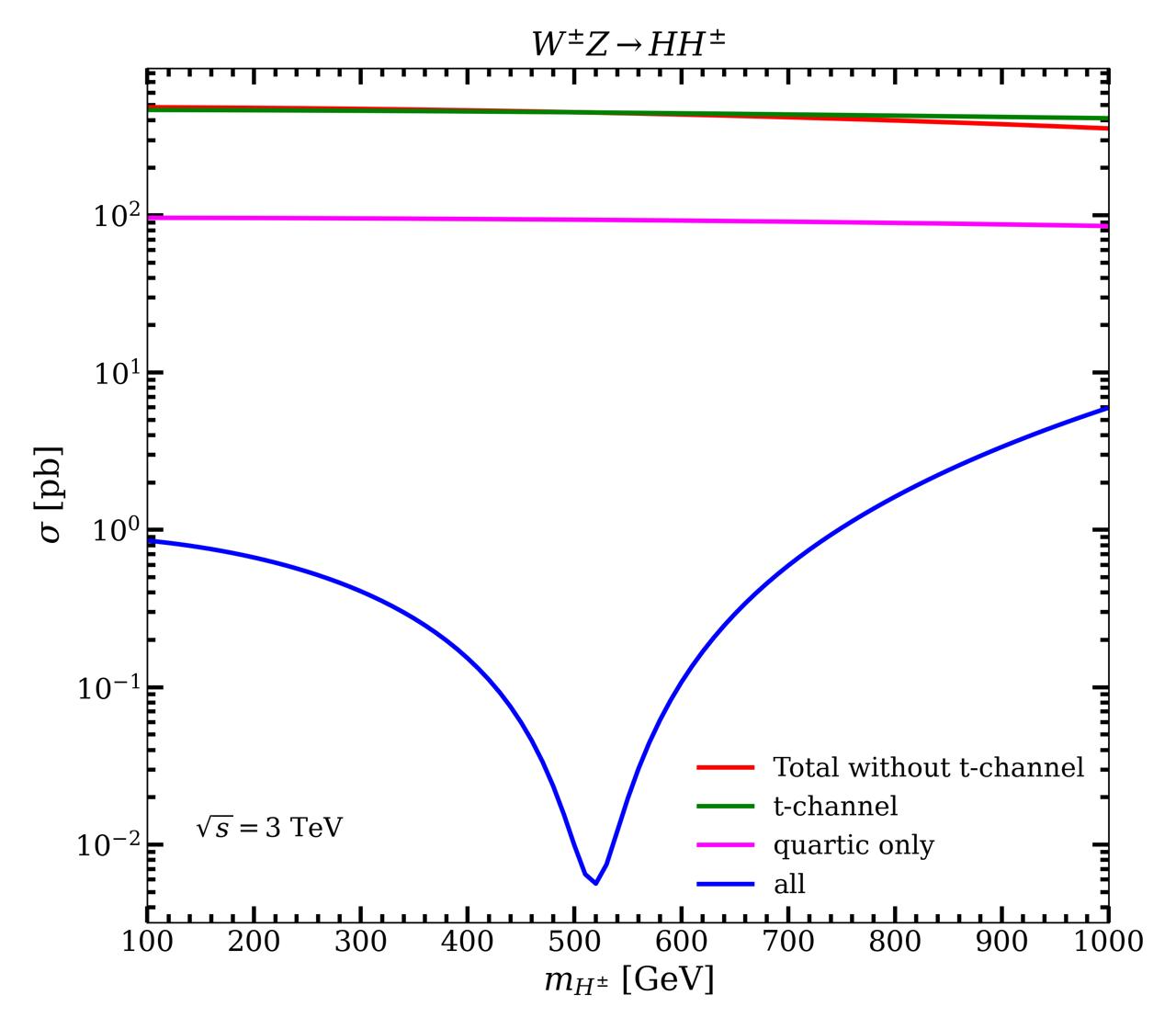}
\includegraphics[width=0.495\textwidth]{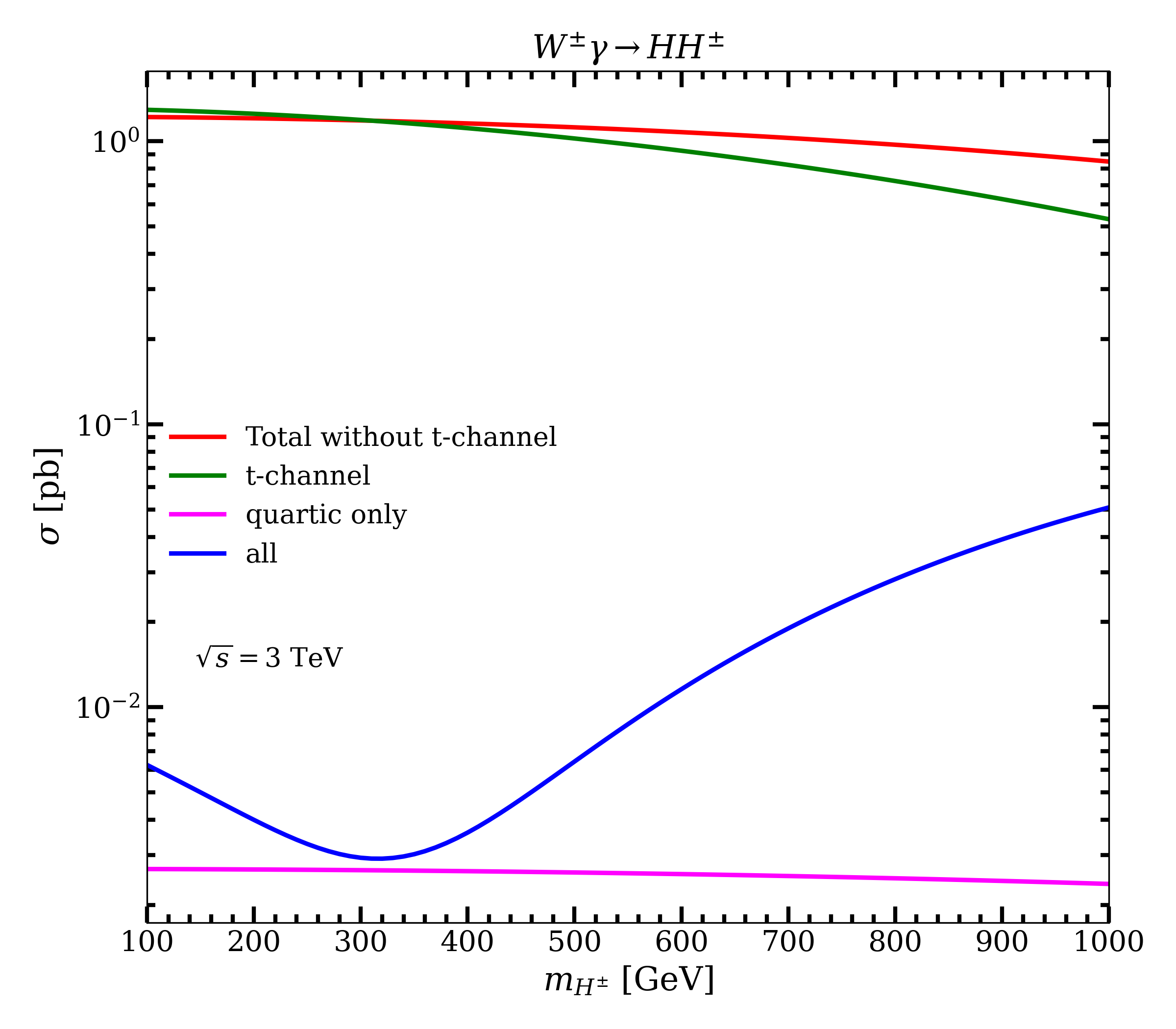}
\caption{Upper panel: partonic cross sections for the subprocesses $W^+V\to HH^-$, $V=Z$, $\gamma$, as a function of the $WV$ center-of-mass energy for $m_{H^\pm}=628$~GeV and $m_H=171$~GeV, shown for different polarisation configurations of the initial-state gauge bosons. Lower panel: dependence of the $W^+Z\to H^+ H$ cross section on the charged Higgs mass $m_{H^\pm}$ at $\sqrt{s}=3$~TeV, illustrating the impact of the individual gauge contributions and their destructive interference.}
\label{mesfig:fig4}
\end{figure}

For numerical illustration, we show results only for $W^+V \to H^+ H$. We expect similar behavior for $W^+V \to H^+A$. We present the polarised partonic cross sections for the subprocess $W^+V \to H^+ H$ in the upper panel of Fig.~\ref{mesfig:fig4}. The cross section is presented individually for the different polarisation states of the incoming $W$ and $V$ bosons. It is evident that the longitudinal configuration $W_L Z_L$ remains the dominant component across the entire energy range under consideration, surpassing both the mixed and purely transverse modes. At the partonic level, we observed that the $W\gamma \to H^+ H$ cross section is notably smaller than the $W Z\to H^+ H$ one. Given that the photon is exclusively transverse, a comparison between $W_{L,T}Z_T$ and $W_{L,T} \gamma_T$ fusion demonstrates that the results are largely consistent. 

In the lower panel of  Fig.~\ref{mesfig:fig4}, we illustrate the partonic cross sections for $W^+Z\to H^+ H$ and $W^+\gamma \to H^+ H$ as a function of the charged Higgs mass
at  $\sqrt{s}=3$~TeV. Similarly to the $WW\to H^+ H^-$ case, it is evident that there is a destructive interference between the contributing Feynman diagrams.

It appears that  $WZ$ fusion and $W\gamma$ fusion contribute approximately  equally to the total cross section. This behavior can likely be explained by the relative weakness of the $\mu^+\mu^- Z$ coupling in comparison to the $\mu^+ \mu^- \gamma$ electromagnetic coupling. Finally, in close analogy with the $W^+W^- \to H^+H^-$ analysis presented above, the bottom panel of Fig.~\ref{mesfig:fig4} illustrates another striking feature common to both subprocesses, namely the strong destructive interference among the individual tree-level amplitudes, which ultimately determines the magnitude of the physical cross section. These cancellations are necessary to ensure the proper high-energy behavior of the amplitudes.

\subsection{$\mu^+\mu^- \to W^\ast W^\ast\to H^+ H^- \nu \bar{\nu}$ and $\mu^+\mu^- \to W^\ast V^\ast\to S_i H^{\pm} \mu^{\pm} \bar{\nu}$, $S_i=H, A$}

The remaining three $2\to 4$ processes constitute the primary focus of the present work and are studied here for the first time within the framework of the IDM. 
These processes share the common feature of producing a charged Higgs boson pair or a charged Higgs boson in association with a neutral inert scalar in the final state, and they receive contributions from a rich set of Feynman topologies involving gauge and scalar mediators, as detailed below. For clarity, throughout this work we refer to these channels as VBF-enhanced processes, since they contain electroweak VBF topologies that become increasingly important at high energies. The quoted production cross sections at leading order, however, are obtained from the complete gauge-invariant $2\to4$ matrix elements, including all relevant $s$-, $t$- and $u$-channel contributions. In particular, the associated channels $\mu^+\mu^- \to HH^\pm\mu^\mp\nu$ and $\mu^+\mu^- \to AH^\pm\mu^\mp\nu$ receive contributions from both $s$-channel annihilation and $t$-channel exchange topologies. 

The leading-order matrix elements for the $2\to4$ processes, phase-space integrations, and the corresponding signal cross sections are computed using \texttt{MadGraph5\_aMC@NLO v2.6.5}~\cite{alwall2014automated} with the IDM model implementation. The calculation is performed directly at the complete $2\to4$ matrix element level, without invoking the effective vector boson approximation (EVBA) or convoluting the subprocess cross sections with electroweak gauge-boson distribution functions inside the incoming muons. Consequently, the full off-shell kinematics of the intermediate gauge bosons and all relevant interference contributions are retained.

\begin{figure}[!h]
\centering\includegraphics[width=0.65\textwidth]{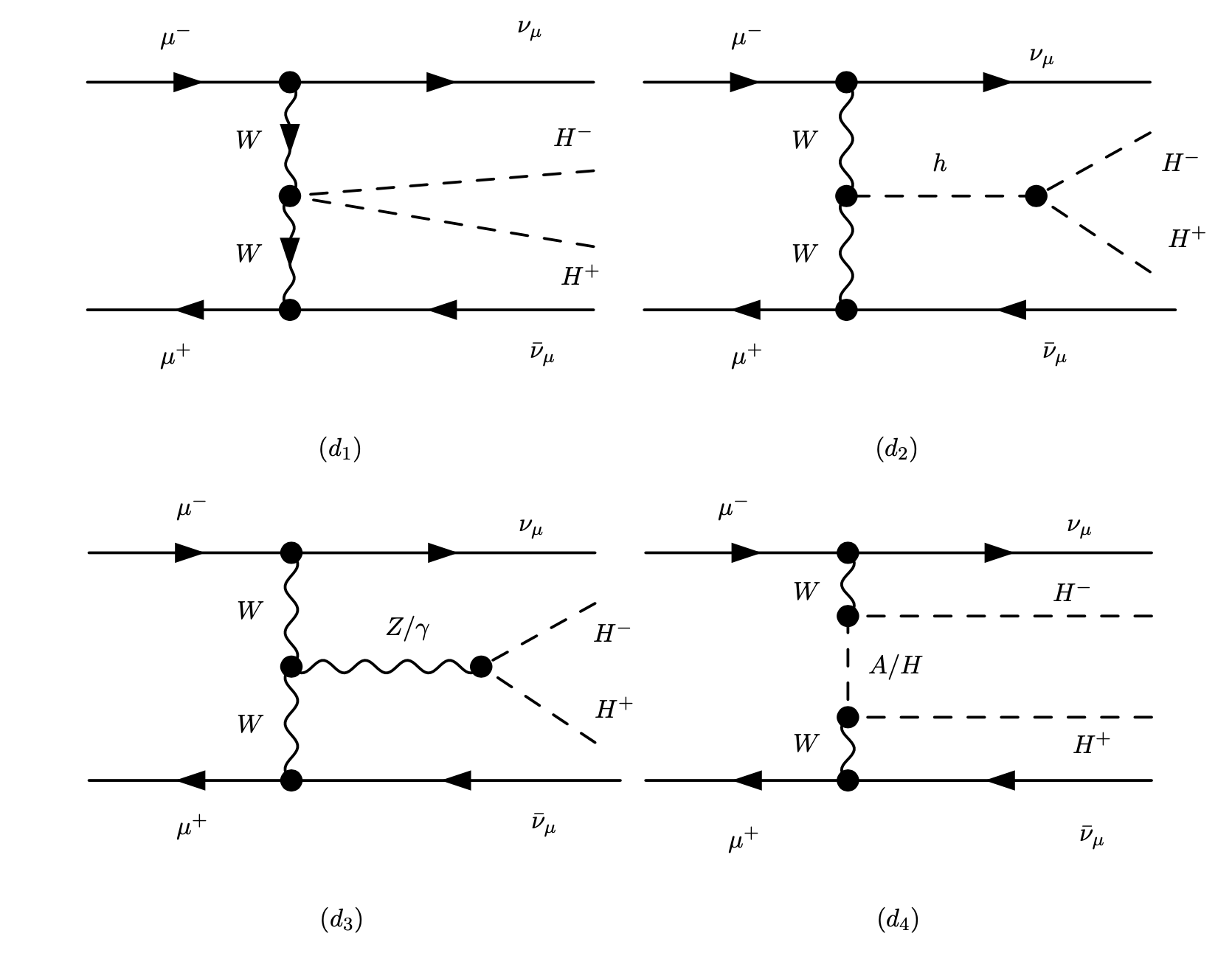}
\caption{Tree-level Feynman diagrams contributing to the VBF-enhanced 
pair-production process $\mu^+\mu^- \to H^+H^-\nu_\mu\bar{\nu}_\mu$ at a future MuC.} 
\label{mesfig:fig5}
\end{figure}

For the process $\mu^+\mu^- \to H^+H^-\nu_\mu\bar{\nu}_\mu$, the contributing tree-level diagrams are shown in Fig.~\ref{mesfig:fig5}. These proceed dominantly via the VBF mechanism, in which two $W$ bosons are radiated collinearly from the incoming muon lines and subsequently fuse to produce the charged Higgs pair. In diagram $(d_1)$, the two $W$ bosons annihilate directly into the $H^+H^-$ pair through the quartic gauge-scalar coupling $WWH^+H^-$. Diagram $(d_2)$ proceeds via $W$-boson fusion with an 
intermediate SM-like Higgs boson $h$ propagating in the $s$-channel, which decays into the charged Higgs pair through the trilinear $hH^+H^-$ vertex, 
 whose coupling strength is proportional to $\lambda_3$, i.e., $g_{hH^+H^-}\propto \lambda_3$. In diagram $(d_3)$, a neutral 
electroweak gauge boson $Z/\gamma$ is exchanged between the two $W$ lines, producing the $H^+H^-$ pair through the $ZH^+H^-$ or $\gamma H^+H^-$ coupling. Finally, diagram $(d_4)$ features a $t$-channel exchange of a neutral inert scalar $A/H$ between the two $W$-emitted charged Higgs lines, probing the trilinear gauge-scalar coupling $WH^\pm A$ or $WH^\pm H$ of the IDM. 

It is worth emphasizing that the VBF topologies involving the $WWH^+H^-$ quartic vertex and the $hH^+H^-$, $WH^\pm S_i$ trilinear couplings are expected to provide the dominant contributions at a high-energy MuC, while the $s$-channel $\gamma/Z$-mediated diagrams remain relevant at lower center-of-mass energies and play an important role in interference effects. Taken together, these processes provide complementary and direct probes of the scalar sector of the IDM, being sensitive to the mass spectrum of the inert scalars $(H^\pm, H, A)$ as well as to the trilinear and quartic scalar-gauge interactions that are characteristic of the IDM.

\begin{figure}[!h]
\centering\includegraphics[width=0.65\textwidth]{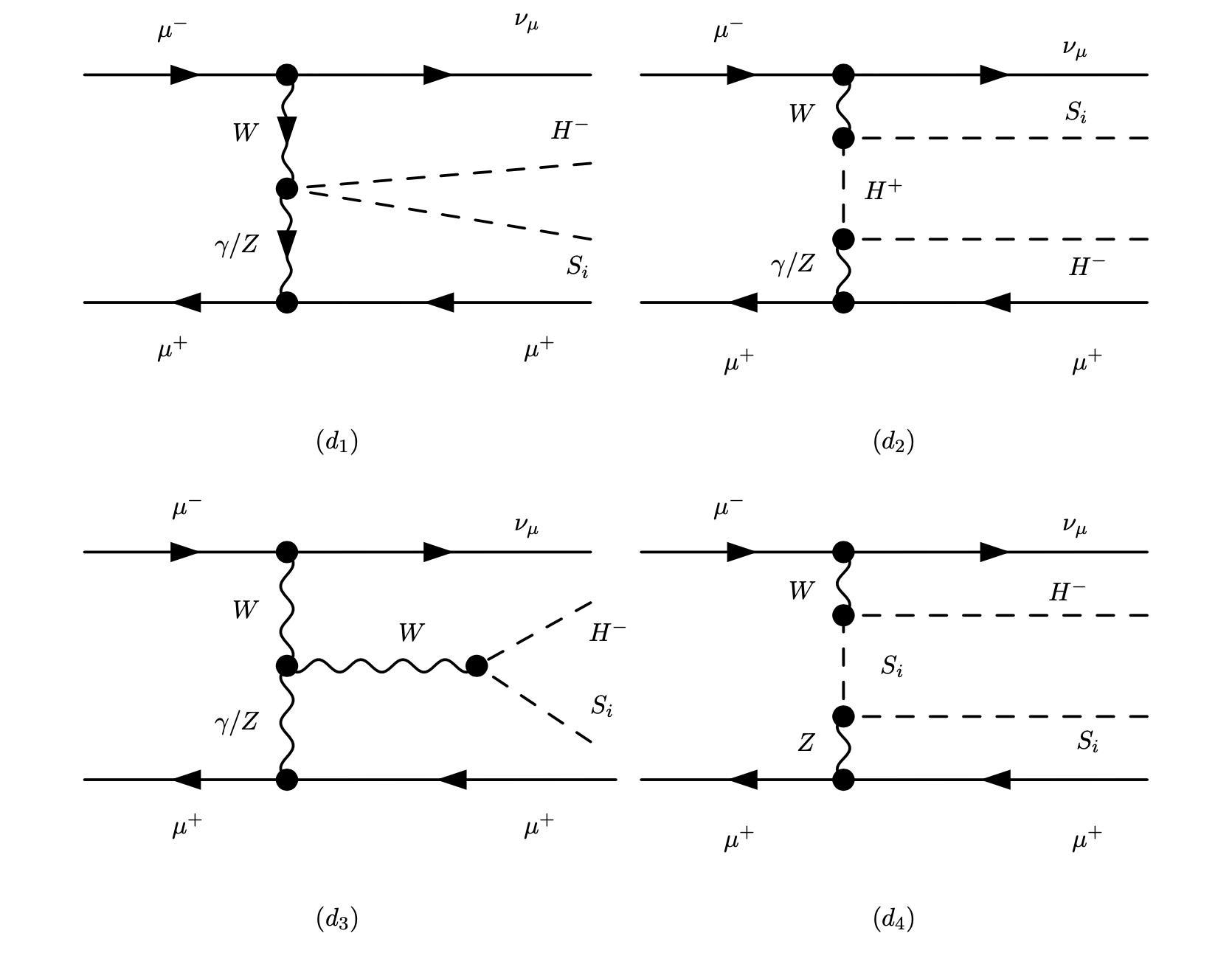}
\caption{Tree-level Feynman diagrams contributing to the VBF-enhanced 
associated production process $\mu^+ \mu^- \to S_i H^+ \mu^- \bar{\nu}$, with 
$S_i=H,A$, at a future MuC.} 
\label{mesfig:fig6}
\end{figure}

For the processes $\mu^+\mu^- \to H H^\pm \mu^\pm \bar{\nu}$ 
and $\mu^+\mu^- \to A H^\pm \mu^\pm \bar{\nu}$, the relevant Feynman diagrams are collected 
in Fig.~\ref{mesfig:fig6}. These processes receive contributions from both $s$-channel annihilation and $t$-channel exchange topologies, involving combinations of gauge and scalar 
mediators. In diagram $(d_1)$, a $W$ boson is radiated off the $\mu^+$ line while a virtual $\gamma/Z$ is exchanged from the $\mu^-$ line; the two gauge bosons meet at a 
quartic gauge-scalar vertex, producing $H^+$ and a neutral inert scalar $S_i$. 
Diagram $(d_2)$ proceeds via a sequential $t$-channel 
topology in which the $W$ boson emitted from the $\mu^+$ line couples through the 
$WH^\pm S_i$ vertex, producing $S_i$ as a final-state particle and an off-shell 
$H^+$ propagator; the latter then interacts with a virtual $\gamma/Z$ radiated from 
the $\mu^-$ line via the $\gamma/Z H^+H^-$ coupling, yielding the final-state $H^+$. 
In diagram $(d_3)$, the $W$ boson radiated from the $\mu^+$ line and the virtual 
$\gamma/Z$ exchanged from the $\mu^-$ line first interact through a triple gauge 
vertex $WW(\gamma/Z)$, generating an internal $W$ propagator, which subsequently 
decays into the $H^-S_i$ pair via the trilinear $WH^\pm S_i$ scalar coupling. 
Diagram $(d_4)$ involves a $W$ boson radiated off the $\mu^+$ line splitting into 
$H^-$ and an intermediate neutral scalar $S_i$ via the $WH^\pm S_i$ vertex; the 
$S_i$ propagates in the $t$-channel and couples to a $Z$ boson radiated from the 
$\mu^-$ line through the $ZS_iS_i$ or $ZHS_i$ vertex, yielding the $S_i H^\pm$ 
system together with the final muon and neutrino.

\section{Numerical Results}
\label{sec:num-res}
We perform random scans over the IDM parameter space within the following ranges:
\begin{eqnarray}
\label{Ei19}
&& m_h = 125.09\,\gev, \hspace*{0.25cm} m_{H},\, m_{A} \in [65,\,10^3]\,\gev, \hspace*{0.25cm} m_{H^\pm}  \in [70,\,10^3]\,\gev, \nonumber\\
&& \lambda_2 = 2\,, \hspace*{0.25cm} \lambda_{345} \in [-1,\,1],\,
\end{eqnarray}
where we ensure that all relevant theoretical and experimental constraints are satisfied for each viable parameter point. The choice of the inert sector self-interaction coupling $\lambda_2 = 2$ throughout the numerical analysis is motivated by the perturbativity bound $\lambda_2 < \frac{4\pi}{3}$~\cite{Belyaev:2016lok} and by the fact that the relative corrections to the triple Higgs couplings are maximised for $\lambda_2\lesssim2$~\cite{Arhrib:2015hoa}.
\begin{figure}[h!]
\centering
\includegraphics[width=0.32\textwidth]{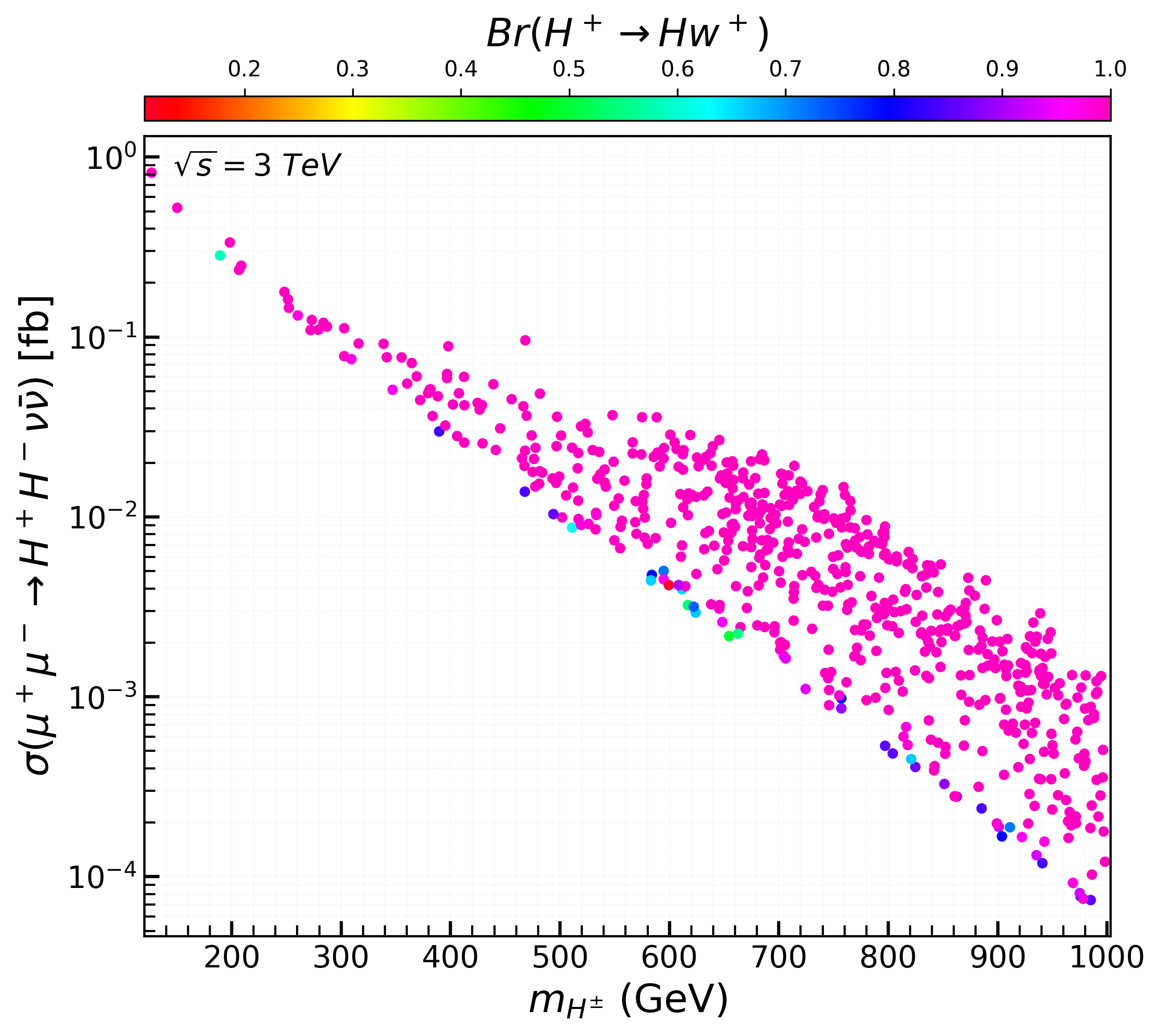}
\includegraphics[width=0.32\textwidth]{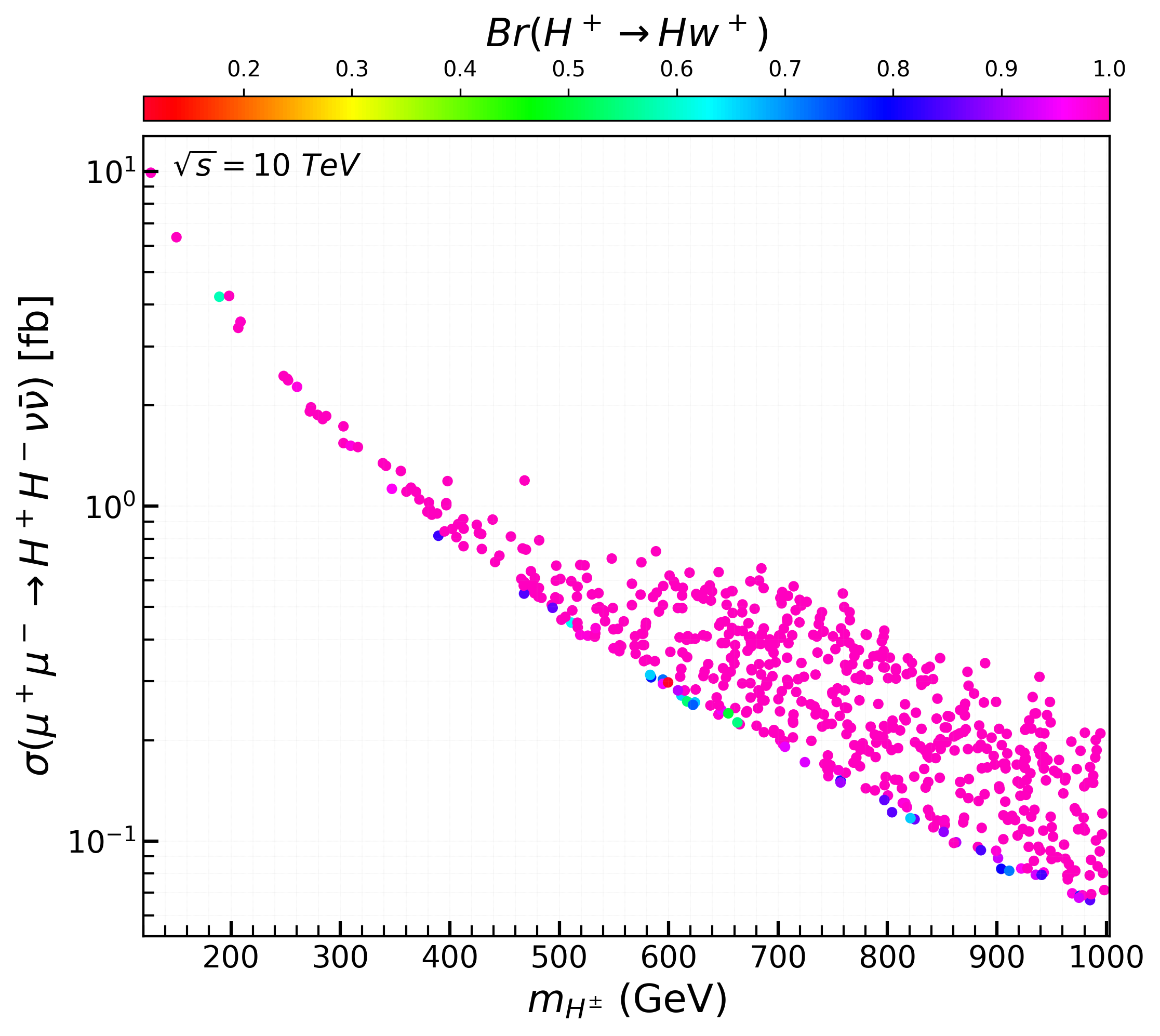}
\includegraphics[width=0.32\textwidth]{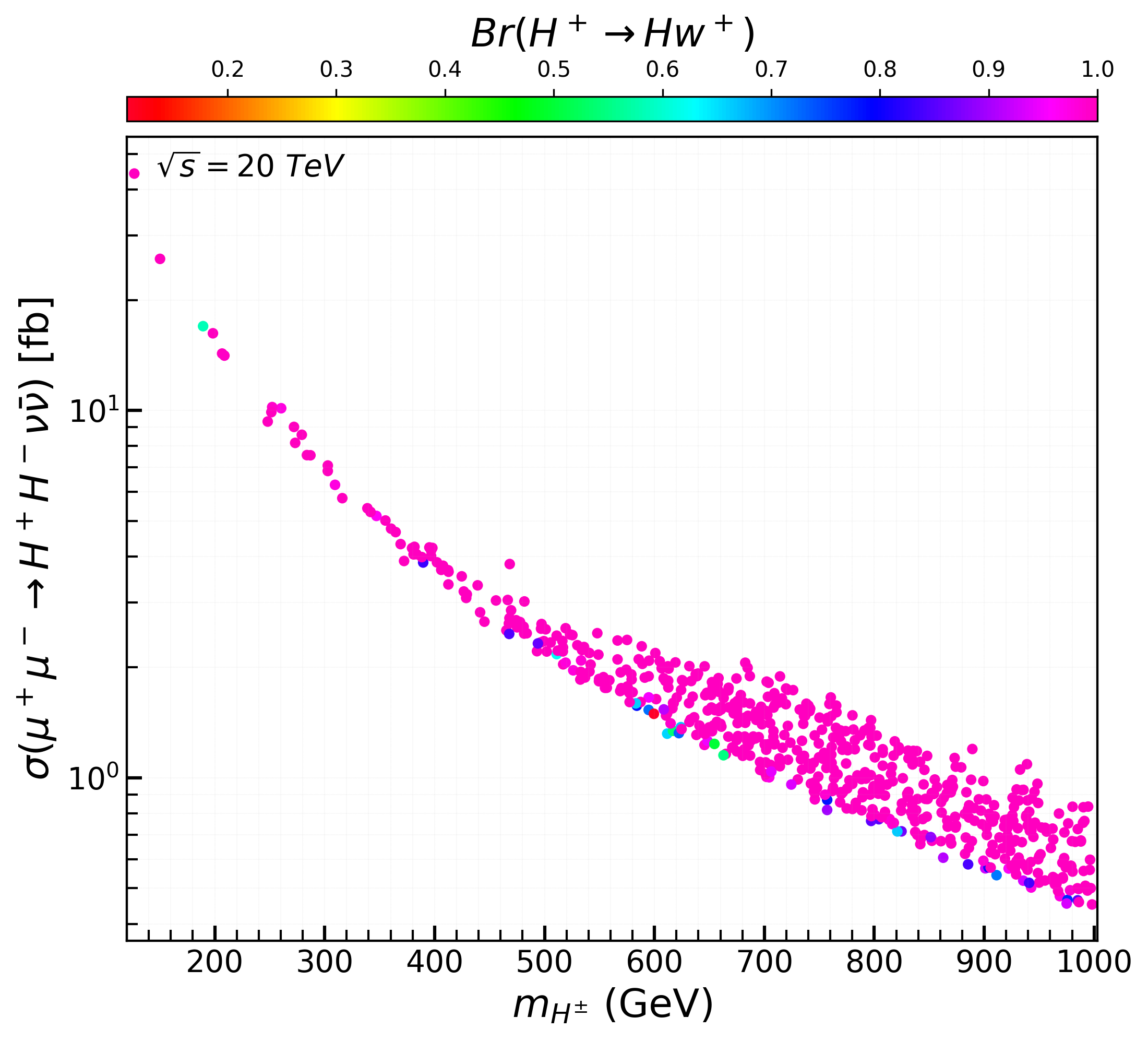}
\caption{The cross sections $\sigma(\mu^+ \mu^- \to H^+ H^- \nu \bar{\nu})$ at different collision energies of the MuC as a function of $m_{H^{\pm}}$, with the color bar indicating $\br(H^+ \to H W^\pm )$.} 
\label{mesfig:fig7}
\end{figure}

Fig.~\ref{mesfig:fig7} presents the production cross section of the process $\mu^+\mu^- \to H^+H^-\,\nu\bar{\nu}$
via the $WW$-fusion mechanism as a function of the charged Higgs mass $m_{H^\pm}$, for three center-of-mass energies, $\sqrt{s}=3$, $10$, and $20$~TeV. The color bar denotes the branching ratio $\br(H^+\to HW^+)$.

A pronounced decrease of the production cross section with increasing charged Higgs mass can be observed in all three panels. This behaviour is mainly due to the phase-space suppression associated with the production of heavier charged scalar pairs. As $m_{H^\pm}$ increases, a larger fraction of the available collision energy is required to produce the final-state particles, thereby reducing the accessible kinematic region and suppressing the cross section. In addition, the kinematic structure of the VBF topology further contributes to this suppression at large charged Higgs masses.
The figure also shows a strong dependence on the collider center-of-mass energy. At $\sqrt{s}=3$~TeV, the production cross section remains below the fb level over most of the parameter space. Increasing the collider energy to $\sqrt{s}=10$~TeV enhances the production rate up to the $\mathcal{O}(10)\,\mathrm{fb}$ level, while at $\sqrt{s}=20$~TeV the cross section can reach several tens of fb for relatively light charged Higgs bosons. This enhancement reflects the increasing efficiency of the $WW$-fusion mechanism at high energies, where the emitted electroweak gauge bosons become more energetic and significantly improve the heavy scalar production probability.
Furthermore, the largest cross sections are generally associated with parameter points characterised by large values of $\br(H^+\to HW^+)$, represented by the red-colored points. This indicates that scenarios in which the decay channel $H^\pm \to HW^\pm$ dominates are also compatible with sizeable charged Higgs pair production rates at a future high-energy MuC.
\begin{figure}[h!]
\centering
\includegraphics[width=0.32\textwidth]{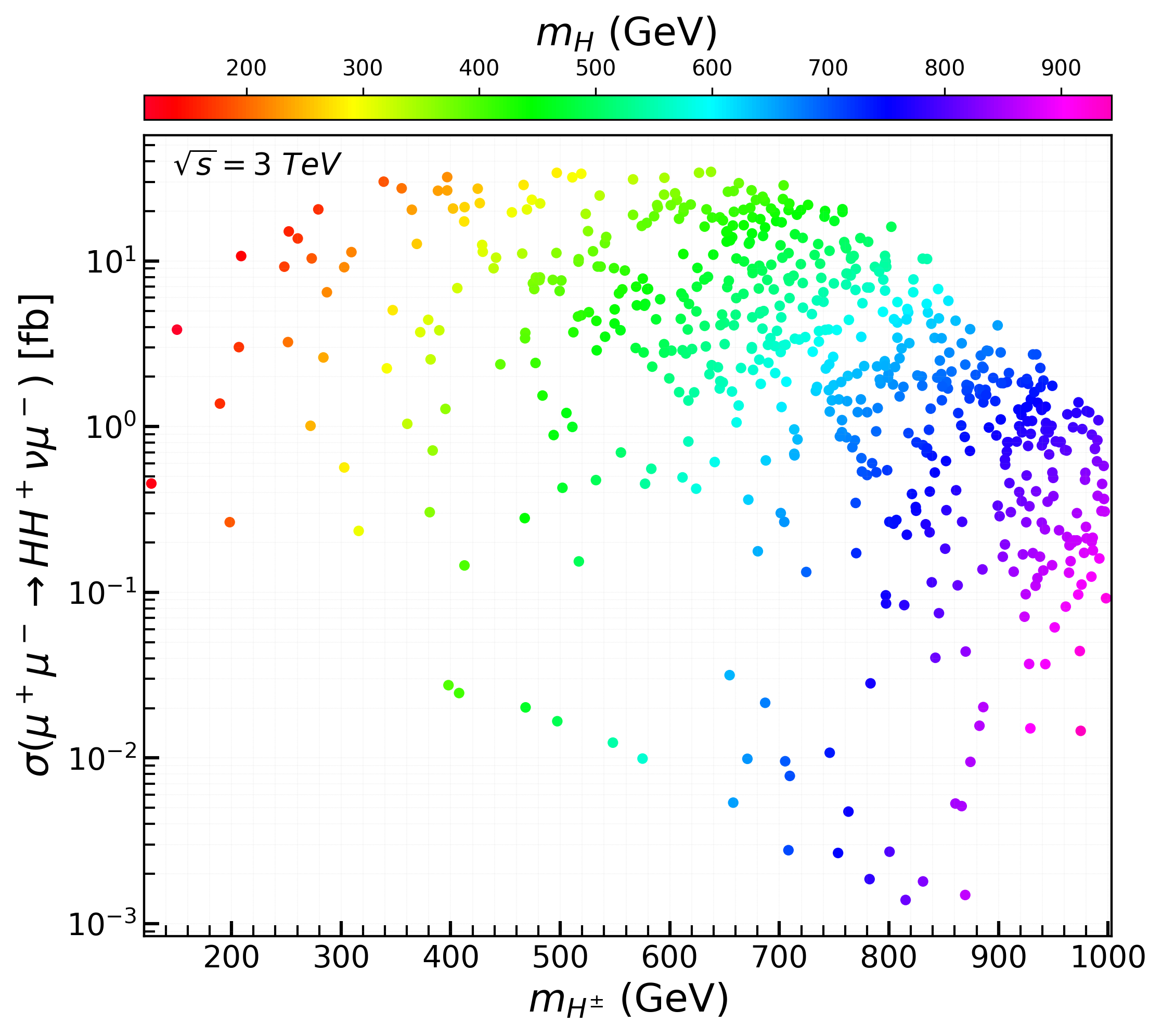}
\includegraphics[width=0.32\textwidth]{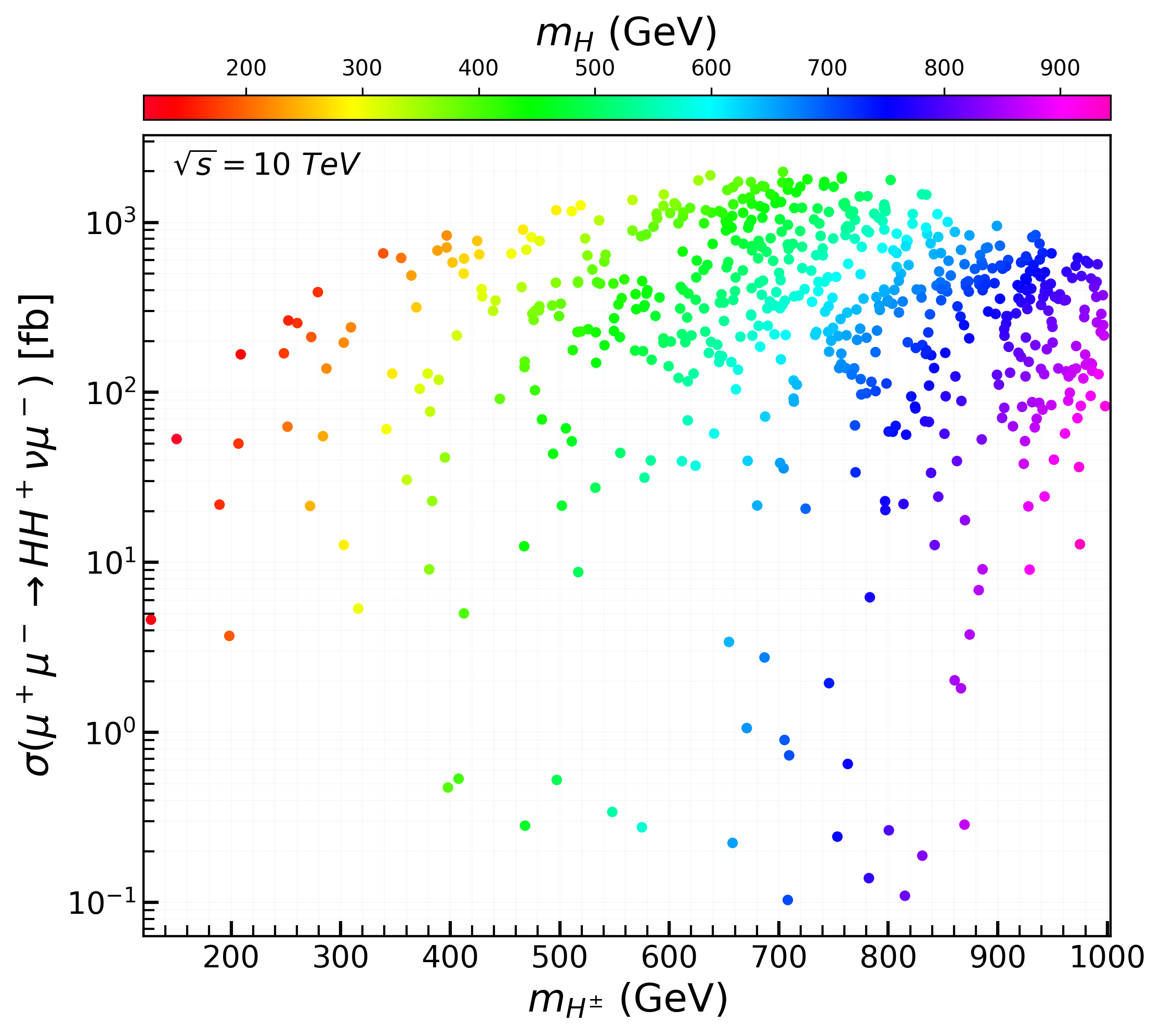}
\includegraphics[width=0.32\textwidth]{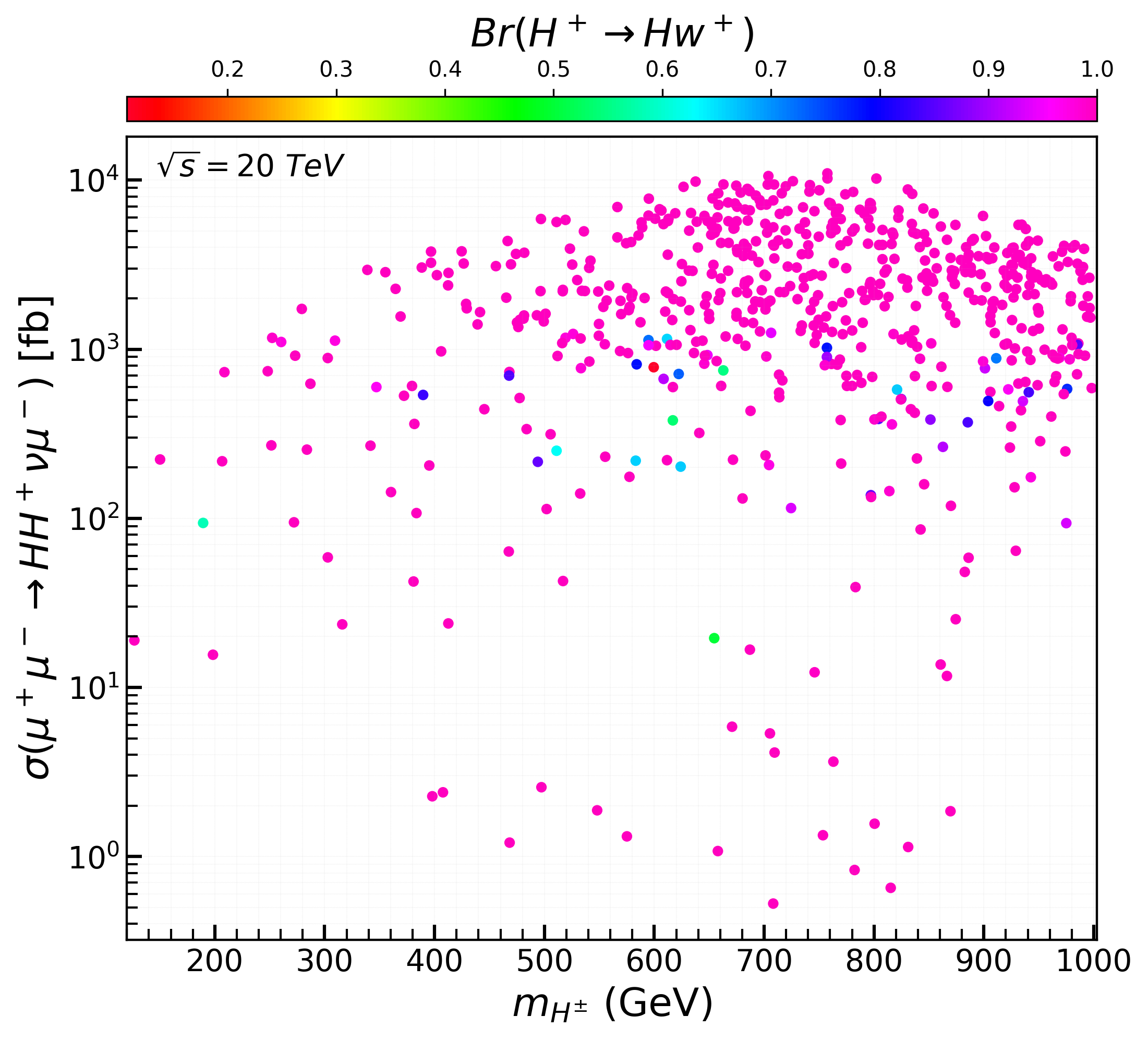}
\caption{The cross sections $\sigma(\mu^+ \mu^- \to H H^{\pm} \mu^{\mp} \nu)$ at different collision energies of the MuC as a function of $m_{H^{\pm}}$, with the color bar indicating $\br(H^+ \to H W^\pm )$.} 
\label{mesfig:fig8}
\end{figure}

Fig.~\ref{mesfig:fig8} shows the production cross section of the process $\mu^+\mu^- \to HH^+\,\nu_\mu \mu^-$
through the $WZ$- and $W\gamma$-fusion mechanisms as a function of the charged Higgs mass $m_{H^\pm}$, for three center-of-mass energies, $\sqrt{s}=3$, $10$, and $20$~TeV. The color scale represents the DM mass $m_H$ and the  branching ratio $\br(H^+\to HW^+)$.

A clear dependence of the production cross section on both the collider energy and the charged Higgs mass can be observed. For all three energies, the cross section generally decreases as $m_{H^\pm}$ increases. This behaviour originates mainly from the reduction of the available phase space for the production of heavier charged Higgs bosons. In addition, within the VBF topology, a larger fraction of the collider energy is required to produce the heavy scalar pair, leading naturally to a suppression of the production rate at large $m_{H^\pm}$.

On the other hand, increasing the center-of-mass energy significantly enhances the cross section. At $\sqrt{s}=3$~TeV, the production rate remains mostly below the $\mathcal{O}(10)\,\mathrm{fb}$ level over the scanned parameter space. For $\sqrt{s}=10$~TeV, the cross section can reach the $\mathcal{O}(10^3)\,\mathrm{fb}$ range, while at $\sqrt{s}=20$~TeV it further increases up to several $10^3\,\mathrm{fb}$. This enhancement reflects the growing importance of the VBF contribution at high energies, where the emitted electroweak gauge bosons become increasingly energetic and favour the production of heavy scalar states.

Furthermore, the figure indicates that the largest cross sections are typically associated with scenarios where $\br(H^+\to HW^+)$ is close to unity, corresponding to the red-colored points. This shows that the parameter regions dominated by the $H^\pm \to HW^\pm$ decay channel are also compatible with sizeable production rates at a high-energy MuC.
\begin{figure}[h!]
\centering
\includegraphics[width=0.32\textwidth]{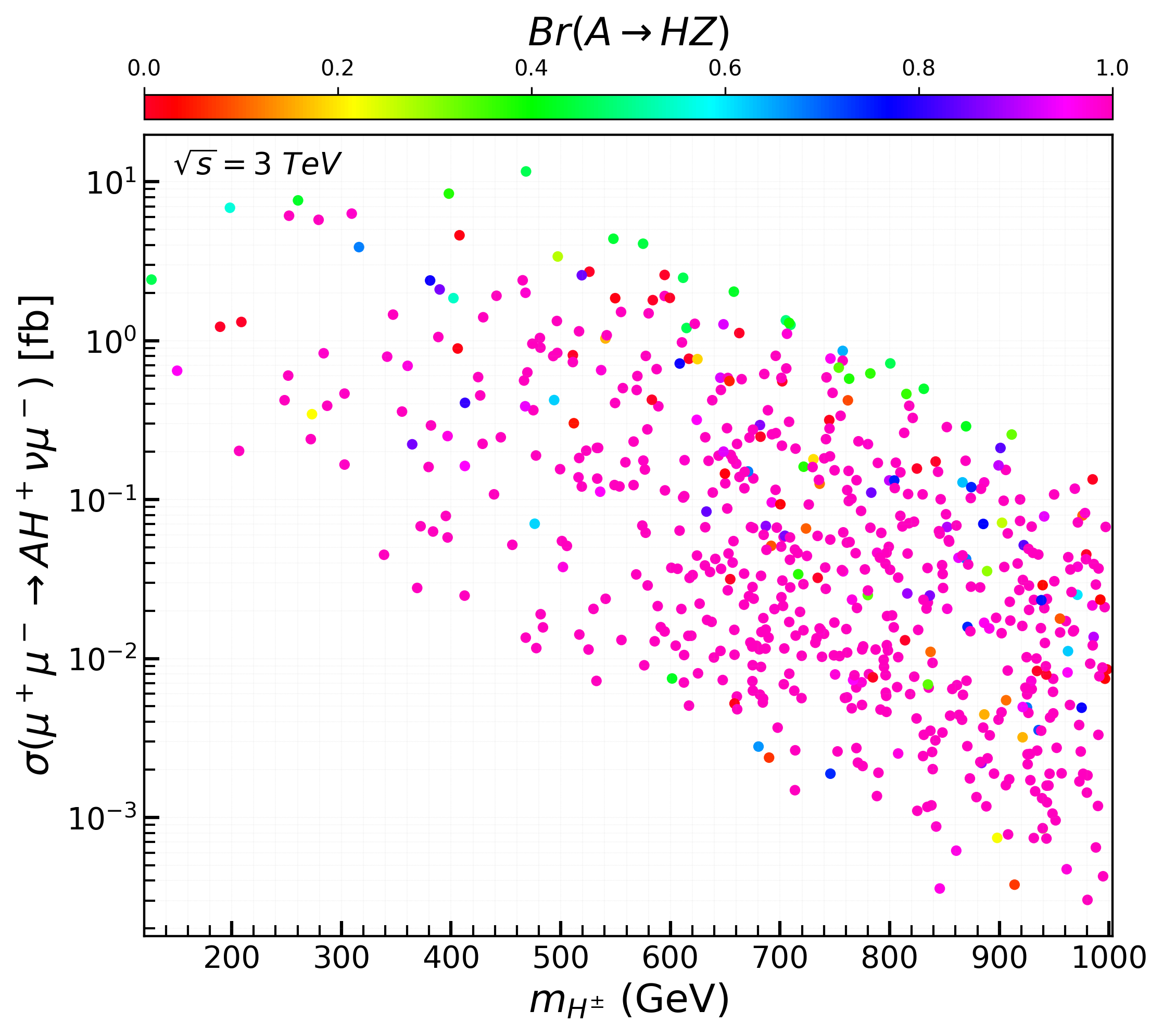}
\includegraphics[width=0.32\textwidth]{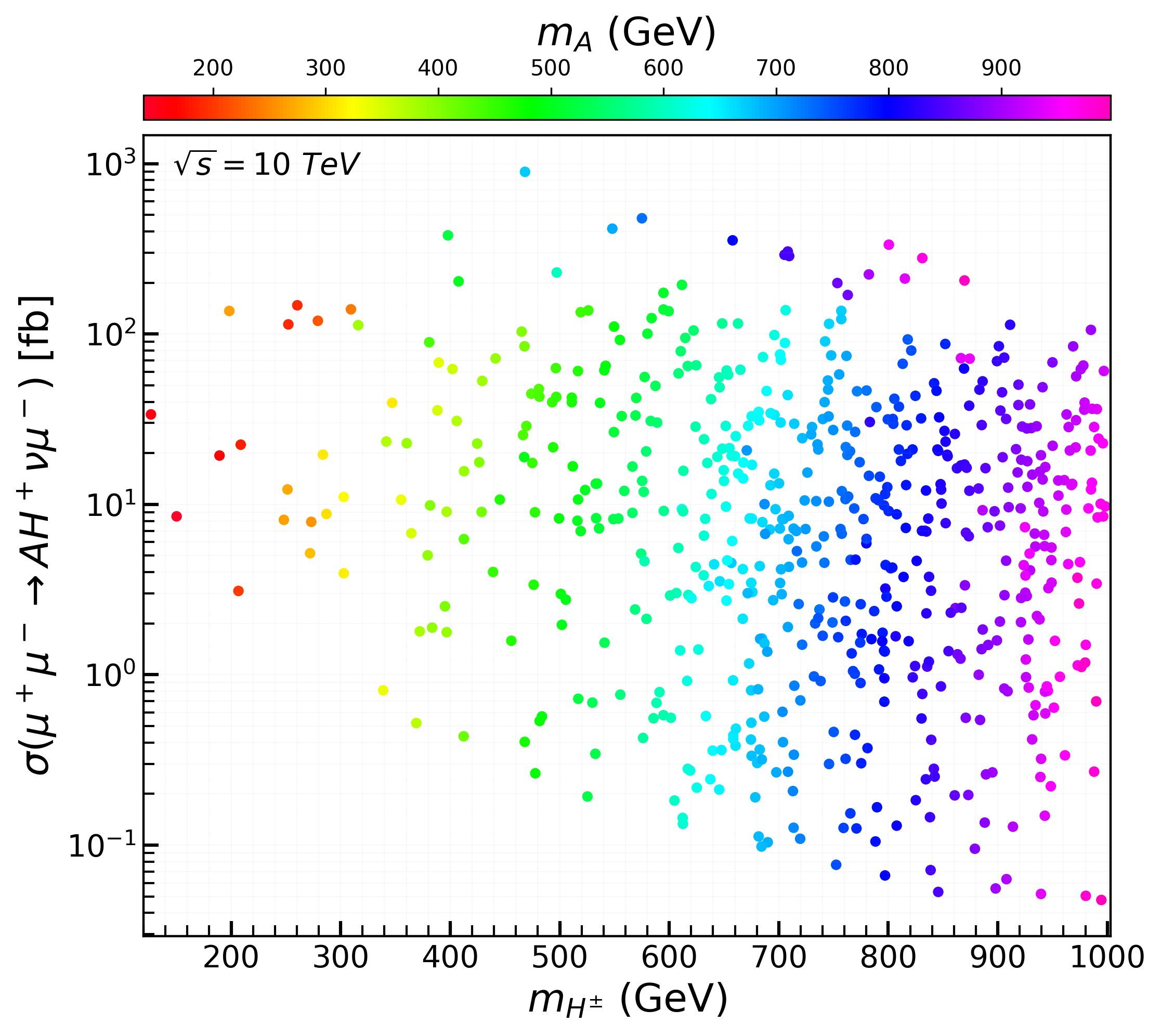}
\includegraphics[width=0.32\textwidth]{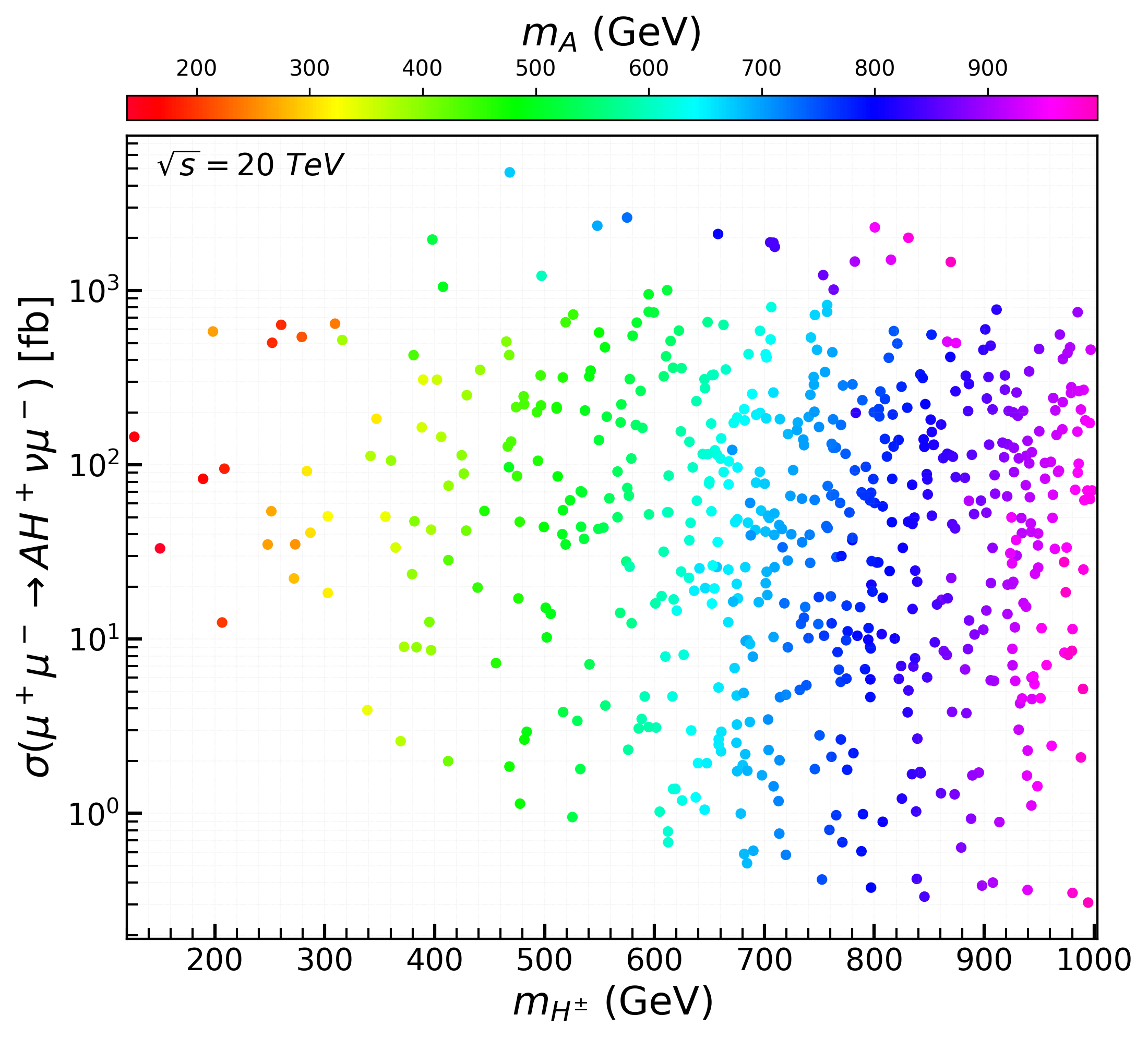}
\caption{The cross sections $\sigma(\mu^+ \mu^- \to A H^{\pm} \mu^{\mp} \nu)$  at different collision energies of the MuCs  as a function of $m_{H^{\pm}}$, with the color bar indicating $\br(A \to ZH  )$ and the CP-odd mass $m_A$.} 
\label{mesfig:fig9}
\end{figure}

Fig.~\ref{mesfig:fig9} displays the production cross section of the process
$\mu^+\mu^- \to AH^+ \,\nu_\mu \mu^-$
through the $WZ$- and $W\gamma$-fusion mechanism as a function of the charged Higgs mass $m_{H^\pm}$, for three center-of-mass energies, $\sqrt{s}=3$, $10$, and $20$~TeV. The color scale represents the branching ratio $\br(H^+\to HW^+)$.
The production cross section exhibits a strong dependence on both the collider energy and the charged Higgs mass. In general, the cross section tends to decrease as $m_{H^\pm}$ increases. This behaviour can be understood from the phase-space suppression associated with the production of heavier scalar states. As the charged Higgs mass becomes larger, a greater amount of the available center-of-mass energy is required to produce the final-state particles, thereby reducing the accessible kinematic region and suppressing the production rate.
A significant enhancement of the cross section is observed when increasing the collider energy. At $\sqrt{s}=3$~TeV, the production rate remains typically at the fb level over most of the parameter space. For $\sqrt{s}=10$~TeV, the cross section can rise up to several hundred fb, while at $\sqrt{s}=20$~TeV it reaches the $\mathcal{O}(10^3)\,\mathrm{fb}$ range for some parameter points. This enhancement reflects the increasing efficiency of the VBF mechanism at high energies, where the emitted electroweak gauge bosons carry sufficiently large energies to efficiently produce heavy scalar states.
Furthermore, the figure shows that the largest production cross sections are mostly associated with parameter points characterised by large values of $\br(H^+\to HW^+)$, represented by the red-colored points. This indicates that scenarios in which the decay channel $H^\pm \to HW^\pm$ dominates are also compatible with sizeable $AH$ production rates at a future high-energy MuC.

\begin{table}[!ht]
\centering
{\footnotesize
\begin{tabular}{|l|ccccc|cc|c|cc|}
\hline
\textbf{Benchmark point} & 
\multicolumn{5}{c|}{\textbf{IDM parameters}}& 
\multicolumn{2}{c|}{\textbf{DM observables}}&
\textbf{Signal}&
\multicolumn{2}{c|}{\textbf{Oblique parameters}}
\\
\hline
\textbf{BP}
& $m_H$ & $m_A$ & $m_{H^\pm}$ & $\lambda_2$ & $\lambda_{345}$
& $\Omega_H h^2$ & $10^{12}\times\sigma_{\rm SI} (pb)$ 
& $\mu_{\gamma\gamma}$
& $S$ & $T$
\\
\hline\hline
BP--1
& $309$
& $478$
& $481$
& $2$
& $0.2648$
& $0.000225$
& $12$
& $0.936309$
& $-0.00997883$
& $0.0055908$
\\
\hline
BP--2
& $353.43$
& $639.89$
& $637.58$
& $2$
& $-0.174$
& $0.000141$
& $2.49$
& $0.930831$
& $-0.0119349$
& $-0.00588866$
\\
\hline
BP--3
& $377.85$
& $488.38$
& $541.93$
& $2$
& $0.042$
& $0.000297$
& $0.26$
& $0.946944$
& $-0.00116314$
& $0.0804230$
\\
\hline
\end{tabular}
}
\caption{Properties of the BPs considered in the detector-level analysis. The IDM parameters and DM observables are shown for each BP.}
\label{tab:benchmarkBP}
\end{table}

Based on the discussion above, we select three benchmark points (BPs) to illustrate distinct phenomenologically relevant regions of the parameter space, featuring sizeable production rates and dominant decay modes relevant to collider signatures, which will be explored in the following section. The relevant scalar masses, relic density, direct-detection cross section, and other key observables are summarised in Tab.~\ref{tab:benchmarkBP} for each BP.

Before turning to the collider analysis, it is important to note that, long before a MuC starts operating, the HL-LHC and FCC-ee will measure the Higgs couplings and the electroweak precision observables with unprecedented accuracy. Although the inert scalars do not couple to fermions, they affect these observables through loops. The most sensitive probes are the $h\to\gamma\gamma$ rate, which receives a charged-scalar contribution controlled by $\lambda_3 v^2/m_{H^\pm}^2$, and the oblique parameter $T$, which grows with the mass splittings within the inert doublet \cite{Barbieri:2006dq}. The corresponding values for our BPs are also given in Tab.~\ref{tab:benchmarkBP} and compared with the projected precisions of the HL-LHC and of the HL-LHC+FCC-ee combination, assuming SM central values. All three BPs predict a $5$--$7\%$ deficit in $h\to\gamma\gamma$, corresponding to a $3$--$3.5\%$ reduction of $\kappa_{\gamma}$. Compared with the projected precisions on $\kappa_{\gamma}$ of $1.9\%$ at the HL-LHC and $1.3\%$ after the FCC-ee, this leads to deviations of $1.4$--$1.9\sigma$ and $2.1$--$2.7\sigma$, respectively. In addition, BP3, which has the largest mass splittings, predicts $T\approx 0.08$, far outside the region expected from the FCC-ee electroweak programme. More importantly, these results show that our BPs correspond to a scenario in which the HL-LHC and FCC-ee observe a deviation from the SM. In this case, a MuC would play a decisive role in identifying its origin, by directly producing the inert scalars and measuring their masses, which indirect measurements alone cannot determine; this is the situation studied in the following sections. If, instead, no deviation is observed, BP3 would be excluded and BP1 and BP2 strongly disfavoured, and the surviving parameter space would consist of heavy, nearly degenerate inert scalars with small $\lambda_3$, for which VBF at a multi-TeV MuC becomes the main direct-production channel.

\section{Signal-Background Analysis for  $[WH][WH]\nu\bar{\nu}$, $[WH]H\mu\nu$ and $[WH][ZH]\mu \nu$}
\label{sec:monteCarlosetup}
In the previous section, we evaluated the parton-level cross sections for the proposed channels aimed at probing a $H^\pm$ within the framework of the IDM. Although these cross sections are not negligible, the actual discovery potential strongly depends on our ability to efficiently discriminate the signal from the overwhelming backgrounds. In this section, we develop dedicated search strategies based on a comprehensive signal-to-background optimization. Our approach incorporates advanced simulation tools, including hard-scattering matrix elements, resonance decays, parton showering, hadronization, hadron decays, and a simplified detector response. 

Focusing on the MuC, we perform a detailed analysis of the following three processes: $\mu^+ \mu^- \to H^+ H^- \nu \bar{\nu}$, $\mu^+ \mu^- \to H H^{\pm} \mu^{\mp} \nu$, and $\mu^+ \mu^- \to A H^{\pm} \mu^{\mp} \nu$, where signal and background events are generated using the same Monte Carlo framework described in the previous section.

The input parameters in the \texttt{MadGraph} cards are adjusted according to the values obtained from the \texttt{IDM} code, while the default run-card settings are adopted, requiring $p_T > 10~\,\gev$, $|\eta| < 2.5$, and $\Delta R(\ell,\ell) \geq 0.4$, with $\Delta R = \sqrt{(\Delta \eta)^2 + (\Delta \phi)^2}$. The generated parton-level events are then passed to \texttt{Pythia\_8.243}~\cite{sjostrand2015introduction} for parton showering, hadronization, and hadron decays, followed by a fast detector simulation using \texttt{Delphes\_3.4.2}~\cite{de2014delphes} with the MuC detector target model. Jets are reconstructed using the anti-$k_T$ algorithm~\cite{cacciari2008anti} with a radius parameter $R = 0.4$. In the present exploratory detector-level analysis, pileup and beam-induced background specific to a MuC are not explicitly simulated. We therefore interpret the resulting sensitivities as idealised assessments of the considered signal channels under the adopted detector and event-selection assumptions. In addition, multiple parton interactions from soft QCD processes are switched off at the \texttt{Pythia} level. Within this framework, the generated signal and background samples will hereafter be referred to as `initial events'.

To quantify the sensitivity of the considered signatures, we evaluate the statistical significance in two different ways, depending on the collider energy. At $\sqrt{s}=3$~TeV, both the signal and background contributions after the full event-selection procedure are retained, and the significance is computed as
\begin{eqnarray}
\mathcal{S}=\frac{N_S}{\sqrt{N_S+N_B}}=\sqrt{\mathcal{L}} \frac{\sigma_s}{\sqrt{\sigma_s+\sigma_b}}
\end{eqnarray}
where $\sigma_s$ and $\sigma_b$ denote the signal and background cross sections after all selection cuts, respectively, while $\mathcal{L}$ is the integrated luminosity. In contrast, at $\sqrt{s}=10$ and $20$~TeV, the background contribution becomes negligible compared with the expected signal yields for the selected BPs after the event pre-selection stage. Consequently, the significance can be approximated by
\begin{eqnarray}
\mathcal{S}=\frac{N_S}{\sqrt{N_S+N_B}}
\simeq \sqrt{N_S}
=\sqrt{\sigma_s.\mathcal{L}}
\end{eqnarray}
and therefore depends only on the signal cross section after event selection and the integrated luminosity.

\begin{table}[!ht]
{\footnotesize
\begin{tabular}{|l|l|l|l|}
\hline
Signal & \multicolumn{2}{c|}{Benchmark point} & Backgrounds \\ \hline\hline
$[WH][WH]\nu \bar{\nu}$ & BP--1 & $m_{h}=125.09 \,\gev$, $m_H=309\,\gev$, $m_A=478\gev$ & $\ttop$, $Wjj$, $Zjj$  \\ 
& & $m_{H^{\pm}}=481\,\gev$, $\lambda_{2}=2$, $\lambda_{345} =0.2648$ & $WWjj$, $WW$ \\ \hline
$[WH]H\mu \nu$   & BP--2 & $m_{h}=125.09\,\gev$, $m_H=353.43\,\gev$, $m_A=639.89\,\gev$ & $tW^\pm b$, $W^+W^-$ \\ 
& & $m_{H^{\pm}}=637.58\,\gev$, $\lambda_{2}=2$, $\lambda_{345} =-0.174$ & $W^{\pm}\mu^\mp \nu_\mu$, $ZZ$  \\ \hline
$[WH][ZH]\mu \nu$  & BP--3 & $m_{h}=125.09\,\gev$, $m_H=377.85\,\gev$, $m_A=488.38\,\gev$ & $tW^\pm b$, $W^+W^-$ , $Zjj$ \\ 
& & $m_{H^{\pm}}=541.93\,\gev$, $\lambda_{2}=2$, $\lambda_{345} =0.042$ & $W^{\pm}\mu^\mp , \nu_\mu$, $ZZ$ , $Wjj$ \\ \cline{2-4}
\hline
\end{tabular} 
}
\caption{Benchmark points for the three signal processes involving a charged Higgs boson at the MuC. The main backgrounds are listed in the last column, with $V=W^{\pm},Z$.}
\label{table:BP:backgrounds}
\end{table}

\subsection{$[WH][WH]\nu \bar{\nu}$}
The $[W^+H][W^-H]$ mode targets the production of a charged Higgs boson pair in association with missing energy, $\mu^+\mu^- \to H^+H^- \nu \bar{\nu}$, followed by the decays $H^\pm \to W^\pm H$.
\bea
\mu^+ \mu^- \to H^+ H^- \nu \bar{\nu} \to W_{qq^{'}} HW_{l^-\nu} H\nu \bar{\nu} \to 2j +l^- + \slashed{E}_T.
\eea

The final state consists of two jets originating from the hadronic decay $W \to qq'$, one charged lepton ($\ell^-$) accompanied by missing transverse energy from $W \to \ell^- \nu$, and substantial missing energy arising from the two stable DM particles $H$ as well as the neutrino pair $\nu\bar{\nu}$. The benchmark point BP–1 in Tab.~\ref{table:BP:backgrounds} yields
\bea
\sigma^{3\,\tev} ( \mu^+ \mu^- \to H^+ H^- \nu \bar{\nu} )=  0.048~\text{fb} , \ \ \ \ \br(H^\pm \to W^\pm H) = 1\\
\sigma^{10\,\tev} ( \mu^+ \mu^- \to H^+ H^- \nu \bar{\nu} )=  0.79~\text{fb} , \ \ \ \ \br(H^\pm \to W^\pm H) = 1\\
\sigma^{20\,\tev} ( \mu^+ \mu^- \to H^+ H^- \nu \bar{\nu} )=  3.01~\text{fb}, \ \ \ \ \br(H^\pm \to W^\pm H) = 1
\textbf{}\eea

\begin{figure*}[h]
\centering
\includegraphics[width=0.45\textwidth]{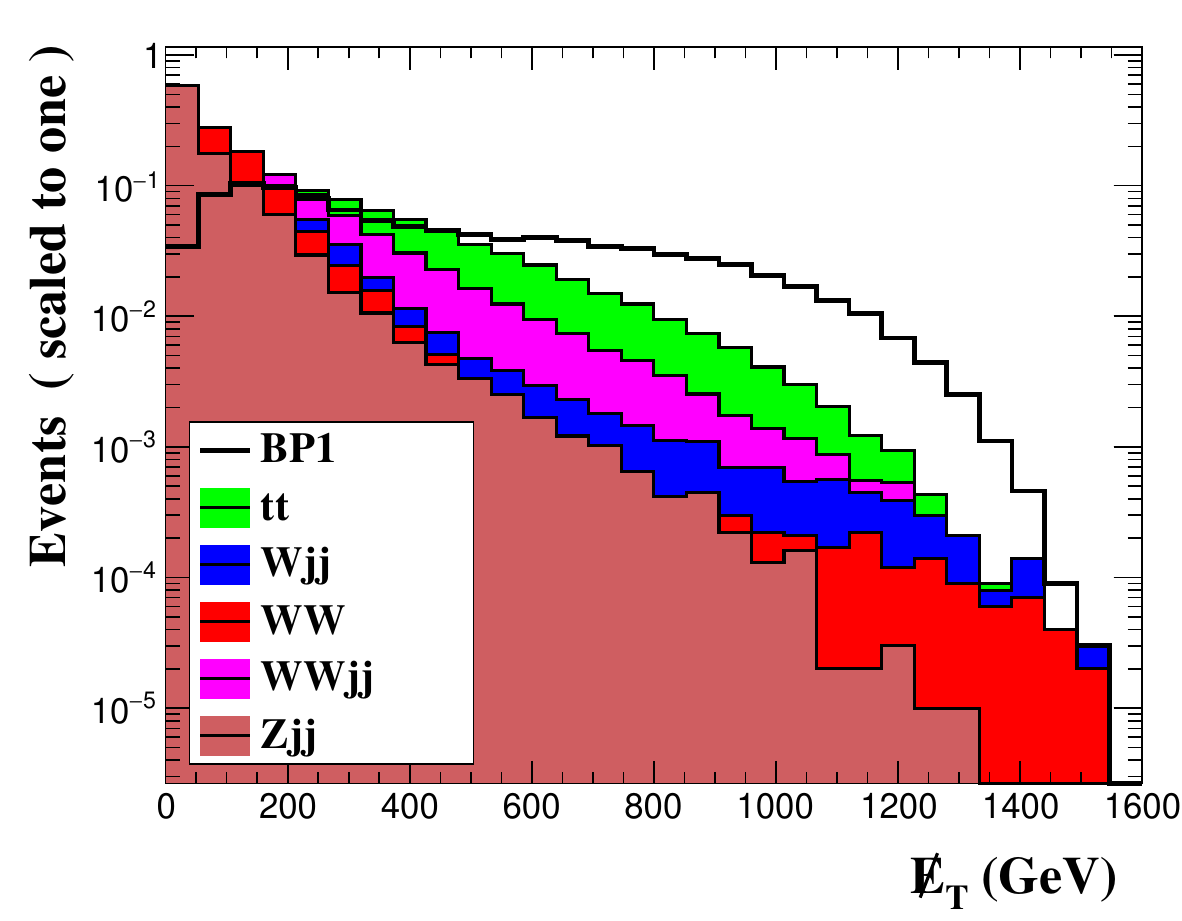}
\includegraphics[width=0.45\textwidth]{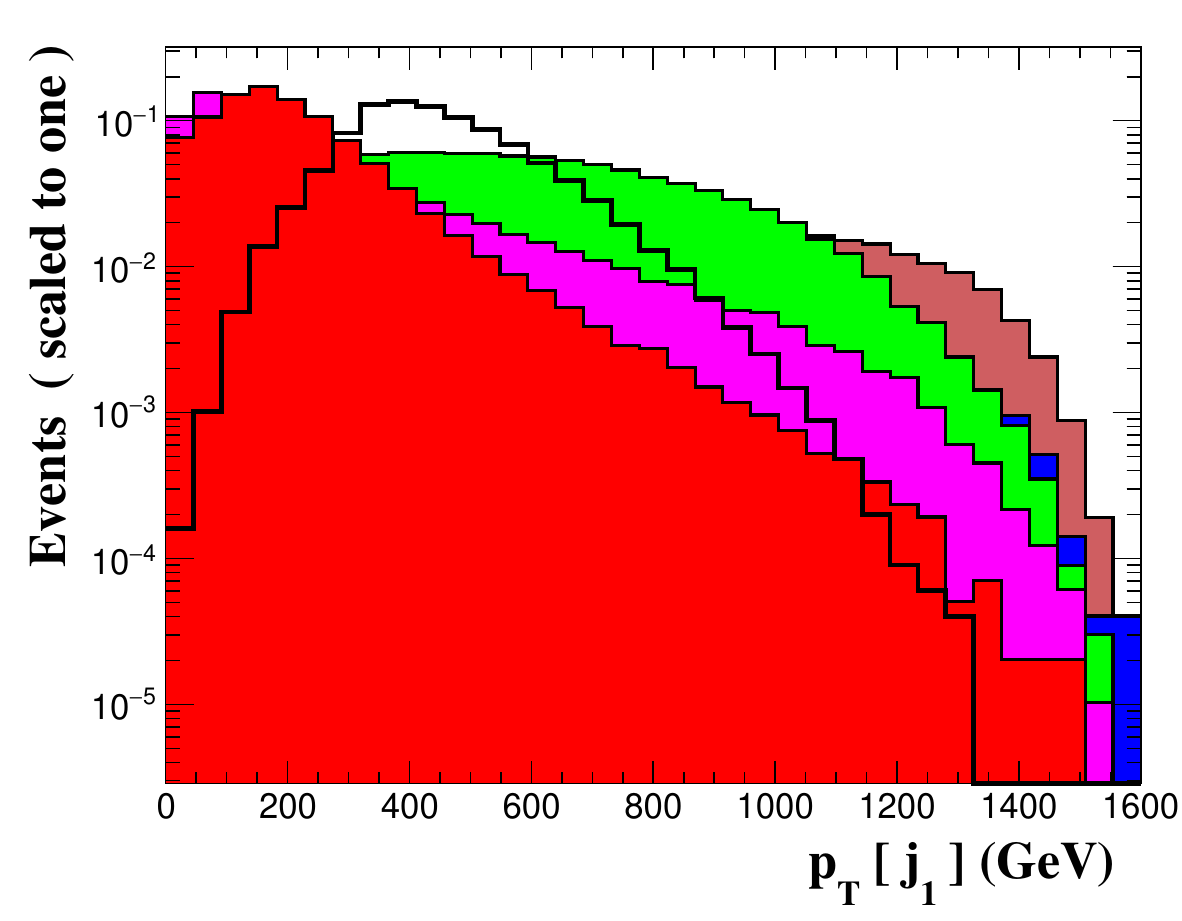}\\
\includegraphics[width=0.45\textwidth]{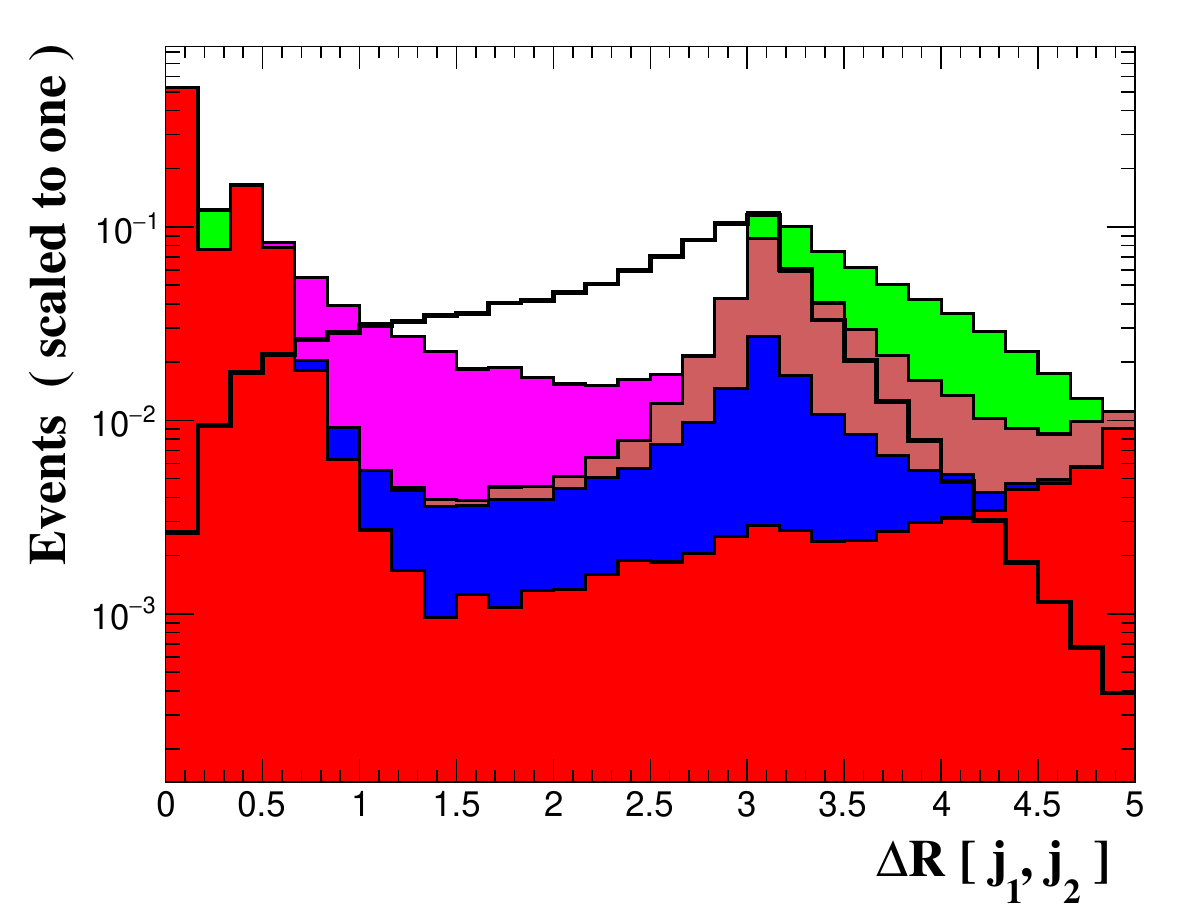}
\includegraphics[width=0.45\textwidth]{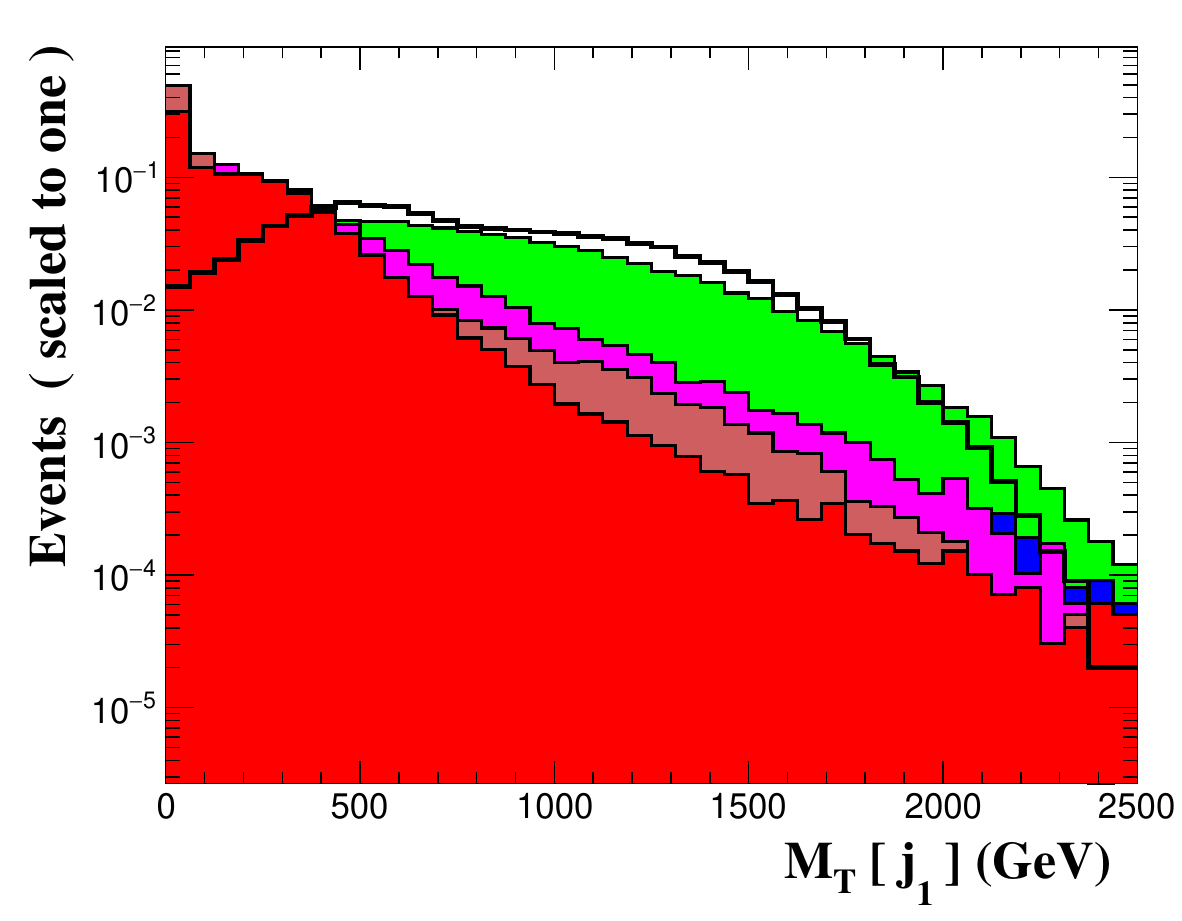}
\caption{Normalized kinematic distributions for the final state  $[WH][WH]\nu \bar{\nu}$, showing the missing transverse energy $\slashed{E}_T$ (top left panel), the transverse momentum of the leading jet $p_T[j_1]$ (top right panel), $\Delta R(j_1, j_2)$ (lower left panel), and the transverse mass
$M_T[j_1]$ (lower right panel) at the 3 TeV MuC.} 
\label{mesfig:fig10}
\end{figure*}

The main SM background contributions come from the processes $t\bar{t}$, $Zjj$, $W^{+}W^-$, $W^{\pm}jj$, and $W^+W^-jj$. For the $t\bar{t}$ production, one of the top quarks is assumed to decay semileptonically, while the other one decays hadronically. Meanwhile, the $Z$ boson is assumed to decay into an $\ell^+ \ell^-$ pair, and one of the $W$ bosons is assumed to decay leptonically, while the other one decays into a pair of light jets. The backgrounds are as follows:
\begin{enumerate}[label=\roman*)]
\item $\mu^+ \mu^- \rightarrow t\bar{t} \ (t  \rightarrow W^+ b , \ W^+ \rightarrow l^+ \nu_{l}),\ (\bar{t} \rightarrow W^- \bar{b},\ W^- \rightarrow j j)$,
\item $\mu^+ \mu^- \rightarrow Zjj, \ (Z \rightarrow \ell^+ \ell^-)$,
\item $\mu^+ \mu^- \rightarrow W^+W^-, \ (W^+ \rightarrow j j),(W^- \rightarrow l^- \nu_{l})$,
\item $\mu^+ \mu^- \rightarrow W^{\pm}jj, \ (W^{\pm} \rightarrow l^{\pm} \nu_{l})$,
\item $\mu^+ \mu^- \rightarrow W^+W^-jj, \ (W^+ \rightarrow j j),(W^- \rightarrow l^- \nu_{l})$. 
\end{enumerate}
At parton level, the signal and background events are required to satisfy the following basic cuts:	
\bea
p_T^{j} > 20\,\gev, \quad p_T^{\ell} > 10\,\gev, \quad |\eta^{j}| < 5, \quad |\eta^{\ell}| < 2.5, \quad \Delta R^{jj,\ell\ell,j\ell} \geq 0.4.
\eea
For the signal and background events passing the basic cuts, we examine several kinematic distributions. In particular, Fig.~\ref{mesfig:fig10} displays the distributions of the missing transverse energy $\slashed{E}_T$ (top-left panel), the angular separation $\Delta R(j_1,j_2)$ (top-right panel), the transverse momentum of the leading jet $p_T^{j_1}$ (bottom-left panel), and its transverse mass $M_T^{j_1}$ (bottom-right panel). The jets $j_1$ and $j_2$ are ordered according to their transverse momenta such that $p_T^{j_1} > p_T^{j_2}$. Here, $M_T^{j_1}$ denotes the transverse mass associated with the leading jet.

\begin{table*}[!t]
\setlength\tabcolsep{12pt}
\centering
{\footnotesize\renewcommand{\arraystretch}{0.6} 
\begin{tabular}{l ||l l l l l |l}
\toprule
\multicolumn{7}{c}{$[WH][WH]\nu \bar{\nu}$ }\\
\toprule
{Cut}   &  $\ttop$ & ~~$Z$jj & $W^+W^-$ & $W^{\pm}$jj &  $W^+W^-$jj   & Signal \\
\toprule
Basic cuts&  0.86& 0.3 & 7.41  & 18.67 &  0.28 & 0.008\\
$N(b) \leq 1$ &0.30 & 0.29 & 7.3 & 18.47 & 0.28 & 0.008\\ 
$\slashed{E}_T>400\,\gev$ & 0.05 & 0.005 & 0.14 & 0.66 &  0.02 & 0.003\\
$260 \,\gev < p_T^{j_1} < 600 \,\gev$ & 0.019 & 0.001 & 0.07 & 0.25 & 0.01 & 0.002\\
$-1  < \eta^{j_1} < 1 $ & 0.01 & 0.0007 & 0.015 & 0.1 & 0.006 & 0.002\\
$1 < \Delta R(j_1,j_2) < 3$ & 0.002 & 0.0002 & 0.0007 & 0.02& 0.001 & 0.001\\
$400 \,\gev <M_T^{j_1} < 2400 \,\gev$ & 0.002 & 0.0001 & 0.0006 & 0.02 & 0.001 & 0.001\\
$M_{j_1 l_1} < 500 \,\gev$ & 0.0008 & 5$\times10^{-5}$ & 7$\times10^{-5}$ & 0.003 & 0.0004 & 0.001\\
\hline \hline
\end{tabular}
}
\caption{Cut-flow chart of the cross section (in fb) of the signal and backgrounds for the channel $\mu^+ \mu^- \to H^+ H^- \nu \bar{\nu} \to [WH][WH]\nu \bar{\nu}$ at the $3$~TeV MuC. More details about the selection cuts are given in the text.}
\label{tab:cutflow:3TeV_BP1}
\end{table*}

Tab.~\ref{tab:cutflow:3TeV_BP1} presents the event yields after applying the selection criteria at the $3$~TeV MuC. The ``Basic cuts" correspond to the trigger requirement imposed on the events. We further require the number of $b$-tagged jets to satisfy $N(b) \leq 1$, rejecting events containing two or more $b$-tagged jets. This requirement efficiently suppresses backgrounds with multiple $b$ jets, in particular the $t\bar{t}$ background, while avoiding an overly restrictive veto that could reduce the signal efficiency.

The missing transverse energy requirement $\slashed{E}_T > 400\,\gev$ provides 
the most significant reduction of backgrounds, suppressing about $96 \%$ of $W^{\pm}jj$ and $99 \%$ of $W^+W^-$, while retaining 
$5\%$ of the signal. Subsequent cuts on the leading jet transverse momentum 
$260\,\gev < p_T^{j_1} < 600\,\gev$ and pseudorapidity $|\eta^{j_1}| < 1$ 
further reject backgrounds by exploiting the more central and boosted topology 
of the signal jets. The angular separation requirement $1 < \Delta R(j_1,j_2) < 3$ 
and the transverse mass window $400\,\gev < M_T^{j_1} < 2400\,\gev$ prove 
particularly powerful in reducing the $t\bar{t}$ and diboson backgrounds to a 
negligible level. Finally, the invariant mass cut $M_{j_1 \ell_1} < 500\,\gev$ 
further selects against combinatorial backgrounds, reducing $W^{\pm}jj$ to 
$0.003$~fb while preserving a signal cross section of $0.001$~fb.

After the full cut sequence, the signal-to-background ratio improves by roughly 
two orders of magnitude compared to the initial selection, with the total 
background reduced from $\sim 27$~fb to $\sim 0.004$~fb, demonstrating the 
effectiveness of the sequential kinematic selection strategy in isolating the 
$H^{\pm}$ pair production signal.

Building upon the event selection strategy developed at $\sqrt{s} = 3$~TeV, 
as detailed in Tab.~\ref{tab:cutflow:3TeV_BP1}, we extend the analysis to the 
higher center-of-mass energy stages of the MuC. Since the signal and 
background processes exhibit qualitatively similar kinematic behaviors at 
$\sqrt{s} = 10$~TeV and $\sqrt{s} = 20$~TeV, the corresponding distributions 
are not shown explicitly. Instead, we directly present the optimized cut-flow 
results for these energies in Tables~\ref{tab:cutflow:6TeV_BP1} and~\ref{tab:cutflow:20TeV_BP1}. Notably, as the center-of-mass energy increases, 
the signal cross section grows significantly from $0.008$~fb at $3$~TeV to 
$0.20$~fb at $10$~TeV and $0.63$~fb at $20$~TeV, while the cut strategies are 
adapted accordingly, requiring fewer selection steps at higher energies to achieve 
adequate background suppression, thanks to the increasingly distinctive kinematic 
topology of the signal.

\begin{table*}[!h]
\setlength\tabcolsep{12pt}
\centering
{\footnotesize\renewcommand{\arraystretch}{0.7} 
\begin{tabular}{l ||l l l l l |l}
\toprule
\multicolumn{7}{c}{$[WH][WH]\nu \bar{\nu}$ }\\
\toprule
{Cut}   &  $\ttop$ & ~~$Z$jj & $W^+W^-$ & $W^{\pm}$jj &  $W^+W^-$jj   & Signal \\
\toprule
Basic cuts&  0.007& 0.01 & 0.03  & 0.49 &  0.03 & 0.20\\
$N(b) \leq 1$ &0.002 & 0.01 & 0.03 & 0.48 & 0.03 & 0.20\\ 
$\slashed{E}_T>900\,\gev$ & 0.0003 & 0.001 & 0.0006 & 0.07 &  0.006 & 0.14\\
$300 \,\gev < p_T^{j_1} < 1800 \,\gev$ & 0.0002 & 0.0001 & 0.0006 & 0.02 & 0.002 & 0.12\\
$0.2 < \Delta R(j_1,j_2) < 2.7$ &7$\times10^{-5}$ & 1$\times10^{-5}$ & 1$\times10^{-5}$ & 0.008& 0.001 & 0.11\\
\hline \hline
\end{tabular}
}
\caption{Cut-flow chart of the cross section (in fb) of the signal and backgrounds for the channel $\mu^+ \mu^- \to H^+ H^- \nu \bar{\nu} \to [WH][WH]\nu \bar{\nu}$ at the $10$~TeV MuC.}
\label{tab:cutflow:6TeV_BP1}
\end{table*}

\begin{table*}[!h]
\setlength\tabcolsep{9pt}
\centering
{\footnotesize\renewcommand{\arraystretch}{0.7} 
\begin{tabular}{l ||l l l l l |l}
\toprule
\multicolumn{7}{c}{$[WH][WH]\nu \bar{\nu}$ }\\
\toprule
{Cut}   &  $\ttop$ & ~~$Z$jj & $W^+W^-$ & $W^{\pm}$jj &  $W^+W^-$jj   & Signal \\
\toprule
Basic cuts&  2.344$\times10^{-5}$& 2.564$\times10^{-3}$ & 3.03244$\times10^{-4}$  & 0.118 &  0.01067 & 0.6327\\
$N(b) \leq 1$ &1.02$\times10^{-4}$ & 0.00247 & 0.000303 & 0.1119 & 0.0105 & 0.6312\\ 
$\slashed{E}_T>800\,\gev$ & 4.04$\times10^{-5}$ & 0.000343 & 0.000122 & 0.0537 &  0.00529 & 0.545\\
$M_{jj} < 500 \,\gev$ & 2.56$\times10^{-5}$ & 0.000303 & 7.98$\times10^{-7}$ & 0.0244 & 0.00436 & 0.542\\
$ M_T^{l_1} > 400 \,\gev$ &1.79$\times10^{-6}$ & 1.74$\times10^{-6}$ & 1.21$\times10^{-8}$ & 0.00572& 0.00269 & 0.339\\
\hline \hline
\end{tabular}
}
\caption{Cut-flow chart of the cross section (in fb) of the signal and backgrounds for the channel $\mu^+ \mu^- \to H^+ H^- \nu \bar{\nu} \to [WH][WH]\nu \bar{\nu}$ at the $20$~TeV MuC.}
\label{tab:cutflow:20TeV_BP1}
\end{table*}

\begin{table}[ht]
\setlength{\tabcolsep}{6pt}
\renewcommand{\arraystretch}{0.7}
\centering
\begin{tabular}{c c c c c c}       
\hline  \hline 
& &BP1& &\\
\hline  \hline 
\hline   
Processes $\ \ \ \ \ $  &&&$\mu^+ \mu^- \to H^+ H^- \nu \bar{\nu} \to [WH][WH]\nu \bar{\nu}$ &\\
\hline  \hline
Luminosity$\ \ $&$\mathcal{L}$=500 fb$^{-1}$&$\mathcal{L}$=1000 fb$^{-1}$& $\mathcal{L}$=1500 fb$^{-1}$&$\mathcal{L}$=10 ab$^{-1}$\\
\hline  \hline 
$\sqrt{s}=3\,\tev$  &0.3&0.43&0.53&1.37 \\
$\sqrt{s}=10\,\tev$  &7.41&10.48&12.84& 33.16\\
$\sqrt{s}=20\,\tev$  &13.01&18.41&22.54&58.22 \\	
\hline \hline
\end{tabular}
\caption{Significance $\mathcal{S}$ for our signal with $\sqrt{s}=3$, $10$, and $20$~TeV and $\mathcal{L}=500$, $1000$, $1500$~fb$^{-1}$, and $10$~ab$^{-1}$.}  
\label{Signi:siBP1}
\end{table}	

The statistical significance $\mathcal{S}$ for the signal process 
$\mu^+\mu^- \to H^+H^-\nu\bar{\nu} \to [WH][WH]\nu\bar{\nu}$ is evaluated 
at three center-of-mass energies and for several benchmark luminosity scenarios. The corresponding results are summarised in Table~\ref{Signi:siBP1}. As can be seen, at $\sqrt{s}=3$~TeV, the significance remains modest across all luminosity values, reaching only 
$\mathcal{S} = 1.37$ even at $\mathcal{L} = 10\,\text{ab}^{-1}$, which reflects 
the limited signal production rate at this energy stage. The situation improves 
dramatically at higher center-of-mass energies. At $\sqrt{s}=10$~TeV, the 
significance already reaches $\mathcal{S} = 7.41$ at $\mathcal{L} = 500\,\text{fb}^{-1}$, 
and reaches $\mathcal{S} = 33.16$ at $\mathcal{L} = 10\,\text{ab}^{-1}$, well 
above the $5\sigma$ discovery threshold. The most promising prospects are found 
at $\sqrt{s} = 20$~TeV, where the significance reaches $\mathcal{S} = 13.01$ 
already at $\mathcal{L} = 500\,\text{fb}^{-1}$, and rises to $\mathcal{S} = 58.22$ 
at $\mathcal{L} = 10\,\text{ab}^{-1}$. These results clearly demonstrate that 
while a $5\sigma$ discovery of the signal is out of reach at the $3$~TeV stage, 
it becomes readily achievable at both the $10$~TeV and $20$~TeV MuC
configurations, even at moderate luminosities, highlighting the crucial role of the higher-energy stages of the MuC in probing this signal topology.

\subsection{$[WH]H\mu \nu$}
The $[W^+H]H$ mode targets the associated production of a charged Higgs boson and the neutral scalar $H$ together with a muon and a neutrino, $\mu^+\mu^- \to H H^{\pm} \mu^{\pm} \nu$, followed by the decays $H^\pm \to W^\pm H$.
\bea
\mu^+ \mu^- \to H H^{\pm} \mu^{\mp} \nu \to W_{qq^{'}} H  H \, \mu^+ \nu \to 2j + \mu^+ + \slashed{E}_T.
\eea
The final state consists of two jets originating from the hadronic decay $W \to qq'$, an additional muon $\mu^{\pm} $, and substantial missing energy arising from the two stable DM particles $H$. The benchmark point BP2 in Table~\ref{table:BP:backgrounds} yields
\bea
&&\sigma^{3\,\tev} ( \mu^+ \mu^- \to H H^{\pm} \mu^{\mp} \nu ) = 34.503 ~\text{fb} , \ \ \ \ \br(H^\pm \to W^\pm H) = 1\\
&&\sigma^{10\,\tev} ( \mu^+ \mu^- \to H H^{\pm} \mu^{\mp} \nu ) = 1890.132~\text{fb} , \ \ \ \ \br(H^\pm \to W^\pm H) = 1\\
&&\sigma^{20\,\tev} ( \mu^+ \mu^- \to H H^{\pm} \mu^{\mp} \nu ) = 9808.655~\text{fb} , \ \ \ \ \br(H^\pm \to W^\pm H) = 1
\eea

The main SM background contributions arise from processes such as $tW^{\pm}b$, $ZZ$, $W^{+}W^{-}$, and $W^{\pm}\mu^{\mp}\nu $. In these cases, additional muons may originate from $Z$ boson decays or leptonic decays of gauge bosons. The dominant backgrounds are listed as follows:
\begin{enumerate}[label=\roman*)]
\item  $\mu^+ \mu^- \rightarrow t W^- \bar{b}, \ (t  \rightarrow W^+ b) , (\ W^+ \rightarrow \mu^+ \nu_{\mu}),\ (W^- \rightarrow j j).$ 
\item   $\mu^+ \mu^- \rightarrow ZZ, \ (Z \rightarrow jj), \ (Z \rightarrow \nu_{\ell} \bar{\nu_{\ell}}),$ fake muons from jet misidentification
\item	$\mu^+ \mu^- \rightarrow W^+W^-,(W^+ \rightarrow \mu^+ \nu_{\mu}), \ (W^- \rightarrow j j).$
\item	$\mu^+ \mu^- \rightarrow W^+ \mu^- \bar{\nu_{\mu}}, \ (W^+ \rightarrow jj).$
\end{enumerate}

\begin{figure}[!hb]	
\centering
\includegraphics[width=0.45\textwidth]{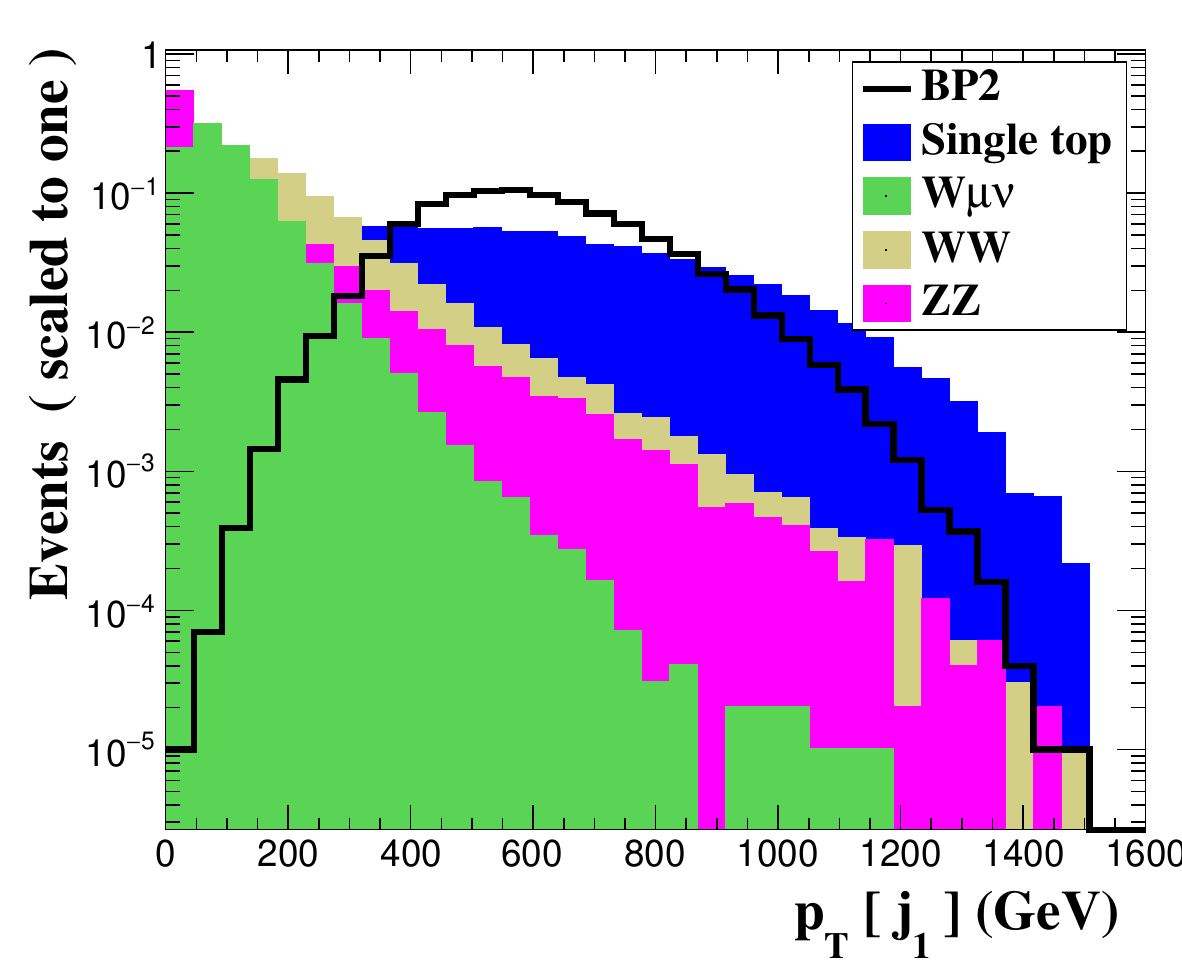}
\includegraphics[width=0.45\textwidth]{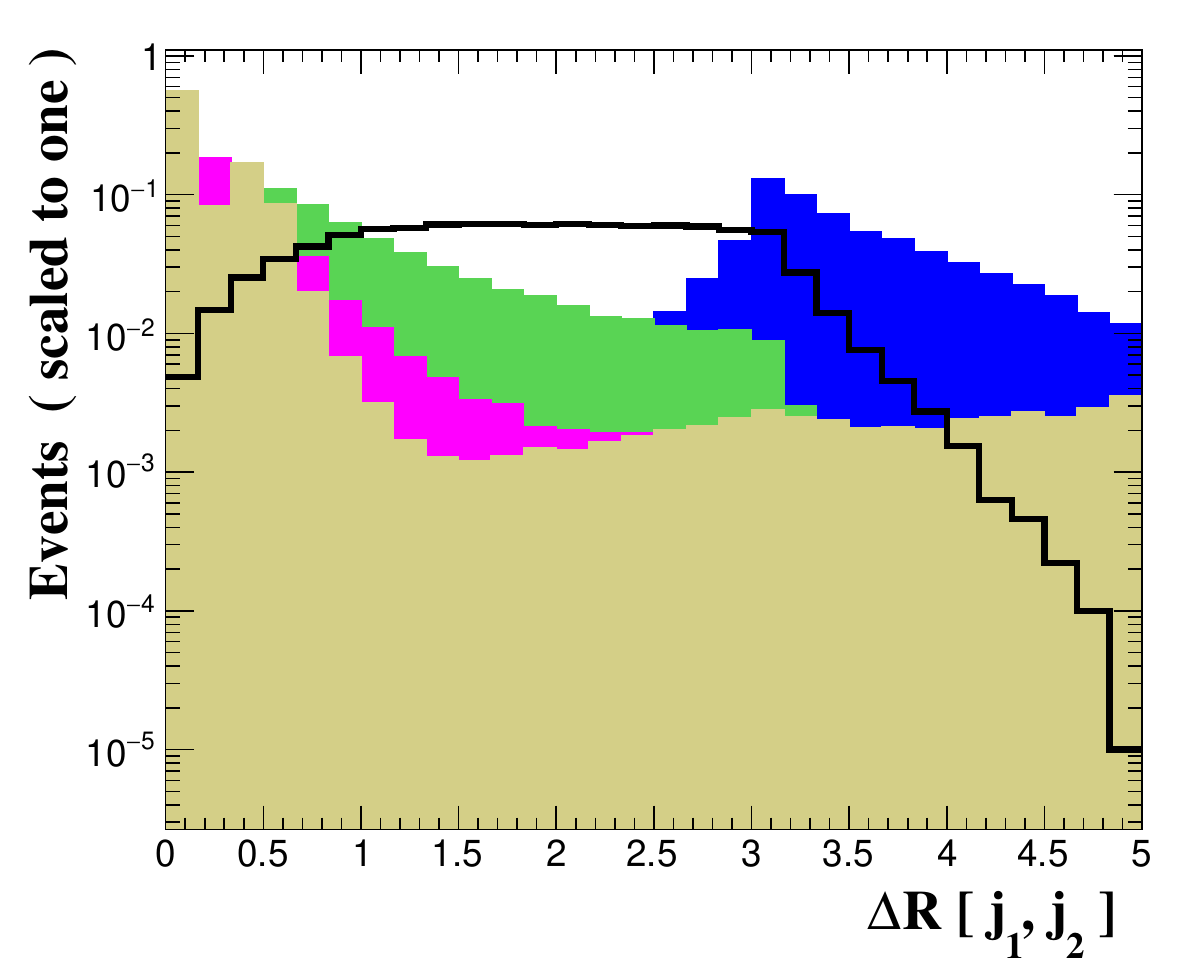}
\caption{Normalized kinematic distributions for the final state $[WH]H\mu \bar{\nu}$, showing the transverse momentum of the leading jet $p_T[j_1]$ (left panel) and $\Delta R(j_1, j_2)$ (right panel) at the 3 TeV MuC.} 
\label{mesfig:fig11}
\end{figure}

At the parton level, the following basic cuts are imposed on both the signal and background events:	
\bea
p_T^{j} > 20\,\gev, \quad p_T^{\ell} > 10\,\gev, \quad |\eta^{j}| < 5, \quad |\eta^{\ell}| < 2.5, \quad \Delta R^{jj,\ell\ell,j\ell} \geq 0.4.
\eea	

For the signal and background events after the basic cuts, we calculate several kinematic observables. Fig.~\ref{mesfig:fig11} shows the distributions of the transverse momentum of the leading jet, $p_T(j_1)$ (left panel), and the angular separation between the two jets, $\Delta R(j_1,j_2)$ (right panel). The results are based on the events passing the basic cuts and $N(b)\leq 1$. The first decisive cut is on the transverse momentum of the leading jet, $p_T(j_1)$. We take $p_T[j_1] > 300\,\gev$ as the first selection, removing about 98$\%$ of $W\mu\nu$, 96$\%$ of $ZZ$, 88$\%$ of $WW$, and 34$\%$ of single top events.  

The right panel in Fig.~\ref{mesfig:fig11} indicates that a $1< \Delta R (j_1, j_2) < 4.8 $ cut would be useful, which constitutes the ``cut~2". It removes about 99$\%$ of $W\mu\nu$ while 82$\%$  of the signal survives. The cut-flow results for the signal and SM backgrounds, expressed in fb, are presented in Table~\ref{tab:cutflow:BP2_3TeV} for $\sqrt{s}=3$~TeV.

\begin{table*}[!h]
\setlength\tabcolsep{18.1pt}
\centering
{\footnotesize\renewcommand{\arraystretch}{0.7} 
\begin{tabular}{l || l l l l |l}
\toprule
\multicolumn{6}{c}{$[WH]H\mu \bar{\nu}$ }\\
\toprule
{Cut}   &  $tW^\pm b$ & ~~$ZZ$ & $W^+W^-$ & $W^{\pm}\mu^\mp \nu_\mu$  & Signal \\
\toprule
Basic cuts &  1.269& 3.589 & 3.707  & 427 &  19.91\\
$N(b) \leq 1$ &0.467 & 3.573 & 3.672 & 425.43 &  19.883\\ 
$ p_T^{j_1} > 300 \,\gev$ & 0.309 & 0.161 & 0.68 & 11.01 &  19.390\\
$ 1 < \Delta R(j_1,j_2) < 4.8$ & 0.245 & 0.0 & 0.0844 & 0.128&  16.02\\
\hline \hline
\end{tabular}
}
\caption{Cut-flow chart of the cross section (in fb) of the signal and backgrounds for the channel $\mu^+ \mu^- \to H H^{\pm} \mu^{\mp} \nu \to [WH]H\mu\nu$  at the $3$~TeV MuC. More details about the selection cuts are given in the text.}
\label{tab:cutflow:BP2_3TeV}
\end{table*}

\begin{table*}[!h]
\setlength\tabcolsep{12pt}
\centering
{\footnotesize\renewcommand{\arraystretch}{0.7} 
\begin{tabular}{l ||l l l l |l}
\toprule
\multicolumn{6}{c}{$[WH]H\mu \bar{\nu}$ }\\
\toprule
{Cut}   &  $tW^\pm b$ & ~~$ZZ$ & $W^+W^-$ & $W^{\pm}\mu^\mp \nu_\mu$  & Signal \\
\toprule
Basic cuts &  0.6539& 0.2235 & 0.01722  & 57.87 &  697.8\\
$N(b) \leq 1$ &0.281 & 0.224 & 0.0169 & 57.620 &  696.6\\ 
$400 \,\gev < p_T^{j_1} < 2.6 \,\tev$& 0.108 & 0.0 & 0.0102 & 2.88 &  632.10\\
$-1  < \eta^{j_1} < 1 $ & 0.036 & 0.0 & $1.66\times10^{-4}$& 1.55&  486.7\\
$\slashed{E}_T<400\,\gev$ & 0.0164 & 0.0 & $8.49\times10^{-5}$ & 0.524&  452.1\\
$ 0.2 < \Delta R(j_1,j_2) < 2.8$ & $3.97\times10^{-3}$ & 0.0 & $2.62\times10^{-5}$ & 0.0498&  412.9\\
\hline \hline
\end{tabular}
}
\caption{Cut-flow chart of the cross section (in fb) of the signal and backgrounds for the channel $\mu^+ \mu^- \to H H^{\pm} \mu^{\mp} \nu \to [WH]H\mu\nu$ at the $10$~TeV MuC. More details about the selection cuts are given in the text.}
\label{tab:cutflow:BP2_10TeV}
\end{table*}

\begin{table*}[!t]
\setlength\tabcolsep{12pt}
\centering
{\footnotesize\renewcommand{\arraystretch}{0.7} 
\begin{tabular}{l ||l l l l  |l}
\toprule
\multicolumn{6}{c}{$[WH]H\mu \bar{\nu}$ }\\
\toprule
{Cut}   &  $tW^\pm b$ & ~~$ZZ$ & $W^+W^-$ & $W^{\pm}\mu^\mp \nu_\mu$  & Signal \\
\toprule
Basic cuts &  0.135& 0.02369 & $0.15\times10^{-3}$  & 14.53 &  2308\\
$N(b) \leq 1$&0.0581 & 0.0237 & $1\times10^{-4}$& 14.47 &  2304.0\\ 
$200 \,\gev < p_T^{j_1} < 3 \,\tev$& 0.0301 & 0.0 & $1\times10^{-4}$ & 2.93 &  2213.51\\
$\slashed{E}_T<380\,\gev$ & 0.0166 & 0.0 & $8\times10^{-5}$ & 0.574&  2030.5\\
$ 0.5 < \Delta R(j_1,j_2) < 2.9$ & $3\times10^{-3}$ & 0.0 & $2\times10^{-6}$ & 0.043&  1577.5\\
\hline \hline
\end{tabular}
}
\caption{Cut-flow chart of the cross section (in fb) of the signal and backgrounds for the channel $\mu^+ \mu^- \to H H^{\pm} \mu^{\mp} \nu \to [WH]H\mu\nu$ at the $20$~TeV MuC. More details about the selection cuts are given in the text.}
\label{tab:cutflow:BP2_20TeV}
\end{table*}

Having established an efficient event selection strategy at $\sqrt{s}=3$~TeV, we extend the analysis to the higher-energy stages of the MuC. As the signal and background processes exhibit similar kinematic features at $\sqrt{s}=10$ and $20$~TeV, the corresponding distributions are not displayed. Instead, we present the optimized cut-flow results for these center-of-mass energies in Tables~\ref{tab:cutflow:BP2_10TeV} and \ref{tab:cutflow:BP2_20TeV}, respectively.

\begin{table}[!h]
\setlength{\tabcolsep}{7pt}
\renewcommand{\arraystretch}{0.7}
\centering
\begin{tabular}{c c c c c c}       
\hline  \hline 		
& &BP2& &\\
\hline  \hline 
\hline   
Process $\ \ \ \ \ $  &&&$\mu^+ \mu^- \to H H^{\pm} \mu^{\mp} \nu \to [WH]H\mu \nu$ &\\
\hline  \hline
Luminosity$\ \ $&$\mathcal{L}$=500 fb$^{-1}$&$\mathcal{L}$=1000 fb$^{-1}$& $\mathcal{L}$=1500 fb$^{-1}$&$\mathcal{L}$=10 ab$^{-1}$\\
\hline  \hline 
$\sqrt{s}=3\,\tev$  &88&124&152&394 \\
$\sqrt{s}=10\,\tev$  &454&642&786& 2031\\
$\sqrt{s}=20\,\tev$  &888&1255&1538&3971 \\	
\hline \hline
\end{tabular}
\caption{Significance $\mathcal{S}$ for our signal with $\sqrt{s}=3$, $10$ and $20$~TeV and $\mathcal{L}=500$, $1000$, $1500$~ fb$^{-1}$, and $10$~$ab^{-1}$.}  \label{Signi:siBP2}
\end{table}	

The statistical significance $\mathcal{S}$ for the signal process 
$\mu^+\mu^- \to HH^{\pm}\mu^{\mp}\nu \to [WH]H\mu\nu$ (BP2) is presented 
in Table~\ref{Signi:siBP2} for three center-of-mass energies and several 
luminosity benchmarks. In striking contrast to BP1, this channel exhibits 
remarkably large significance values across all energy stages and luminosity 
scenarios considered. At $\sqrt{s} = 3$~TeV, the significance already reaches 
$\mathcal{S} = 88$ at $\mathcal{L} = 500\,\text{fb}^{-1}$, and rises to 
$\mathcal{S} = 394$ at $\mathcal{L} = 10\,\text{ab}^{-1}$, far reaching the 
$5\sigma$ discovery threshold even at the lowest luminosity considered. At 
$\sqrt{s} = 10$~TeV, the significance increases further, reaching 
$\mathcal{S} = 454$ at $\mathcal{L} = 500\,\text{fb}^{-1}$ and reaching 
$\mathcal{S} = 2031$ at $\mathcal{L} = 10\,\text{ab}^{-1}$. The most 
impressive prospects are found at $\sqrt{s} = 20$~TeV, where the significance 
reaches $\mathcal{S} = 888$ at $\mathcal{L} = 500\,\text{fb}^{-1}$ and 
rises to $\mathcal{S} = 3971$ at $\mathcal{L} = 10\,\text{ab}^{-1}$. These 
exceptionally large significance values reflect the favorable signal-to-background 
ratio achieved after the optimized event selection, and confirm that the BP2 
signal topology is highly accessible across the full range of MuC
energy and luminosity configurations explored in this study, making it one 
of the most promising channels for the discovery of charged Higgs bosons at 
a future MuC.

\subsection{$[WH][ZH]\mu \nu$}
The $[WH][ZH]\mu\nu$ mode targets the production of a charged Higgs boson and a pseudoscalar Higgs boson in association with a muon and a neutrino, $\mu^+\mu^- \to A H^- \mu^+ \nu$, followed by the decays $H^- \to W^- H$ and $A \to ZH$. This process yields a distinctive final state containing a $W$ gauge boson, a $Z$ gauge boson, two $CP$-even Higgs bosons, a charged lepton, and missing energy arising from the neutrino as follows,
\bea
\mu^+ \mu^- \to A H^{\pm} \mu^{\mp} \bar{\nu} \to (Z_{qq'} H)(W_{\ell^-\nu} H)\mu^{\mp} \bar{\nu} \to 2j + \ell^- + \mu^{\mp} + \slashed{E}_T.
\eea
The corresponding parton-level cross sections for the three center-of-mass energies read 
\bea
&&\sigma^{3\,\tev} ( \mu^+ \mu^- \to A H^{\pm} \mu^{\mp} \nu )= 1.08~\text{fb},\,\,  \br(A \to Z H) = \br(H^\pm \to W^\pm H) = 1\\
&&\sigma^{10\,\tev} ( \mu^+ \mu^- \to A H^{\pm} \mu^{\mp} \nu )=  65.05~\text{fb},\,\,  \br(A \to Z H) = \br(H^\pm \to W^\pm H) = 1\\
&&\sigma^{20\,\tev} ( \mu^+ \mu^- \to A H^{\pm} \mu^{\mp} \nu )=  347.31~\text{fb},\,\, \br(A \to Z H) = \br(H^\pm \to W^\pm H) = 1
\eea

The main SM background contributions come from the processes $tWb$, $WW$, $W\mu \nu$, $Wjj$, and $Zjj$. Backgrounds such as $ZZ$ and $WWZ$ are minor. The backgrounds are as follows: 
\begin{enumerate}[label=\roman*)]
\item   $\mu^+ \mu^- \rightarrow t W^{\pm} \bar{b}, \ (t  \rightarrow W^+ b) , (\ W^{\pm} \rightarrow \mu^{\pm} \nu_{\mu}),\ (W^- \rightarrow j j).$ 
\item	$\mu^+ \mu^- \rightarrow W^+W^-,(W+ \rightarrow \mu^{+} \nu_{\mu}), (W^- \rightarrow  j j )$
\item	$\mu^+ \mu^- \rightarrow W^+ \mu^- \bar{\nu_{\mu}}, \ (W^+ \rightarrow jj).$
\item	$\mu^+ \mu^- \rightarrow Wjj,(W \rightarrow \mu^{\pm} \nu_{\mu}).$
\item	$\mu^+ \mu^- \rightarrow Zjj,(Z \rightarrow \mu^+ \mu^-).$
\end{enumerate}

For event selection, we take the following steps. The basic cuts consist of	
\bea
p_T^{j} > 20\,\gev, \quad p_T^{\ell} > 10\,\gev, \quad |\eta^{j}| < 5, \quad |\eta^{\ell}| < 2.5, \quad \Delta R^{jj,\ell\ell,j\ell} \geq 0.4.
\eea

\begin{figure}[!hb]
\centering		
\includegraphics[width=0.45\textwidth]{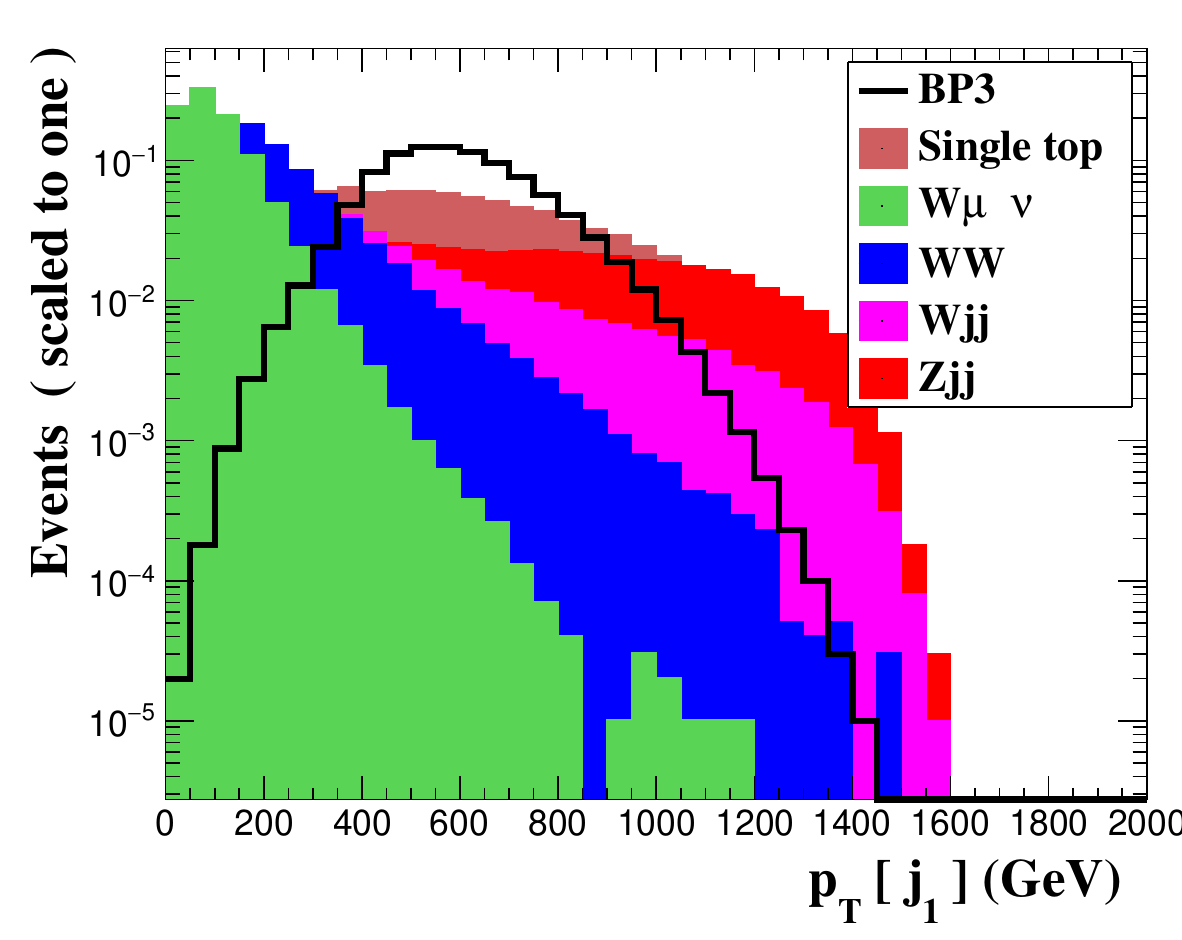}
\includegraphics[width=0.45\textwidth]{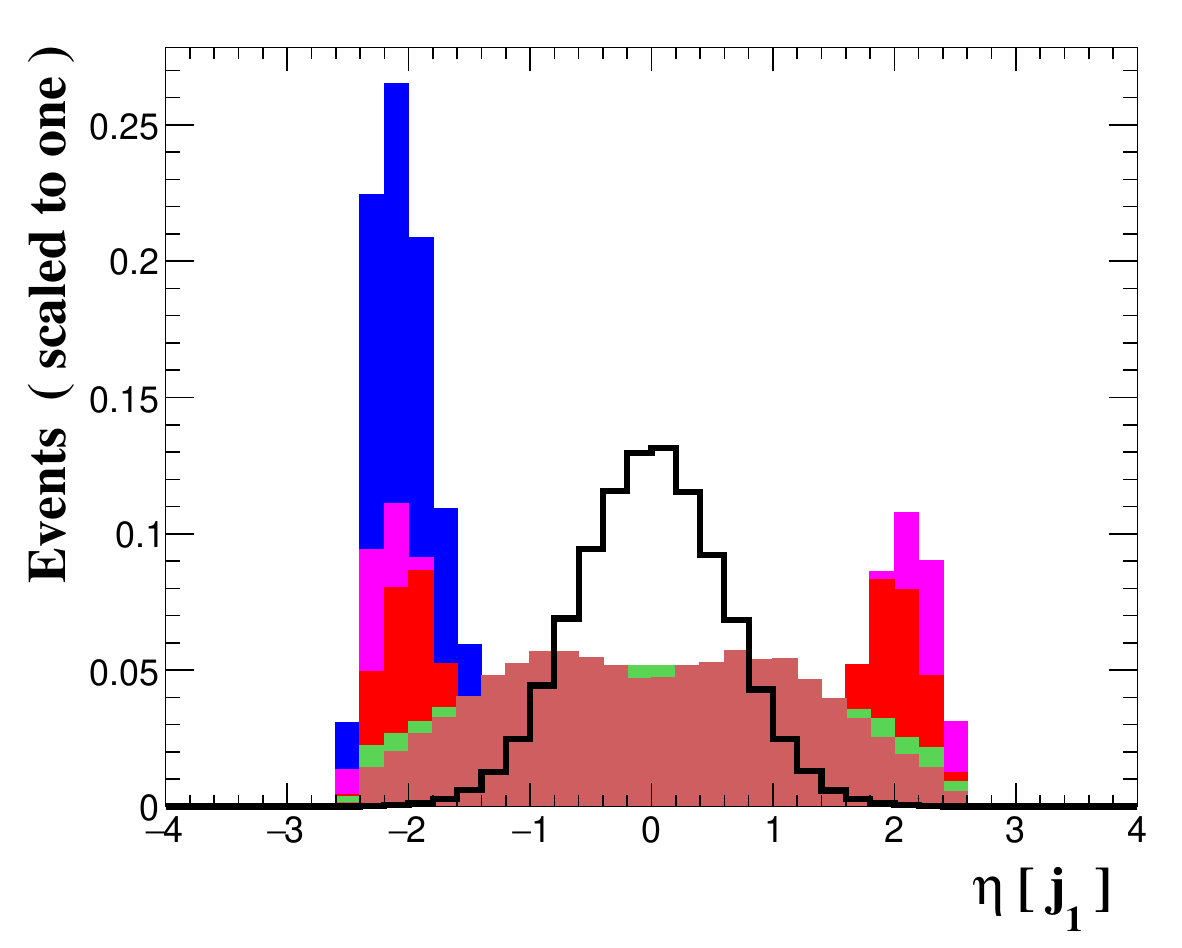}
\includegraphics[width=0.45\textwidth]{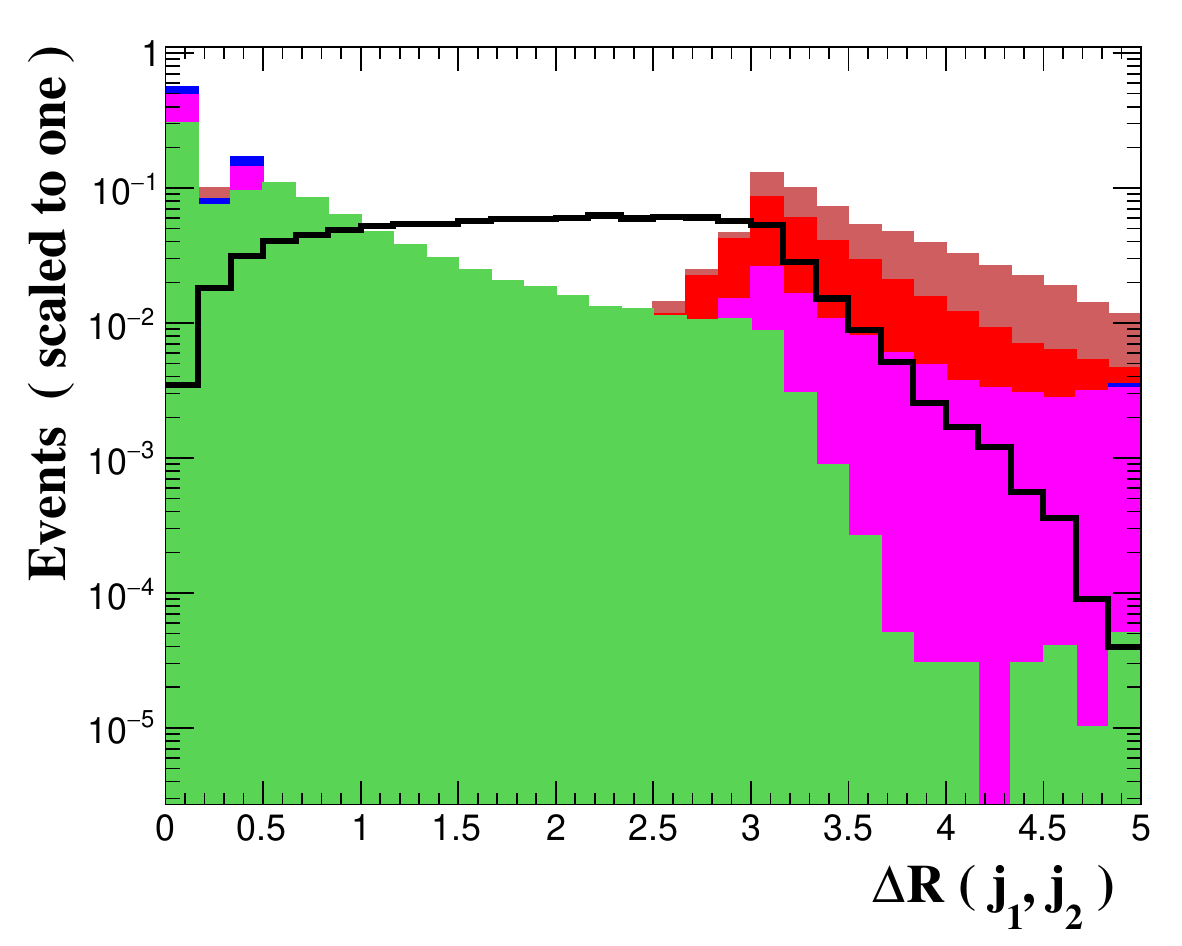}
\caption{Normalized kinematic distributions for the final state  $[WH][ZH]\mu \nu$, showing the transverse momentum of the leading jet $p_T[j_1]$ (top left panel), the leading-jet pseudorapidity $\eta(j_1)$ (top right panel), and  $\Delta R(j_1, j_2)$ (lower panel) at the 3 TeV MuC.} 
\label{mesfig:fig12}
\end{figure}

After imposing the requirements on the $b$-jet multiplicity, $N(b)\leq 1$, a strong discrimination between the signal and the background is achieved. The signal remains largely preserved, with an acceptance times efficiency of approximately $42\%$. In contrast, the combined background contribution is drastically suppressed, reaching an acceptance times efficiency of only $0.03\%$. To develop a more refined event selection strategy, the kinematic properties of the signal and background events are investigated. For this purpose, we show in Fig.~\ref{mesfig:fig12} the distributions of the leading-jet transverse momentum, $p_T(j_1)$ (top-left panel), the leading-jet pseudorapidity, $\eta(j_1)$ (top-right panel), and the separation between the two leading jets, $\Delta R(j_1,j_2)$ (bottom panel). These observables provide valuable discriminating power for enhancing the signal sensitivity while suppressing the SM backgrounds.

\begin{table*}[!t]	
\setlength\tabcolsep{14pt}
\centering
{\footnotesize\renewcommand{\arraystretch}{0.7} 
\begin{tabular}{l ||l l l l l | l}
\toprule
\multicolumn{6}{c}{$[WH][ZH]\mu \nu$ }\\
\toprule
{Cut}   &  $tW^\pm b$ & $WW$ & $W^{\pm}\mu^\mp \nu_\mu$ &$Wjj$ &$Zjj$& Signal \\
\toprule
Basic cuts  & 1.26 &  3.7  &427&9.33&0.15 & 0.22\\
$N(b) \leq 1$& 0.46 &3.6  & 425  &9.2 &0.14& 0.22\\ 
$300\,\gev < p_T^{j_1} < 900 \,\gev$& 0.26& 0.66 & 10.97  &2.31 &0.04 & 0.2\\
$-0.8 < \eta [j_1]<0.8$ & 0.09 & 0.04 &  6.13 & 0.5 & 0.0143&0.17\\
$ 1 < \Delta R(j_1,j_2) < 3$ & 0.02 & 0.008 &  0.02 & 0.17 & 0.005 & 0.12\\
\hline \hline
\end{tabular}
}
\caption{Cut-flow chart of the cross sections (in fb) of the signal and backgrounds for the channel $\mu^+ \mu^- \to A H^{\pm} \mu^{\mp} \nu \to [WH][ZH]\mu\nu$ at the $3$~TeV MuC. More details about the selection cuts are given in the text.}
\label{tab:cutflow:BP3_3TeV}
\end{table*}

\begin{table*}[!t]
\setlength\tabcolsep{9pt}
\centering
{\footnotesize\renewcommand{\arraystretch}{0.8} 
\begin{tabular}{l ||l l l l l |l}
\toprule
\multicolumn{6}{c}{$[WH][ZH]\mu \nu$ }\\
\toprule
{Cut}   &  $tW^\pm b$ & $WW$ & $W^{\pm}\mu^\mp \nu_\mu$ &$Wjj$ &$Zjj$& Signal \\
\toprule
Basic cuts& 0.65 &  0.01  &57.87 &0.24&5.9$\times10^{-3}$ &  7.76\\
$N(b) \leq 1$ & 0.28 & 0.01 & 57.62 &0.24 &  5$\times10^{-3}$&7.7\\ 
$700 \,\gev < p_T^{j_1} < 2 \,\tev$&0.06  &0.003  &  0.35 & 0.08&1$\times10^{-3}$&5.77\\
$-0.8 < \eta [j_1]<0.8$  & 0.01 & 8$\times10^{-5}$ & 0.16  & 0.01&1$\times10^{-4}$ &3.72\\
$ 0.5 < \Delta R(j_1,j_2) < 2.7$ & 3$\times10^{-3}$ & 2$\times10^{-5}$ & 1$\times10^{-3}$ &6$\times10^{-3}$ & 2$\times10^{-5}$ & 2.7 \\
\hline \hline
\end{tabular}
}
\caption{Cut-flow chart of the cross sections (in fb) of the signal and backgrounds for the channel $\mu^+ \mu^- \to A H^{\pm} \mu^{\mp} \nu \to [WH][ZH]\mu\nu$ at the $10$~TeV MuC.}
\label{tab:cutflow:BP3_10TeV}
\end{table*}

\begin{table*}[!t]
\setlength\tabcolsep{9pt}
\centering
{\footnotesize\renewcommand{\arraystretch}{0.8} 
\begin{tabular}{l ||l l l l l |l}
\toprule
\multicolumn{6}{c}{$[WH][ZH]\mu \nu$ }\\
\toprule
{Cut}   &  $tW^\pm b$ & $WW$ & $W^{\pm}\mu^\mp \nu_\mu$ &$Wjj$ &$Zjj$& Signal \\
\toprule
Basic cuts&   0.13  &0.001 &14.5&0.054&0.0014 &  16.77\\
$N(b) \leq 1$ & 0.05 & 1$\times10^{-4}$  & 14.4 &0.05 &  1$\times10^{-3}$&16.7\\ 
$1 \,\tev < p_T^{j_1} < 2.9 \,\tev$&0.007  &3$\times10^{-5}$  &  0.02 & 0.01&1$\times10^{-4}$&11.91\\
$-1.2 < \eta [j_1]<1.2$  & 0.002 & 1$\times10^{-6}$ & 0.01 & 0.003&1$\times10^{-5}$ &8.61\\
$ 0.2 < \Delta R(j_1,j_2) < 2.5$ & 5$\times10^{-4}$ & 1$\times10^{-7}$ & 5$\times10^{-4}$ &1$\times10^{-3}$ & 2$\times10^{-6}$ & 7.64 \\
\hline \hline
\end{tabular}
}
\caption{Cut-flow chart of the cross section (in fb) of the signal and backgrounds for the channel $\mu^+ \mu^- \to A H^{\pm} \mu^{\mp} \nu \to [WH][ZH]\mu\nu$ at the $20$~TeV MuC.}
\label{tab:cutflow:BP3_20TeV}
\end{table*}

To enhance the signal significance, a series of selection cuts was applied based on the behavior of the kinematic distributions. In particular, a requirement on the number of $b$-tagged jets, $N(b) \leq 1$, was imposed. This criterion plays an important role in suppressing the SM backgrounds while preserving a large fraction of the signal events, thereby improving the overall signal-to-background discrimination. The first selection cut applied is the transverse momentum of the leading jet, $700\,\gev < p_T[j_1] < 2\,\tev$, which removes about 98$\%$ of the $W^{\pm}\mu^\mp \nu_\mu$, 82$\%$ of the $WW$, and 44$\%$ of the  $tW^\pm b$, while the survival rate for the signal is more than 93$\%$. The requirement
$-0.8 < \eta(j_1) < 0.8$ further enhances the signal-to-background separation, proving especially effective in suppressing the $WW$ and $Wjj$ backgrounds.
The final selection considered is on the angular separation between the two leading jets, $\Delta R(j_1,j_2)$. The selection cut
$1 < \Delta R(j_1,j_2) < 3$ provides an efficient reduction of the residual background events, leading to a further improvement in the signal purity. Table~\ref{tab:cutflow:BP3_3TeV} shows the cut flows on the cross sections (in fb) for both the signal and the SM backgrounds at $\sqrt{s}=3$~TeV. 

Having established an efficient event selection strategy at $\sqrt{s}=3$~TeV, we extend the analysis to the higher-energy stages of the MuC. Since the signal and background processes exhibit qualitatively similar kinematic behaviour at $\sqrt{s}=10$ and $20$~TeV, the corresponding distributions are not shown. Instead, we directly present the optimized cut-flow results for these energies in Tables~\ref{tab:cutflow:BP3_10TeV} and ~\ref{tab:cutflow:BP3_20TeV}.

\begin{table}
\setlength{\tabcolsep}{7pt}
\renewcommand{\arraystretch}{0.8}
\centering
\begin{tabular}{c c c c c c}       
\hline  \hline 	
& &BP3& &\\
\hline  \hline 
\hline   
Process $\ \ \ \ \ $  &&&$\mu^+ \mu^- \to A H^{\pm} \mu^{\mp} \nu \to [WH][ZH]\mu \nu$ &\\
\hline  \hline
Luminosity$\ \ $&$\mathcal{L}$=500 fb$^{-1}$&$\mathcal{L}$=1000 fb$^{-1}$& $\mathcal{L}$=1500 fb$^{-1}$&$\mathcal{L}$=$10\ ab^{-1}$\\
\hline  \hline 
$\sqrt{s}=3\,\tev$  & 4.5 & 6.4 & 7.9 &20.4 \\
$\sqrt{s}=10\,\tev$  &36&51.9& 63.6 & 164 \\
$\sqrt{s}=20\,\tev$  & 61.8 & 87.4 & 107 & 276 \\	
\hline \hline
\end{tabular}	
\caption{Significance $\mathcal{S}$  for our signal with $\sqrt{s}=3$, $10$ and $20$~TeV and $\mathcal{L}=500$, $1000$, $1500$~fb$^{-1}$ and $10$~$ab^{-1}$.}  \label{Signi:si3}
\end{table}	

The statistical significance $\mathcal{S}$ for the signal process 
$\mu^+\mu^- \to HH^{\pm}\mu^{\mp}\nu \to [WH][ZH]\mu\nu$ (BP3) is reported in Table~\ref{Signi:si3} for three center-of-mass energies and four luminosity 
benchmarks. This channel yields significance values that are intermediate between 
those of BP1 and BP2, reflecting its distinct signal topology and production rate. 
At $\sqrt{s} = 3$~TeV, the significance reaches $\mathcal{S} = 4.5$ at 
$\mathcal{L} = 500\,\text{fb}^{-1}$, falling just below the $5\sigma$ discovery 
threshold, but exceeds it at $\mathcal{L} = 1000\,\text{fb}^{-1}$ with 
$\mathcal{S} = 6.4$, and rises to $\mathcal{S} = 20.4$ at 
$\mathcal{L} = 10\,\text{ab}^{-1}$. The prospects improve substantially at 
higher energies. At $\sqrt{s} = 10$~TeV, the significance reaches 
$\mathcal{S} = 36$ at $\mathcal{L} = 500\,\text{fb}^{-1}$ and rises to 
$\mathcal{S} = 164$ at $\mathcal{L} = 10\,\text{ab}^{-1}$, well above the 
discovery threshold across all luminosity scenarios. The most favorable 
sensitivity is achieved at $\sqrt{s} = 20$~TeV, where the significance 
reaches $\mathcal{S} = 61.8$ at $\mathcal{L} = 500\,\text{fb}^{-1}$ and 
rises to $\mathcal{S} = 276$ at $\mathcal{L} = 10\,\text{ab}^{-1}$. Overall, 
these results demonstrate that the BP3 signal channel is discoverable at the 
$5\sigma$ level at $\sqrt{s} = 3$~TeV with sufficient luminosity, and becomes 
highly significant at both the $10$~TeV and $20$~TeV MuC stages even at moderate luminosities, further consolidating the physics case for higher-energy MuC configurations in the search for charged Higgs bosons.

\section{Conclusion}
\label{sec:conclusion}
We have investigated the potential of an upcoming MuC to probe the inert scalar sector of the IDM, focusing on three $2\to4$ VBF production modes: $\mu^+\mu^-\to H^+H^-\nu\bar{\nu}$, $\mu^+\mu^-\to HH^\pm\mu^\mp\nu$ and $\mu^+\mu^-\to AH^\pm\mu^\mp\nu$,
generated respectively through $WW$ and mixed $WZ$/$W\gamma$ fusion topologies. 
We have computed the amplitudes of the subprocesses $V_1V_2\to H^+ H^-, H^+H, H^+A$ in the Feynman gauge and shown that 
there is a strong destructive interference among them. Such cancellations are indeed necessary to ensure the proper high-energy behavior of the amplitudes.
Additionally, we have shown, numerically, that the subprocesses $V_1V_2\to H^+ H^-, H^+H, H^+A$ are dominated by the longitudinal $W$ contributions.

After imposing the relevant constraints, we performed a dedicated parameter scan and evaluated the parton-level cross sections for all three channels at $\sqrt{s}=3$,$10$, and $20$~TeV. The three production modes show a clear enhancement with the collider energy as expected for VBF-induced processes at a multi-TeV MuC. While the charged pair mode reaches a rate of up to $3.01$~fb at $20$~TeV, the associated production channels are significantly larger: the $HH^\pm\mu^\mp\nu$ channel reaches the pb level, and the $AH^\pm\mu^\mp\nu$ channel shows a rate of up to several hundred~fb in the same energy regime. To translate these rates into discovery prospects, the selected BPs are chosen so that the targeted bosonic decay chains $H^\pm \to W^\pm H$ and $A \to Z H$ fully control the signal topology. After including relevant background processes, we performed a signal-background analysis for each final state. The resulting significances reveal a clear improvement with the collider energy: the associated $[WH]H\mu\nu$ channel is by far the most sensitive among the three, already yielding $\mathcal{S}\simeq 88$ at $3$~TeV for $500~\mathrm{fb}^{-1}$, increasing to $\mathcal{S}\simeq 888$ at 20 TeV, and exceeding $\mathcal{S}\simeq 3971$ for the high-luminosity $10$~ab$^{-1}$ scenario. The $[WH][ZH]\mu\nu$ mode gives an intermediate but important complementary sensitivity, growing from $\mathcal{S}\simeq 4.5$ at $3$~TeV to $\mathcal{S}\simeq 61.8$ at $20$~TeV for $\mathcal{L}=500$~fb$^{-1}$, and reaching $\mathcal{S}\simeq 276$ at $10$~ab$^{-1}$. Finally, the significance of the charged-pair channel $[WH][WH]\nu\bar{\nu}$ remains below the discovery level at $3$~TeV, but rises to $7.41~\sigma$ at $10$~TeV and $13.01~\sigma$ at $20$~TeV for $\mathcal{L}=500$~fb$^{-1}$. 
Overall, our results demonstrate and reinforce that a multi-TeV MuC is a powerful probe of the IDM inert sector through VBF production. The $3$~TeV stage can already test favourable associated mode scenarios, while the $10$ and $20$~TeV stages provide a stronger, more viable path to the discovery of the charged and neutral inert states. 

These channels probe the electroweak gauge structure of the inert doublet and yield distinctive signatures involving missing transverse momentum. Their study at a multi-TeV MuC provides a complementary perspective on the IDM phenomenology, particularly through VBF production mechanisms. While a quantitative comparison with the projected sensitivities of other future colliders is beyond the scope of the present analysis, our results provide a useful basis for assessing the potential reach of these channels within the broader future collider programme. In particular, the results at $\sqrt{s}=3$~TeV offer a useful reference point for future comparative studies with CLIC~\cite{CLICPhysicsWorkingGroup:2004qvu,aicheler2014multi}.

Further refinements, including systematic background uncertainties and machine-learning-based event selection, would sharpen the sensitivity. Still, the present analysis already establishes VBF production at a MuC as a highly promising direction for testing DM-motivated scalar sectors. 
	
\section*{Acknowledgements}
\label{sec:acknowledgements}
AA gratefully acknowledges the hospitality of LAPTh, Annecy, where part of this work was carried out during his sabbatical leave. The authors would like to thank Fawzi Boudjema and Brahim Ait-Ouazghour for useful discussions, and Thomas Hahn for helpful advice on the FormCalc calculations.
	
\appendix
\section{Feynman Rules}
\label{app:feyn-rules}
We summarise here the Feynman rules relevant to the processes investigated in this work. 
\begin{table}[!h]
\begin{center}
\renewcommand{\arraystretch}{2.0}
\caption{Feynman rules for the $SSVV$ interaction vertices in the IDM.}
\label{tab:ssvv_rules}
\begin{tabular}{c@{\;\;}l@{\hspace{3.85cm}}c@{\;\;}l}
\ssvv{H}{H^+}{\gamma}{W^-}{scalar}{charged}  & $=-\dfrac{ie^2}{2s_W}g_{\mu\nu},$  &  
\ssvv{H^+}{H}{Z}{W^-}{charged}{scalar}  &  $=-\dfrac{ie^2}{2c_W}g_{\mu\nu},$ \\[5mm]
\ssvv{H^+}{H^-}{W^+}{W^-}{charged}{chargedrev}  & $=\dfrac{ie^2}{2s_W^2}g_{\mu\nu},$  &  
\ssvv{H^+}{H^-}{Z}{Z}{charged}{chargedrev}  & $=\dfrac{ie^2c_{2W}^2}{2s_W^2c_W^2}g_{\mu\nu},$ \\[5mm]
\ssvv{H^+}{A}{W^-}{\gamma}{charged}{scalar}  & $=-\dfrac{e^2}{2s_W}g_{\mu\nu},$  &  
\ssvv{H^+}{A}{W}{Z}{charged}{scalar}  & $=-\dfrac{e^2}{2 c_W}g_{\mu\nu}.$
\end{tabular}
\end{center}
\end{table}	
\begin{table}[!h]
\begin{center}
\renewcommand{\arraystretch}{2.0}
\caption{Feynman rules for the $SSS$, $VVV$ and $SSV$ interaction vertices in the IDM.}
\label{tab:sss_svv_rules}
\begin{tabular}{c@{\;\;}l@{\hspace{0.5cm}}c@{\;\;}l}
\threeS{h}{charged}{H^+}{chargedrev}{H^-}  & $=-i\dfrac{2m_Ws_W}{e}\lambda_3,$  &  
\threeS{h}{charged}{A}{chargedrev}{A}  & $=-i\dfrac{2m_Ws_W}{e}\bar{\lambda}_{345},$  \\[5mm]
\threeS{h}{charged}{H}{chargedrev}{H}  & $=-i\dfrac{2m_Ws_W}{e}\lambda_{345},$  &  
\ssv{A}{H^+}{W^-}{scalar}{charged}  &  $=\dfrac{e}{2s_W}(p_1-p_2)_\mu,$ \\[5mm]
\ssv{H^-}{H^+}{\gamma}{chargedrev}{charged}  & $=-ie(p_1-p_2)_\mu,$  &  
\ssv{H^-}{H^+}{Z}{chargedrev}{charged}  & $=i\dfrac{e(c_W^2-s_W^2)}{2c_Ws_W}(p_1-p_2)_\mu,$ \\[5mm]
\ssv{H}{A}{Z}{scalar}{scalar}  &  $=\dfrac{e}{2c_Ws_W}(p_1-p_2)_\mu,$  & 
\ssv{H}{H^+}{W^-}{scalar}{charged}   &  $=i\dfrac{e}{2s_W}(p_1-p_2)_\mu.$ \\[5mm]
\vvv{\gamma}{W^+}{W^-}  &  $=-ie\Gamma_{\mu\nu\rho}(p_\gamma,p_{+},p_{-}),$  & 
\vvv{Z}{W^+}{W^-}  &  $=-i\dfrac{ec_W}{s_W}\Gamma_{\mu\nu\rho}(p_Z,p_{+},p_{-}).$
\end{tabular}
\end{center}
\label{tab:sss_vvv}
\end{table}	

\noindent
with
\begin{eqnarray}
\Gamma_{\mu\nu\rho}(p_\gamma,p_{+},p_{-}) &=& g_{\mu\nu}(p_\gamma-p_+)_\rho + g_{\nu\rho}(p_+-p_-)_\mu + g_{\rho\mu}(p_+-p_\gamma)_\nu \\
\Gamma_{\mu\nu\rho}(p_Z,p_{+},p_{-}) &=& g_{\mu\nu}(p_Z-p_+)_\rho + g_{\nu\rho}(p_+-p_-)_\mu + g_{\rho\mu}(p_+-p_Z)_\nu
\end{eqnarray}

For convenience, the interaction vertices are grouped according to their Lorentz structure, namely the quartic $SSVV$ vertices given in Tab.~\ref{tab:ssvv_rules} and the trilinear $SSV$, $SSS$, and $VVV$ vertices given in Tab.~\ref{tab:sss_svv_rules}.

\section{Kinematics for $V_1V_2 \to S_1S_2$}
\label{app:kine_VBF}
We provide below the kinematics for the scattering process of two vector bosons into two scalar particles, $V_1(p_1)\,V_2(p_2) \to S_1(k_1)\,S_2(k_2)$, where $p_{1,2}$ denote the four-momenta of the incoming gauge bosons, while $k_{1,2}$ are the four-momenta of the outgoing scalars. The incoming gauge bosons are taken to propagate along the $z$-axis, and the scattering is  assumed to occur in the $(x-z)$ plane. In the center-of-mass (CM) frame, the four-momenta are given by
\begin{eqnarray}
&&p_{1,2}=\frac{\sqrt{s}}{2} (\frac{s\pm (m_{V_1}^2-m_{V_2}^2)}{s},0,0,\pm \kappa_{V_1V_2})\nonumber\\
&& k_{1,2}=\frac{\sqrt{s}}{2} ((\frac{s\pm (m_{S_1}^2-m_{S_2}^2)}{s},\pm \kappa_{S_1S_2} \sin\theta,0,\pm   \kappa_{S_1S_2} \cos\theta),\nonumber
\end{eqnarray}
where $\sqrt{s}/2$ denotes the beam energy, $\theta$ is the scattering angle between $S_1$ and $V_1$, and $\kappa_{ij}$ is given by: 
\[\kappa_{ij}^2= (s-(m_i+m_j)^2)(s-(m_i-m_j)^2 )/s^2.\]
One can readily verify that $p_1^2=m_{V_1}^2$, $p_2^2=m_{V_2}^2$, $ k_1^2=m_{S_1}^2$, and $ k_2^2=m_{S_2}^2$.\\
The Mandelstam variables are
\begin{eqnarray}
s=(p_1+p_2)^2&=&(k_1+k_2)^2\nonumber\\
 t=(p_1-k_1)^2 &=& (p_2-k_2)^2=-\frac{s}{2}  -\frac{1}{2s}  (m_{S_1}^2-  m_{S_2}^2) (m_{V_1}^2 -m_{V_2}^2 )  + \nonumber\\ &&
\frac{1}{2}( m_{S_1}^2 + m_{S_2}^2 + 
 m_{V_1}^2 + m_{V_2}^2)  + \frac{s}{2} \kappa_{S_1S_2} \kappa_{v_1V_2} \cos\theta\nonumber\\
 u=(p_1-k_2)^2 &=& (p_2-k_1)^2= -\frac{s}{2}  + \frac{1}{2s}  (m_{S_1}^2-  m_{S_2}^2) (m_{V_1}^2 -m_{V_2}^2 )  + \nonumber\\ &&
\frac{1}{2}( m_{S_1}^2 + m_{S_2}^2 + 
 m_{V_1}^2 + m_{V_2}^2)  - \frac{s}{2} \kappa_{S_1S_2} \kappa_{v_1V_2} \cos\theta
\nonumber\\
 s+t+u &=& m_{V_1}^2 + m_{V_2}^2 + m_{S_1}^2 + m_{S_2}^2 
\end{eqnarray}

\noindent
In the case of identical incoming and outgoing particles like $W^+W^- \to H^+H^-$, we get
\begin{eqnarray}
&&p_{1,2}=\frac{\sqrt{s}}{2} (1,0,0,\pm \kappa_W)\nonumber\\
&& k_{1,2}=\frac{\sqrt{s}}{2} (1,\pm \kappa_{H^\pm} \sin\theta,0,\pm \kappa_{H^\pm} \cos\theta)\nonumber
\end{eqnarray}
where $\theta$, now, is the scattering angle between the $W^+$ and $H^+$, and $\kappa_X= \sqrt{1- 4\frac{m_X^2}{s}}$.
The Mandelstam variables become
\begin{eqnarray}
&& s=(p_1+p_2)^2=(k_1+k_2)^2\nonumber\\
&& t=(p_1-k_1)^2 = (p_2-k_2)^2=-\frac{s}{2}+m_{H^\pm}^2+m_W^2+\frac{s}{2} \kappa_{H^\pm} \kappa_W \cos\theta\nonumber\\
&& u=(p_1-k_2)^2 = (p_2-k_1)^2=-\frac{s}{2}+m_{H^\pm}^2+m_W^2-\frac{s}{2} \kappa_{H^\pm} \kappa_W \cos\theta\nonumber\\
&& s+t+u=2m_{H^\pm}^2+2m_W^2
\end{eqnarray}

\section{Amplitude for $V_1V_2 \to S_1S_2$}
\label{app:ampl_VBF}
In what follows, we use the Feynman gauge for computing the corresponding scattering amplitude for all $V_1V_2 \to S_1S_2$ processes. 

\subsection{$W^+W^- \to H^+H^-$}
\label{app:VBF_HpHm}
The total tree-level scattering amplitude for the  subprocess $W^+W^- \to H^+H^-$ is obtained by adding all four contribution types, already shown in Fig.\ref{mesfig:fig1}, and reads
\begin{eqnarray}
\mathcal{M}^{WW \to H^+H^-} = \big[\mathcal{M}_Q^{WW}+\mathcal{M}_h^{WW}+\mathcal{M}_{\gamma}^{WW}+\mathcal{M}_Z^{WW}+\mathcal{M}_{H}^{WW}+\mathcal{M}_A^{WW} \big]  \epsilon_W^\mu(p_1) \epsilon_W^{*\nu}(p_2), \label{eq: total_amp_1}
\end{eqnarray} 
where $\epsilon$'s are the polarisations of the incoming $W^\pm$ gauge bosons, and 
\begin{eqnarray}
\mathcal{M}_Q^{WW}&=&\frac{e^2}{2 s_W^2} g_{\mu \nu}   \nonumber\\
\mathcal{M}_h^{WW}&=& 2 \lambda_3 \frac{m_W^2}{(s-m_h^2)}g_{\mu \nu}   \nonumber\\
 \mathcal{M}_{Qh}^{WW}&=& g_{Qh} g_{\mu \nu}   \nonumber\\
\mathcal{M}_{\gamma}^{WW}&=& \frac{e^2}{s} (k_1-k_2)^\delta \big( (-p_1+p_2)^\rho  g_{\mu \nu}  + (p_1+k_1+k_2)_\nu  g_{\mu}^{\rho} + (-p_2-k_1-k_2)_\mu g_{\nu}^{\rho} \big) g_{\rho \delta}  \\
\mathcal{M}_{Z}^{WW}&=&\frac{e^2 (c_W^2-s_W^2)}{2 s_W^2 (s-m_Z^2) } (k_1-k_2)^\delta \big( (-p_1+p_2)^\rho  g_{\mu \nu} + (p_1+k_1+k_2)_\nu  g_{\mu}^{\rho} + (-p_2-k_1-k_2)_\mu g_{\nu}^{\rho} \big) g_{\rho \delta}   \nonumber\\
\mathcal{M}_{\gamma Z}^{WW}&=& g_{\gamma Z} (k_1-k_2)^\delta \big( (-p_1+p_2)^\rho  g_{\mu \nu}  + (p_1+k_1+k_2)_\nu  g_{\mu}^{\rho} + (-p_2-k_1-k_2)_\mu g_{\nu}^{\rho} \big) g_{\rho \delta}   \nonumber\\
\mathcal{M}_{H}^{WW}&=& \frac{e^2 }{4 s_W^2}   \frac{1 }{t-m_H^2 } \big(p_2-2k_2\big)_\nu \big(-p_2-k_1+k_2\big)_\mu    \nonumber\\
\mathcal{M}_{A}^{WW}&=&  \frac{e^2 }{4 s_W^2 }   \frac{1}{t-m_A^2 } \big(p_2-2k_2\big)_\nu \big(-p_2-k_1+k_2\big)_\mu   \nonumber\\
\mathcal{M}_{HA}^{WW}&=&  g_{HA} \big(p_2-2k_2\big)_\nu \big(-p_2-k_1+k_2\big)_\mu   \nonumber
\end{eqnarray}
\begin{figure}[!h]
\centering
\includegraphics[scale=0.40]{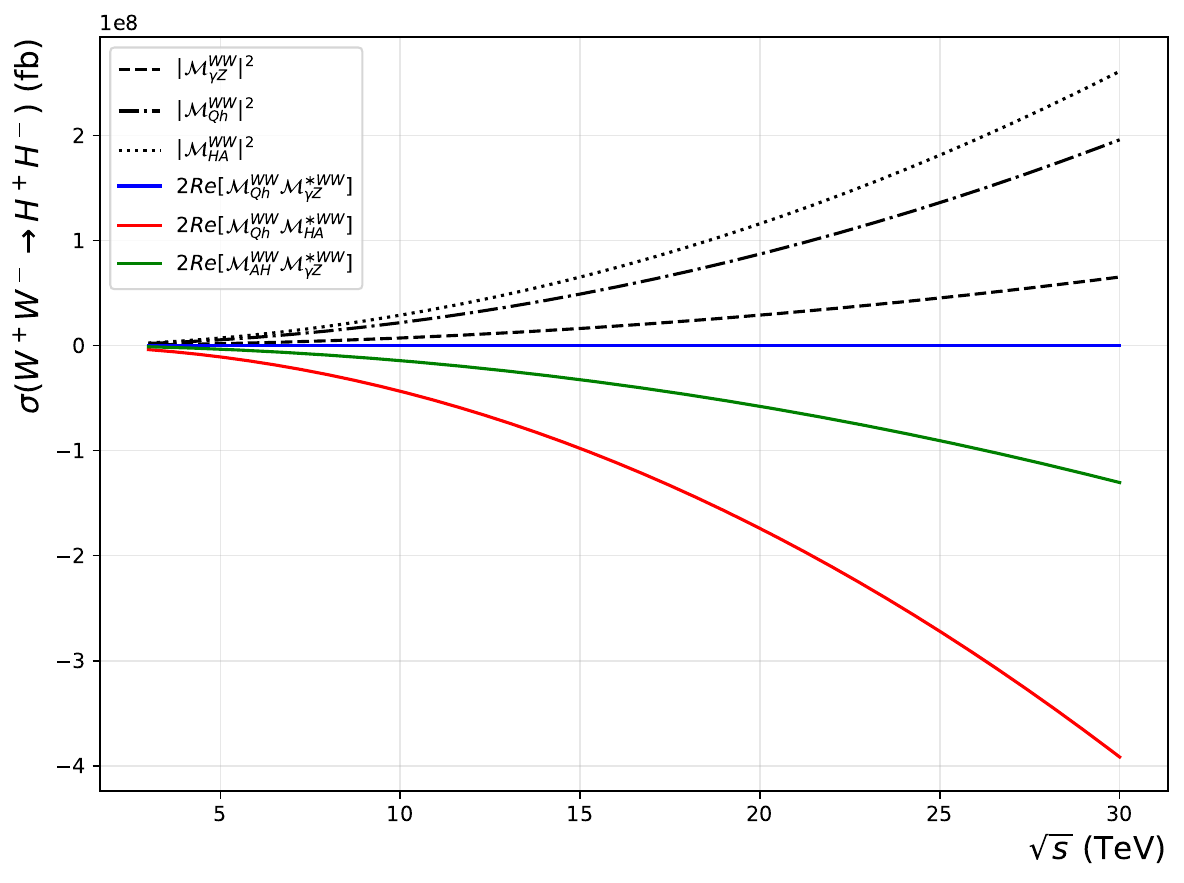}
\includegraphics[scale=0.39]{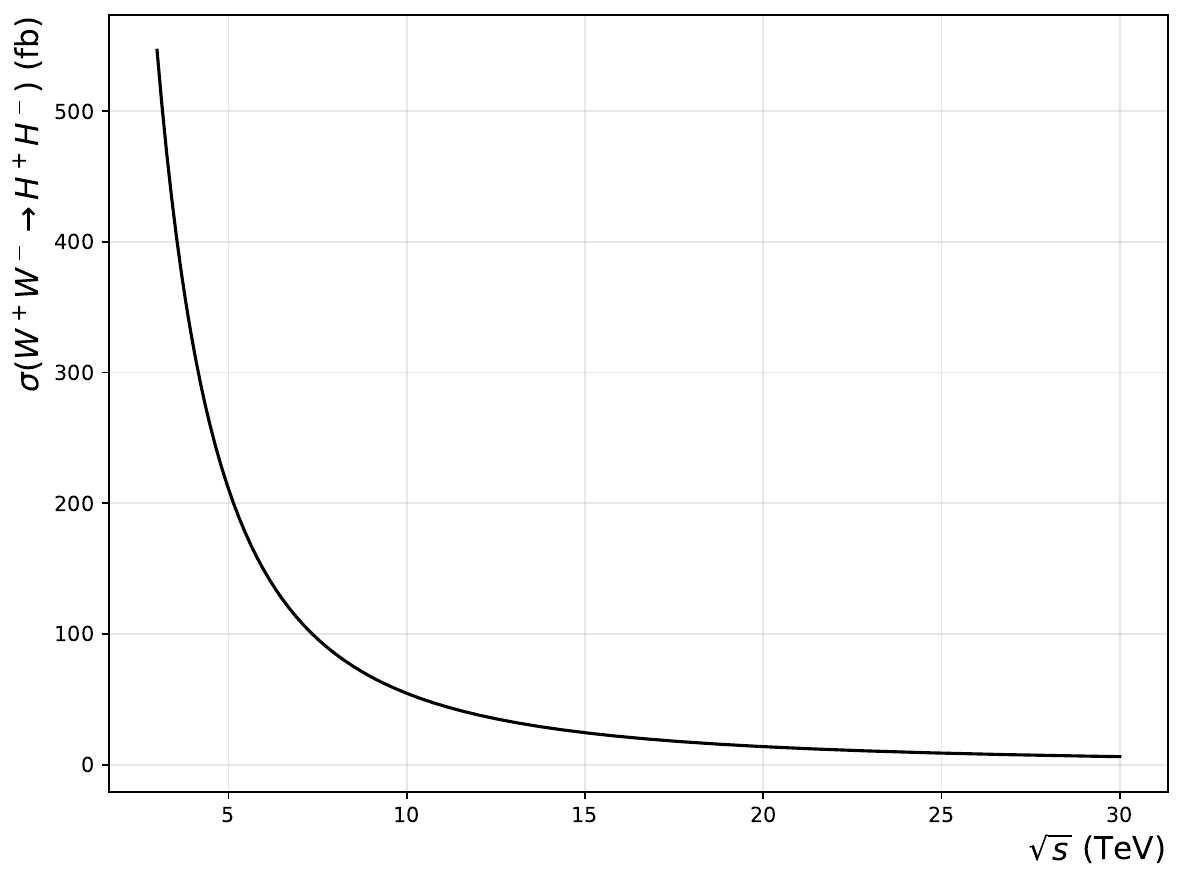}
\caption{Left: individual contributions to the $W^+W^- \rightarrow H^+H^-$ scattering cross section as a function of the center-of-mass energy $\sqrt{s}$, together with the corresponding interference terms. Right: resulting total cross section after summing all contributions and interference effects.}
\label{mesfig:fig13}
\end{figure}  
where
\begin{eqnarray}
&& g_{Qh} =  \frac{e^2}{2 s_W^2} + 2 \lambda_3 \frac{m_W^2}{(s-m_h^2)} \nonumber\\
&& g_{\gamma Z} =  \frac{e^2 (c_W^2-s_W^2)}{2 s_W^2 (s-m_Z^2) }+  \frac{e^2}{s}  \nonumber\\
&& g_{HA} = \frac{e^2 }{4 s_W^2 (t-m_{H^\pm}^2) }  + \frac{e^2 }{4 s_W^2 (t-m_A^2) } \nonumber
\end{eqnarray}

We stress that $ \mathcal{M}_{Qh}^{WW}$ is the sum of $\mathcal{M}_Q^{WW}$ and $\mathcal{M}_h^{WW}$, $\mathcal{M}_{\gamma Z}^{WW}$ is the sum of 
$\mathcal{M}_{\gamma}^{WW}$ and $\mathcal{M}_{Z}^{WW}$, and $\mathcal{M}_{HA}^{WW}$ is the sum of $ \mathcal{M}_{H}^{WW}$ and $\mathcal{M}_{A}^{WW}$.

Fig.~\ref{mesfig:fig13} shows the pure quartic+$h$ contribution $|\mathcal{M}_{Qh}^{WW}|^2$, the gauge contribution $|\mathcal{M}_{\gamma Z}^{WW}|^2$, and the inert scalar exchange contribution $|\mathcal{M}_{HA}^{WW}|^2$, together with their interference terms. As can be seen, large cancellations among the various amplitudes are clearly observed, especially between the quartic/scalar and inert scalar exchange contributions, ensuring a proper high-energy behaviour of the total amplitude. Hence, the total cross section decreases with increasing $\sqrt{s}$ as exhibited in the right panel of Fig.~\ref{mesfig:fig13}.

\subsection{$W^+Z \to H^+ S_i$ \, ($S_i=H$ or $A$) }
\label{app:VBF_WZ}
For the tree-level subprocess $W^+Z \to H^+ H,\,H^+ A$, the scattering amplitude can be written as
\begin{eqnarray}
\mathcal{M}^{W^+Z \to H^+H}= \big[ \mathcal{M}_Q^{WZ}+\mathcal{M}_G^{WZ}+\mathcal{M}_W^{WZ}+\mathcal{M}_{A}^{WZ}+\mathcal{M}_{H^\pm}^{WZ} \big] \epsilon_W^\mu(p_1) \epsilon_Z^{*\nu}(p_2),
\end{eqnarray}
where the involved contributions read
\begin{eqnarray}
\mathcal{M}_Q^{WZ}&=&-\frac{e^2}{c_W} g_{\mu \nu} \,,   \nonumber\\
\mathcal{M}_G^{WZ}&=& - \frac{e^2 (m_H^2-m_{H\pm}^2)}{2 s_W (s-m_W^2)}g_{\mu \nu}  \,,\nonumber\\
\mathcal{M}_{QG}^{WZ}&=& \bigg[-\frac{e^2}{2c_W} + \frac{-e^2  (m_H^2-m_{H\pm}^2)}{2 s_W (s-m_W^2)}\bigg]  g_{\mu \nu}  =b_{QG}  g_{\mu \nu}  \,,\nonumber\\
\mathcal{M}_{W}^{WZ}&=& \frac{e^2 c_W}{2 s_W^2(s-m_W^2)} (k_1-k_2)^\delta \Big[ \big(p_1-p_2\big)^\rho  g_{\mu \nu}  - \big(p_1+k_1+k_2\big)_\nu  g_{\mu}^{\rho} + \big(p_2+k_1+k_2\big)_\mu g_{\nu}^{\rho} \Big] g_{\rho \delta}  \,,  \nonumber\\
\mathcal{M}_{A}^{WZ}&=& \frac{e^2}{4 c_W s_W^2 (t-m_{A}^2)} \big(p_2-2 k_2\big)_\nu \big(p_2+k_1-k_2\big)_\mu \,, \nonumber\\
\mathcal{M}_{H^\pm}^{WZ}&=&  \frac{e^2(c_W^2-s_W^2)}{4 c_W s_W^2(u-m_{H^\pm}^2)}   \big(p_2-2k_2\big)_\nu \big(-p_2-k_1+k_2\big)_\mu \,.
\end{eqnarray}
We note that $\mathcal{M}_{QG}^{WZ}$ is the sum of $\mathcal{M}_{Q}^{WZ}$ and $\mathcal{M}_{G}^{WZ}$.

\begin{figure}[!h]
\centering
\includegraphics[scale=0.40]{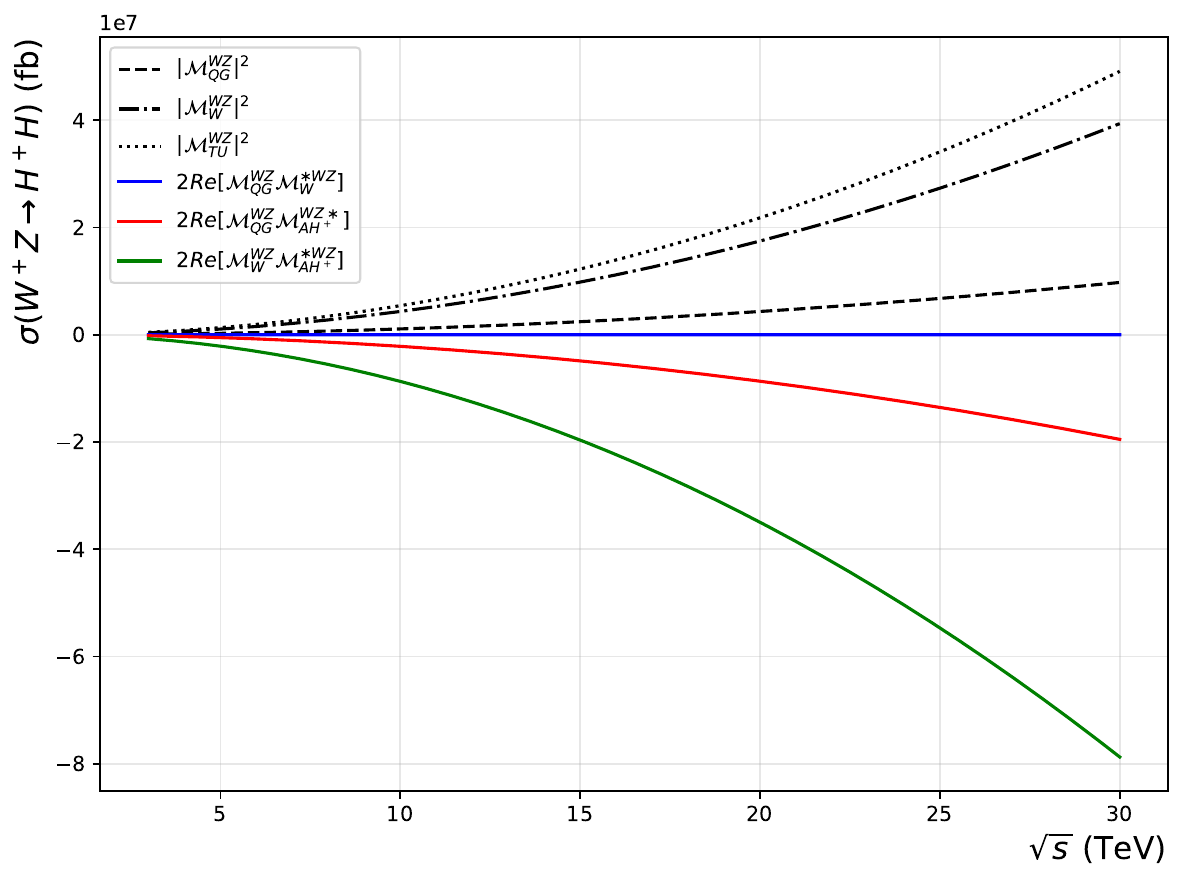}
\includegraphics[scale=0.39]{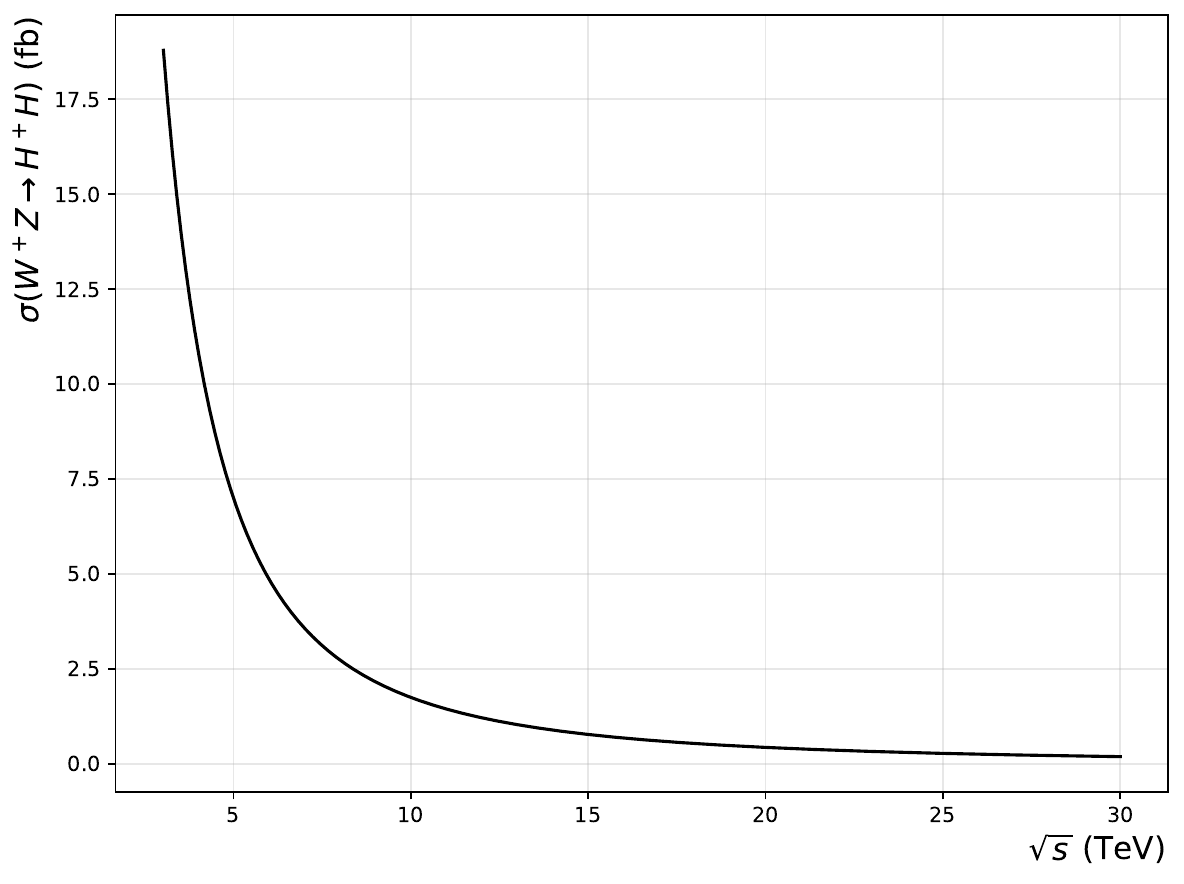}
\caption{Left: individual contributions to the $W^+Z \rightarrow H^+H$ scattering cross section as a function of the center-of-mass energy $\sqrt{s}$, together with the corresponding interference terms. Right: the resulting total cross section after summing all contributions and interference effects.}
\label{mesfig:fig14}
\end{figure}  

Fig.~\ref{mesfig:fig14} illustrates the individual contributions to the $W^+Z \rightarrow H^+H$ scattering cross section together with their interference terms as a function of the center-of-mass energy. As can be seen from the left panel, the quartic+$G^+$-exchange contribution $|\mathcal{M}_{QG}^{WZ}|^2$, the $W$-exchange contribution $|\mathcal{M}_{W}^{WZ}|^2$, and the $A/H^+$ $t/u$-channel contributions $|\mathcal{M}_{AH^+}^{WZ}|^2$ all increase rapidly with energy when considered separately. However, these large positive contributions are accompanied by sizeable destructive interference terms, in particular between the quartic+$G^+$ and $A/H^+$ amplitudes, as well as between the $W$ and $A/H^+$ amplitudes. The interference involving the quartic+$G^+$ and $W$ diagrams remains comparatively small over the entire energy range. As a consequence of these cancellations, the large individual energy-growing terms almost completely compensate each other, leading to the moderate total cross section displayed in the right panel of Fig.~\ref{mesfig:fig14}. The latter decreases monotonically with increasing $\sqrt{s}$ from above $300$~fb at $\sqrt{s}=3$~TeV to $\approx 1$~fb at $\sqrt{s}=30$~TeV.

\begin{figure}[!h]
\centering
\includegraphics[scale=0.40]{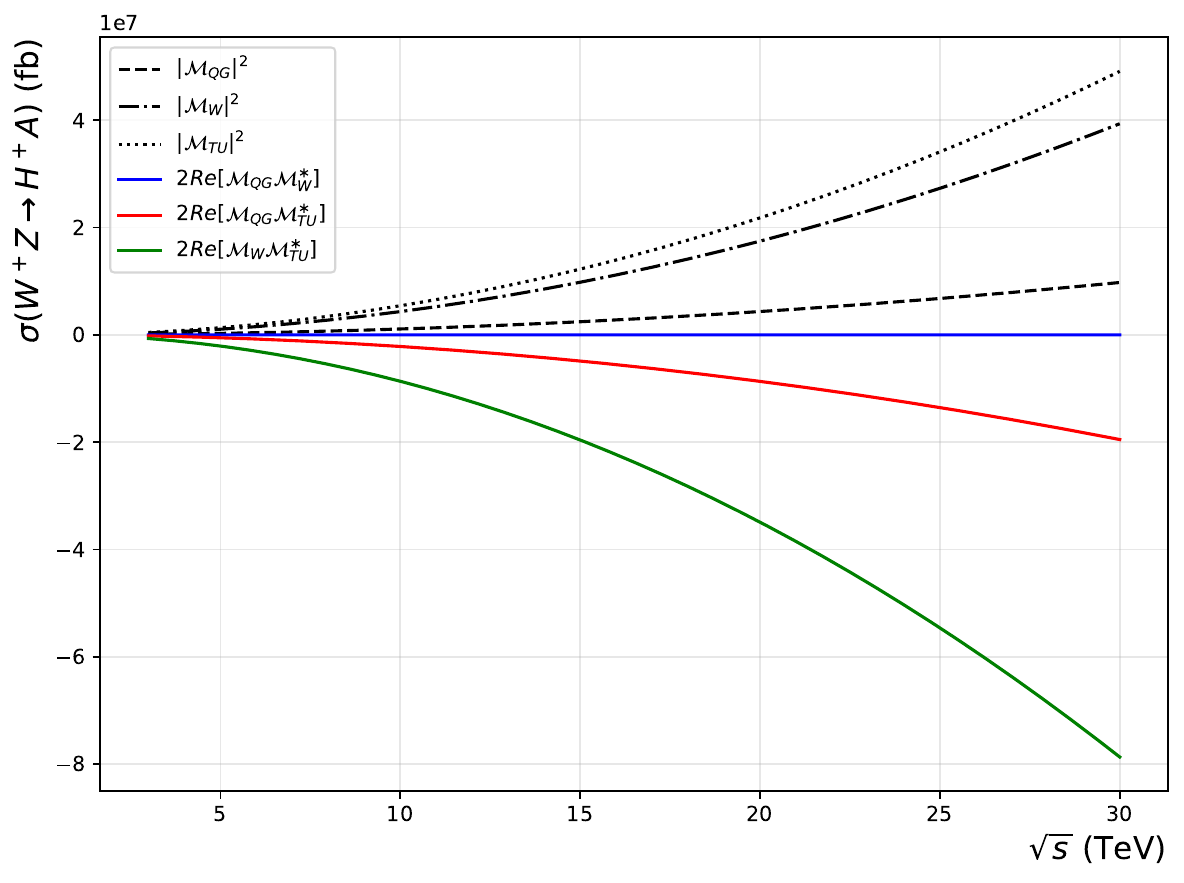}
\includegraphics[scale=0.39]{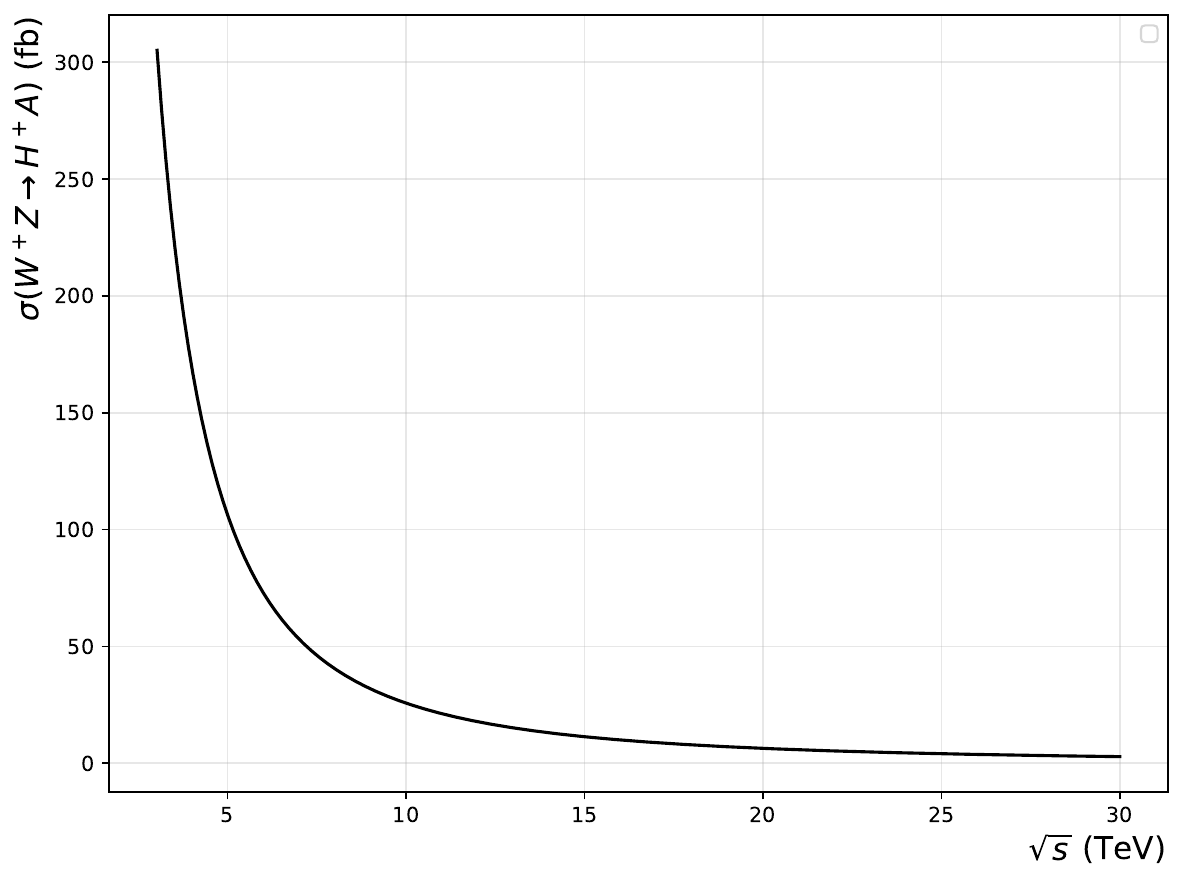}
\caption{Left: individual contributions to the $W^+Z \rightarrow H^+A$ scattering cross section as a function of the center-of-mass energy $\sqrt{s}$, together with the corresponding interference terms. Right: the resulting total cross section after summing all contributions and interference effects.}
\label{mesfig:fig15}
\end{figure}

Next, we turn to the $W^+Z \rightarrow H^+A$ process, where the analytical calculation of the squared amplitudes and 
interference terms follows the same procedure as that presented for the above $W^+Z \rightarrow H^+H$ process. Hence, the explicit expressions for these quantities can be obtained straightforwardly by multiplying by an overall factor $(-i)$ and performing the following replacements: $m_H \longrightarrow m_A$ and $1/(t - m_H^2) \longrightarrow 1/(t - m_A^2)$,
together with the corresponding substitutions in the propagators and kinematic invariants wherever appropriate.
The corresponding tree-level amplitude is expressed by
\begin{eqnarray}
\mathcal{M}^{W^+Z \to H^+ A}= \big[\mathcal{M}_Q^{WZ}+\mathcal{M}_G^{WZ}+\mathcal{M}_W^{WZ}+\mathcal{M}_{H}^{WZ}+\mathcal{M}_{H^\pm}^{WZ} \big] \epsilon_W^\mu(p_1) \epsilon_Z^{*\nu}(p_2)
\end{eqnarray}
where the individual amplitudes read
\begin{eqnarray}
\mathcal{M}_Q^{WZ}&=& -i\frac{-e^2}{2c_W} g_{\mu \nu}    \nonumber\\
\mathcal{M}_G^{WZ}&=& -i \frac{-e^2 (m_A^2- m_{H\pm}^2)}{c_W (s-m_W^2)}g_{\mu \nu}   \nonumber\\
\mathcal{M}_{QG}^{WZ}&=& \bigg[-\frac{e^2}{2c_W} + \frac{-e^2 (m_A^2- m_{H\pm}^2) }{ c_W (s-m_W^2)} \bigg]  g_{\mu \nu}  \nonumber\\
\mathcal{M}_{W}^{WZ}&=& i \frac{c_W e^2}{2 s_W^2(s-m_W^2)} (k_1-k_2)^\delta \Big[ (p_2-p_1)^\rho  g_{\mu \nu}  + \big(p_1+k_1+k_2\big)_\nu  g_{\mu}^{\rho} - \big(p_2+k_1+k_2\big)_\mu g_{\nu}^{\rho} \Big]  g_{\rho \delta}    \nonumber\\
\mathcal{M}_{H}^{WZ}&=& i\frac{e^2 }{4 c_W s_W^2(t-m_{H}^2)} \big(p_2-2 k_2\big)_\nu \big(-p_2-k_1+k_2\big)_\mu \nonumber\\
\mathcal{M}_{H^\pm}^{WZ}&=&  -i\frac{e^2(c_W^2-s_W^2)}{4 c_W s_W^2 (u-m_{H^\pm}^2)}\big(p_2-2k_2\big)_\nu \big(-p_2-k_1+k_2\big)_\mu 
\end{eqnarray}
where $\mathcal{M}_{QG}^{WZ}=\mathcal{M}_{Q}^{WZ}+\mathcal{M}_{G}^{WZ}$. The individual contributions for this process, as well as the total cross section, are shown in Fig.~\ref{mesfig:fig15}.

\subsection{$W^+\gamma \to H^+ S_i$ \, ($S_i=H$ or $A$) }
\label{app:VBF_Wga}
The tree-level amplitude for the subprocess $W^+\gamma \rightarrow H^+H$ can be written as
\begin{eqnarray}
\mathcal{M}^{W^+\gamma \to H^+H}=\big[ \mathcal{M}_Q^{W\gamma}+\mathcal{M}_G^{W\gamma}+\mathcal{M}_W^{W\gamma}+\mathcal{M}_{H^\pm}^{W\gamma} \big] \epsilon_W^\mu(p_1)\,\epsilon_\gamma^\nu(p_2),
\end{eqnarray}
where the individual contributions are given by
\begin{eqnarray}
\mathcal{M}_Q^{W\gamma} &=& -\frac{e^2}{2s_W} g_{\mu \nu} , \nonumber\\
\mathcal{M}_G^{W\gamma} &=& - \frac{e^2 (m_H^2-m_{H\pm}^2)}{ 2 s_W (s-m_W^2)}g_{\mu \nu}, \nonumber\\
\mathcal{M}_{QG}^{W\gamma}&=& \bigg[-\frac{e^2}{2s_W} - \frac{e^2 (m_H^2-m_{H\pm}^2)}{ 2 s_W (s-m_W^2)} \bigg]  g_{\mu \nu} =c_{QG} g_{\mu \nu}   \nonumber\\
\mathcal{M}_W^{W\gamma}
&=& \frac{e^2}{2s_W(s-m_W^2)} (k_1-k_2)^\delta \Big[(p_1-p_2)^\rho g_{\mu\nu} -(p_1+k_1+k_2)_\nu g_\mu^{\ \rho}+(p_2+k_1+k_2)_\mu g_\nu^{\ \rho} \Big] g_{\rho\delta},\nonumber\\
\mathcal{M}_{H^\pm}^{W\gamma} &=& -\frac{e^2} {2s_W(u-m_{H^\pm}^2)} (2k_1-p_2)_\nu (p_2-k_1-k_2)_\mu = c_{H^\pm}\,(2k_1-p_2)_\nu (p_2-k_1-k_2)_\mu .
\end{eqnarray}
where $\mathcal{M}_{QG}^{W\gamma}=\mathcal{M}_{Q}^{W\gamma}+\mathcal{M}_{G}^{W\gamma}$.

Similarly, the tree-level amplitude for the subprocess $W^+\gamma \to H^+A$ is given by
\begin{eqnarray}
\mathcal{M}^{W^+\gamma \to H^+A}=\big[ \mathcal{M}_Q^{W\gamma}+\mathcal{M}_G^{W\gamma}+\mathcal{M}_W^{W\gamma}+\mathcal{M}_{H^\pm}^{W\gamma} \big] \epsilon_W^\mu(p_1)\,\epsilon_\gamma^\nu(p_2),
\end{eqnarray}
where 
\begin{eqnarray}
\mathcal{M}_Q^{W\gamma} &=& -i\frac{-e^2}{2 s_W} g_{\mu \nu}, \nonumber\\
\mathcal{M}_G^{W\gamma} &=& -i\frac{-e^2 (m_A^2-m_{H\pm}^2)  }{ 2 s_W (s-m_W^2)}g_{\mu \nu}, \nonumber\\
\mathcal{M}_W^{W\gamma}
&=& \frac{i\,e^2}{2s_W(s-m_W^2)} (k_1-k_2)^\delta \Big[(p_1-p_2)^\rho g_{\mu\nu} -(p_1+k_1+k_2)_\nu g_\mu^{\ \rho}+(p_2+k_1+k_2)_\mu g_\nu^{\ \rho} \Big] g_{\rho\delta},\nonumber\\
\mathcal{M}_{H^\pm}^{W\gamma} &=& -i\,\frac{e^2} {2s_W(u-m_{H^\pm}^2)} (2k_1-p_2)_\nu (p_2-k_1-k_2)_\mu .
\end{eqnarray}
The amplitude is obtained from the $W^+\gamma \to H^+ H$ one by replacing $m_H$ with $m_A$ and multiplying by $-i$, to take into account the $CP$-odd nature of the inert Higgs boson $A$.

\begin{figure}[!h]
\centering
\includegraphics[scale=0.4]{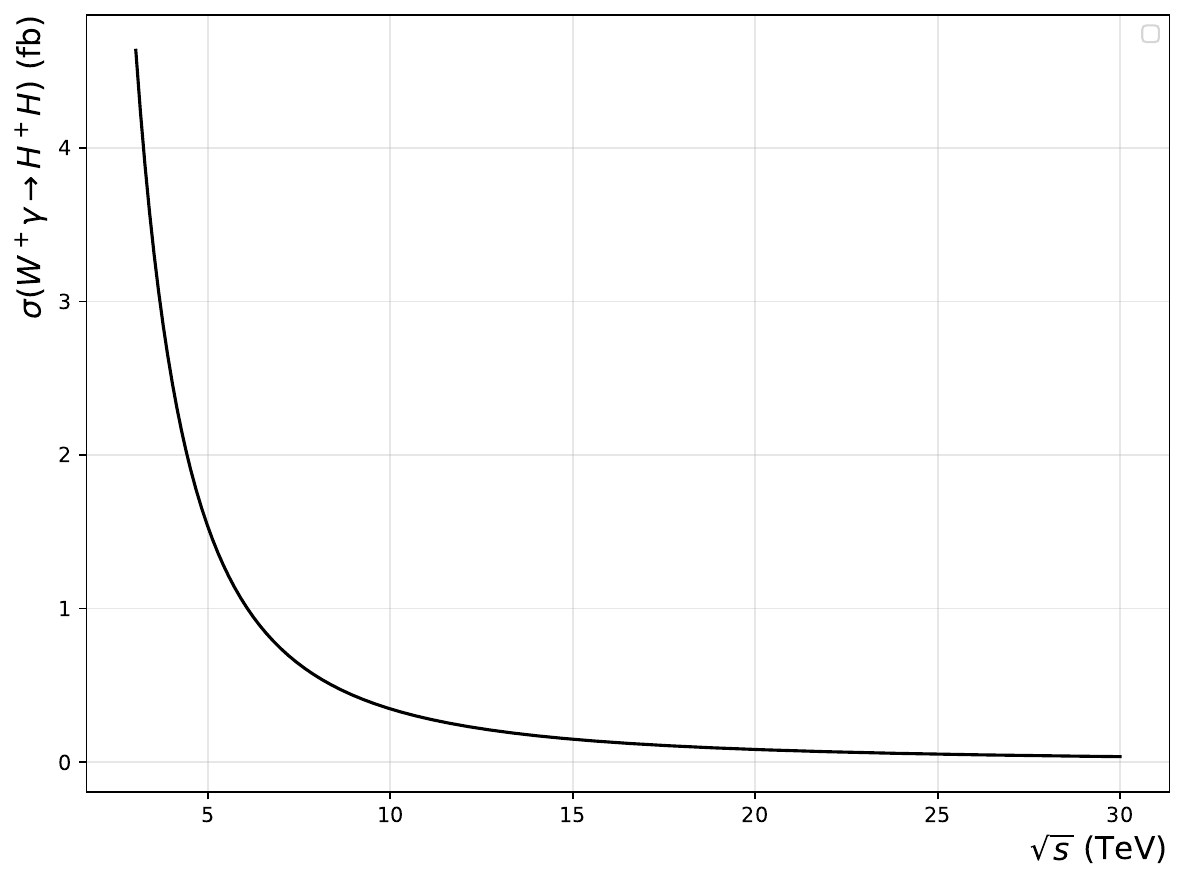}
\includegraphics[scale=0.4]{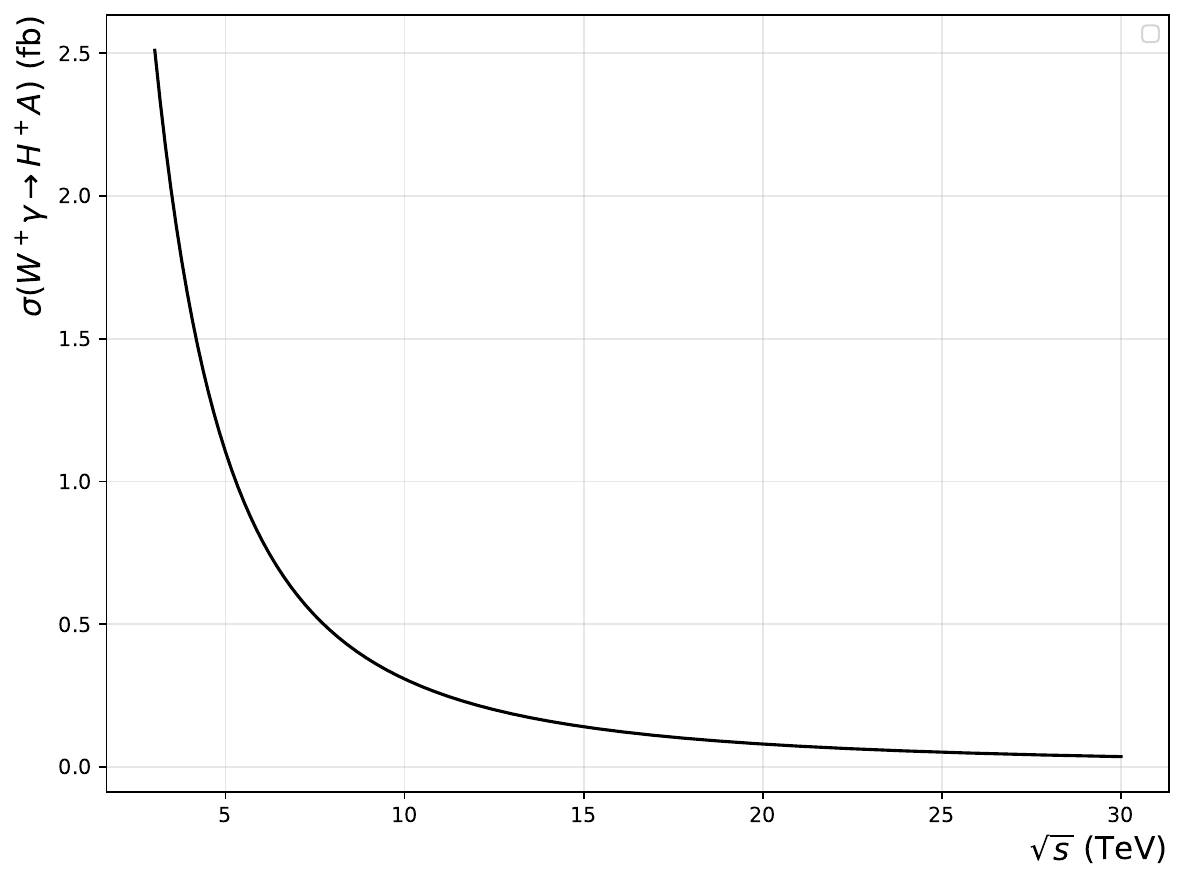}
\caption{Total tree-level cross section for the subprocesses $W^+\gamma \to H^+H$ (left) and $W^+\gamma \to H^+A$ (right) as a function of the center-of-mass energy $\sqrt{s}$.}
\label{mesfig:fig16}
\end{figure}  

Fig.~\ref{mesfig:fig16} shows the energy dependence of the tree-level cross sections for the subprocesses $W^+\gamma \to H^+H$ and $W^+\gamma \to H^+A$. In both cases, the production rates decrease rapidly as the center-of-mass energy increases, reaching values well below $0.1$~fb at multi-TeV energies, indicating that the total amplitude does not grow with energy after the gauge cancellations among the contact, gauge-boson exchange, and charged-Higgs exchange diagrams are taken into account. The larger rate observed for the $H^+H$ final state compared with $H^+A$ originates from the different electroweak couplings entering the corresponding amplitudes and their interference.

\clearpage
\bibliographystyle{JHEP}
\bibliography{bibliography}
\end{document}